\documentclass[10pt,aps,showpacs,nofootinbib,prd,aps,epsf,floats,
               amsmath,amssymb,amsfonts,axodraw,floatfix,graphicx,twocolumn]{revtex4}
\usepackage{amsmath, amssymb}
\newcommand{\mathsym}[1]{{}}

\usepackage{graphicx}% Include figure files
\usepackage{amsmath}
\usepackage{amssymb}
\usepackage{bm}% bold math
\newcommand{\be}{\begin{equation}}
\newcommand{\ee}{\end{equation}}
\newcommand{\bea}{\begin{eqnarray}}
\newcommand{\eea}{\end{eqnarray}}

\newsavebox{\PSLASH}
 \sbox{\PSLASH}{$p$\hspace{-1.8mm}/}
 
\renewcommand{\theequation}{\thesection.\arabic{equation}}
\newcounter{saveeqn}
\newcommand{\add}{\addtocounter{equation}{1}}
\newcommand{\alpheqn}{\setcounter{saveeqn}{\value{equation}}%
\setcounter{equation}{0}%
\renewcommand{\theequation}{\mbox{\thesection.\arabic{saveeqn}{\alph{equation}}}}}
\newcommand{\reseteqn}{\setcounter{equation}{\value{saveeqn}}%
\renewcommand{\theequation}{\thesection.\arabic{equation}}}

 \newsavebox{\notrightarrow}
 \sbox{\notrightarrow}{$\to$\hspace{-4mm}/}
 
 \newsavebox{\PARTIALSLASH}
 \sbox{\PARTIALSLASH}{$\partial$\hspace{-1.6mm}/}
 
 \newsavebox{\ASLASH}
 \sbox{\ASLASH}{$A$\hspace{-2.1mm}/}
 
 \newsavebox{\KSLASH}
 \sbox{\KSLASH}{$k$\hspace{-1.8mm}/}
 
 \newsavebox{\LSLASH}
 \sbox{\LSLASH}{$\ell$\hspace{-1.8mm}/}
 
 \newsavebox{\QSLASH}
 \sbox{\QSLASH}{$q$\hspace{-1.8mm}/}
 
 \newsavebox{\DSLASH}
 \sbox{\DSLASH}{$D$\hspace{-2.2mm}/}
 
 \newsavebox{\DbfSLASH}
 \sbox{\DbfSLASH}{${\mathbf D}$\hspace{-2.8mm}/}
 
 \newsavebox{\DELVECRIGHT}
 \sbox{\DELVECRIGHT}{$\stackrel{\rightarrow}{\partial}$}
 
 \newcommand{\blue}{\IfColor{\textCadetBlue}{}}
\newcommand{\black}{\IfColor{\textBlack}{}}
\newcommand{\red}{\IfColor{\textRed}{}}
\newcommand{\green}{\IfColor{\textOliveGreen}{}}
\newcommand{\lila}{\IfColor{\textRedViolet}{}}

\newcommand{\rd}[1]{\mathop{{d}#1}}

\newcommand{\deriv}[2]{\frac{{d}#1}{{d}#2}}
\newcommand{\derivv}[1]{\frac{\mathop{{d}}}{\rd#1}}

\newcommand{\at}[2][]{#1|_{#2}}
\newcommand*{\derivn}[3][]{\ensuremath{\frac{{d}^{#1} #2}{{d} #3^{#1}}}}

\newcommand{\inv}[1]{\frac{1}{#1}}

\newcommand{\tder}[1]{\mathfrak{D}#1}

\begin{document}
\title{Novel self-similar rotating solutions of
non-ideal transverse magnetohydrodynamics}
\author{M. Shokri}\email{m\_shokri@physics.sharif.ir}
\author{N. Sadooghi}\email{sadooghi@physics.sharif.ir}
\affiliation{Department of Physics, Sharif University of Technology,
P.O. Box 11155-9161, Tehran, Iran}

%%%%%%%%%%%%%%%%%%%%%%%%%%%%%%%%%%%%%%
\begin{abstract}
The evolution of electromagnetic and thermodynamic fields in a non-ideal fluid are studied in the framework of ultrarelativistic transverse magnetohydrodynamics (MHD), which is essentially characterized by electric and magnetic fields being transverse to the fluid velocity. Extending the method of self-similar solutions of relativistic hydrodynamics to the case of non-conserved charges, the differential equations of non-ideal transverse MHD are solved, and two novel sets of self-similar solutions are derived. The first set turns out to be a boost-invariant and exact solution, which is characterized by non-rotating electric and magnetic fields. The second set is a non-boost-invariant solution, which is characterized by rotating electric and magnetic fields. The rotation occurs with increasing rapidity $\eta$, as the angular velocity is defined by $\omega_{0}\equiv\frac{\partial\zeta}{\partial\eta}=\frac{\partial\phi}{\partial\eta}$, with $\zeta$ and $\phi$ being the angles of electric and magnetic vectors with respect to a certain axis in the local rest frame of the fluid. For both sets of solutions, the electric and magnetic fields are either parallel or anti-parallel to each other.
Performing a complete numerical analysis, the effects of finite electric conductivity as well as electric and magnetic susceptibilities of the medium on the evolution of rotating and non-rotating MHD solutions are explored, and the interplay between the angular velocity $\omega_{0}$ and these quantities is scrutinized. The lifetime of electromagnetic fields and the evolution of the temperature of the electromagnetized fluid are shown to be affected by $\omega_{0}$.
\end{abstract}
\maketitle

%%%%%%%%%%%%%%%%%%%%%%%%%%%%%%%%%%%%%%
\section{Introduction}\label{Introduction}\label{sec1}
%%%%%%%%%%%%%%%%%%%%%%%%%%%%%%%%%%%%%%
One of the most significant achievements of relativistic hydrodynamics (RHD) in recent years, is in the ability to reproduce experimental data from relativistic Heavy Ion Collisions (HICs) at RHIC\footnote{Relativistic Heavy Ion Collider (RHIC).} and LHC\footnote{Large Hadron Collider (LHC).}. In particular, the elliptic flow data of low to intermediate transverse momenta for almost all particle species and for various centralities, beam energies and colliding nuclei are successfully described by RHD model calculations, performed with realistic initial conditions and the Equation of State (EoS) for relativistic HICs \cite{teaney2001,kolb2001}. These studies have already led to the RHIC discovery that the Quark-Gluon Plasma (QGP) created in relativistic HICs is a strongly-coupled nearly perfect fluid \cite{shuryak2003,mclerran2004,gyulassy2005,romatschke2007} (for a review see, e.g., \cite{teany2009}).
\par
The non-linear differential equations describing RHD are remarkably complex. Their major characteristic is, however, that they do not contain any internal scale. Using this special feature, a large class of exact, self-similar solutions of relativistic hydrodynamics have been found in recent years. Motivated by the seminal work by Landau \cite{landau1953, landau1956} and Khalatnikov \cite{khalatnikov1954}, who presented the first exact one-dimensional implicit solution of RHD, R. C. Hwa \cite{hwa1974} and J. D. Bjorken \cite{bjorken1982} found independently an explicit analytic solution for RHD equations in the ultrarelativistic limit.
This solution, referred to as Bjorken-flow, represents a one-dimensional, longitudinally boost-invariant solution of relativistic ideal hydrodynamics (RIHD).
Other analytic and self-similar exact solutions of RIHD are presented in \cite{bondorf1978} and \cite{csorgo1998,biro2000}, where, in particular, a three-dimensional expanding Gaussian fireball is described, which exhibits a Hubble-type linear radial flow. These solutions are then generalized to one- and three-dimensional solutions, exhibiting various cylindrical, spheroidal and ellipsoidal symmetries. They describe, in particular, the evolution of the fireball with or without rotation and acceleration (for a recent review see \cite{kodama2015} and references therein). In combination with the EoS, arising from lattice QCD\footnote{Quantum Chromodynamics (QCD).}, the main goal is, among others, to describe various physical observables in relation with HIC experiments by these exact analytical solutions. They shall therefore reflect various symmetry properties of HICs before and at hadro-chemical freeze-out stage \cite{csorgo2016}.
\par
An important feature of non-central HICs is the generation of very strong magnetic fields in the early stage of the collisions. Depending on the impact parameter and collision energy, their strengths are estimated to be $eB\sim 1.5 m_{\pi}^{2}$ at RHIC and $eB\sim 15 m_{\pi}^{2}$ at LHC \cite{warringa2007, skokov2009}, with the pion mass $m_{\pi}=0.14$ GeV.\footnote{$eB=1$ GeV$^{2}$ corresponds to a magnetic field strength $B\sim 1.7\times  10^{20}$ Gau\ss.} The magnetic field created at HICs is time-dependent, and rapidly decays after $\tau\sim 1-2$ fm/c. However, as it is argued in \cite{tuchin2013,rajagopal2014, zakharov2014}, due to the relatively large electric conductivity of the QGP medium, the external magnetic field is essentially frozen, and its decay is thus substantially delayed.
Most theoretical studies deal therefore with the idealized limit of constant and homogeneous magnetic fields.\footnote{See \cite{fayazbakhsh2010,fayazbakhsh2011} for a complete analysis of the effect of constant magnetic fields on QCD phase diagram, including magnetic catalysis \cite{klimenko1992,miransky1995} and inverse magnetic catalysis effects \cite{bali2011}, and \cite{taghinavaz2016,munshi2016,ayala2016} in relation with the effect of constant magnetic fields on various particle production rates in HICs. See also \cite{andersen2014,shovkovy2015} for recent reviews on the effect of constant magnetic fields on Quark Matter.}
\par
One of the possibilities to explore the dynamics of external electromagnetic fields is the relativistic magnetohydrodynamics (MHD). Recently, MHD methods are used to study the effect of magnetic fields created in HICs on the evolution of the energy density of QGP.
A one-dimensional, longitudinally boost-invariant solution of ultrarelativistic ideal MHD is presented in \cite{rischke2015, rischke2016}. Here, the external magnetic field is assumed to be transverse to the fluid velocity. In \cite{rischke2015}, it is found that in the ideal MHD limit, where, in particular, the electric conductivity of the medium is assumed to be infinitely large, the (proper) time evolution of the energy density is the same as in the case without any magnetic field. This remarkable result can be best understood by the well-known ``frozen-flux theorem'' \cite{plasma}, which states that the ratio $B/s$, with $B$ the magnetic field and $s$ the entropy density, is conserved, and the magnetic field is thus advected with the fluid, and evolves therefore as $B(\tau)\propto \tau^{-1}$ with $\tau\equiv (t^{2}-z^{2})^{1/2}$ being the proper time. The deviation from the frozen-flux theorem is also imposed in \cite{rischke2015, rischke2016} by a parameterized power-law \textit{Ansatz} for the evolution of the magnetic field, $B(\tau)\propto \tau^{-a}$, where $a$ is an arbitrary free parameter. It is shown that the decay of the energy density depends on whether $a>1$ or $a<1$. The additional effect of a constant (temperature-independent) magnetic susceptibility on the energy density of QGP is studied within the same ideal transverse MHD framework in \cite{rischke2016}.
 In \cite{pu2016}, the aforementioned power-law decay Ansatz of $B$ is generalized to a power-law decay of magnetic fields with spatial inhomogeneity, characterized by a Gaussian distribution in one of the transverse directions.
\par
It is the purpose of this paper to study the dynamics of electromagnetic and thermodynamic fields within a one-dimensional ultrarelativistic non-ideal MHD framework with electric and magnetic fields being transverse to the fluid velocity (hereafter non-ideal transverse MHD).
We will present novel non-boost-invariant, self-similar solutions for electromagnetic and thermodynamic fields, appearing in non-ideal transverse MHD with finite electric conductivity $\sigma$ and electric as well as magnetic susceptibilities, $\chi_{e}$ and $\chi_{m}$. Using the method presented in, e.g., \cite{csorgo2002}, where a certain self-similar, non-accelerating exact solution of RIHD is presented, we will first show that the boost-invariant solution $B(\tau)\propto \tau^{-1}$, derived in \cite{rischke2015,rischke2016}, is a self-similar exact solution which naturally arises in ideal transverse MHD. Here, apart from the exact solution for $B$, satisfying the aforementioned frozen flux theorem, self-similar, exact and non-boost-invariant solutions for thermodynamic fields, such as temperature $T$, entropy and number densities, $s$ and $n$, arise.
To go beyond the ideal limit of infinite electric conductivity \cite{rischke2015,rischke2016}, we will extend the method used in \cite{csorgo2002} to the case of non-conserved charges. We will solve the corresponding MHD equations, combined with homogeneous and inhomogeneous Maxwell equations. By appropriately parameterizing these equations in terms of the magnitudes of the electromagnetic fields, $E$ and $B$, as well as the angles $\zeta$ and $\phi$ of $\mathbf{E}$ and $\mathbf{B}$ with respect to a certain axis in the local rest frame (LRF) of the fluid, we will show that two series of solutions arise, which are particularly characterized by vanishing and non-vanishing angular velocity of $\mathbf{E}$ and $\mathbf{B}$, $\omega_{0}$, defined by $\omega_{0}\equiv \frac{d\zeta}{d\eta}=\frac{d\phi}{d\eta}$. Here, $\eta\equiv \frac{1}{2}\ln\frac{t+z}{t-z}$ is the rapidity.
\par
For vanishing angular velocity, non-rotating, boost-invariant, self-similar and analytic solutions for $\mathbf{B}$ and $\mathbf{E}$ will be derived. They are shown to be either parallel or anti-parallel with respect to each other. Their magnitude are given by $B(\tau)\propto \tau^{-1}$ and $E\propto \tau^{-1}\exp\left(-f(\sigma,\chi_{e})\tau\right)$, where $f$ is a certain (positive) function of $\sigma$ and $\chi_{e}$. As concerns the case of non-vanishing angular velocity of the magnetic and electric vectors, we will derive approximate analytical as well as numerical solutions for $\mathbf{B}$ and $\mathbf{E}$. This will be done by solving a second-order and quadratic differential equation for a certain function ${\cal{M}}(\tau)$, describing, in particular, the deviation of the dynamics of the magnetic field from the frozen flux theorem. It arises in our method of self-similar solutions for non-conserved charges (see below). We will show, that a power-law solution $B(\tau)\propto \tau^{-a}$, similar to the one previously used in \cite{rischke2015,rischke2016},\footnote{In contrast to the power-law solution which will be derived in the present paper, the power-law decay Ansatz used in \cite{rischke2015,rischke2016} turns out to be a solution of transverse MHD, where, in particular, $\sigma\to \infty$, and thus $E\to 0$.} naturally arises as one of the approximate analytical solutions to this differential equation, where, in particular, the ratio $E/B$ is assumed to be constant in $\tau$.
Here, the power $a$ in $B(\tau)\propto \tau^{-a}$ will be shown to be expressed in terms of the angular velocity $\omega_{0}$, which, by its part, turns out to be a function of $\sigma,\chi_{e}$ and $\chi_{m}$. A second series of approximate analytical solution to the above mentioned non-linear differential equation for ${\cal{M}}$ will be also derived. It eventually leads to slowly rotating $\mathbf{B}$ and $\mathbf{E}$ fields. We will present the corresponding self-similar and non-boost-invariant solutions to the temperature $T$ in these approximations. As concerns the non-boost-invariance of the solutions for $\mathbf{E}$ and $\mathbf{B}$, it will be shown that in contrast to the non-boost-invariance of $T$, which reflects itself in the appearance of an $\eta$-dependent scale factor, the non-boost-invariance of electromagnetic fields is particularly characterized by the dependence of the angles $\zeta$ and $\phi$ on the rapidity $\eta$.
\par
We will also numerically solve the aforementioned differential equation for ${\cal{M}}$. The aim is to quantitatively study the effects of free parameters $\sigma,\chi_{e},\chi_{m}$ and $\sigma_{0}\equiv \frac{B_{0}^{2}}{\epsilon_{0}}$ on $B,E$ and $T$. Here, $B_{0}$ and $\epsilon_{0}$ are the magnetic field and energy density of the fluid at the initial (proper) time. The effects of the angular velocity $\omega_{0}$ on the evolution of $B,E,T$, and the interplay between $\omega_{0}$ and other free parameters will be further scrutinized.
We will, in particular, show that the evolution of thermodynamic fields $T$ are strongly affected by rotating and non-rotating solutions to non-ideal transverse MHD equations, as well as the magnetic susceptibility of the medium.
\par
The organization of this paper is as follows: In Sec. \ref{sec2}, we will first apply the method presented in \cite{csorgo2002} on ideal transverse MHD, and derive self-similar, boost-invariant and exact solutions for the number density $n$, temperature $T$, energy density $\epsilon$ and magnetic field $B$. Then, we will  extend the method from \cite{csorgo2002} to non-ideal transverse MHD, where we are, in particular, faced with inhomogeneous continuity equations with the generic form $\partial_{\mu}(fu^{\mu})=f\tder{\lambda}$ for $B$, $E$ and $T$. The corresponding inhomogeneity functions for $f=\{B,E,T\}$ in $\partial_{\mu}(fu^{\mu})=f\tder{\lambda}$ will be denoted in the rest of this paper by $\lambda=\{{\cal{M}},{\cal{N}},{\cal{L}}\}$, respectively.
In Sec. \ref{sec3}, after presenting the necessary definition of the non-ideal transverse MHD with finite $\sigma$, $\chi_{e}$ and $\chi_{m}$, we will derive the corresponding differential equations of MHD and Maxwell equations in terms of variable $B,E$ and $\zeta,\phi$ as well as free parameters $\sigma,\chi_{e}$ and $\chi_{m}$ (Sec. \ref{sec3A}). Using the formal self-similar solutions to the inhomogeneous continuity equations for $B,E$ and $T$, arising from our generalized self-similar method for non-conserved charges (Sec. \ref{sec3B}), we will then combine, in Sec. \ref{sec3C}, the aforementioned differential equations for non-ideal transverse MHD, and arrive, in particular, at the corresponding differential equations to ${\cal{M}}$. We will show that ${\cal{M}}$ satisfies either $\frac{d{\cal{M}}}{du}=0$ with $u\equiv \ln\left(\frac{\tau}{\tau_{0}}\right)$ or a second-order non-linear differential equation. Solutions to these differential equations play a major role in determining ${\cal{N}}$ and ${\cal{L}}$, and thus in determining rotating and non-rotating solutions for $B,E$ and $T$.  In Secs. \ref{sec4A} and \ref{sec4B}, we will introduce the exact and approximate analytical, self-similar non-rotating and rotating solutions of $B,E$ and $T$. Numerical solutions of the second-order, non-linear differential equation corresponding to ${\cal{M}}$ will be presented in Sec. \ref{sec5}.  In Sec. \ref{sec5A}, we will qualitatively compare the space-time evolutions of $B,E$ and $T$ in ideal MHD with their non-rotating and rotating solutions in non-ideal MHD. In Sec. \ref{sec5B}, the reliability of approximate analytical solutions from Sec. \ref{sec3C} will be quantitatively studied. The effects of free sets of parameters $\sigma, \chi_{e},\chi_{m}$ as well as $\omega_{0},\sigma_{0}$ and $\beta_{0}\equiv E_{0}/B_{0}$ on the evolution of $B,E$ and $T$ will be studied in Sec. \ref{sec5C}. Here, $E_{0}$ and $B_{0}$ are the electric and magnetic fields at initial proper time. Although for the choice of free set of parameters, we have strongly oriented ourselves to sets which may be relevant for QGP, the considerations in this paper are quite general, and can be applied to every magnetized fluid with finite electric and magnetic susceptibilities.
Section \ref{sec6} is devoted to a summary of the results and a number of concluding remarks. A general analysis of the solutions of the non-trivial differential equation for ${\cal{M}}$ will be presented in an Appendix.
%%%%%%%%%%%%%%%%%%%%%%%%%%%%%%%%%%%%%%%%%%%%%%%%%%%%%%%
\section{Self-similar solutions of relativistic ideal hydrodynamics and the method of non-conserved charges}\label{sec2}
\setcounter{equation}{0}
%%%%%%%%%%%%%%%%%%%%%%%%%%%%%%%%%%%%%%%%%%%%%%%%%%%%%%%%
As aforementioned, self-similar solutions of RHD are generalizations of the Hwa-Bjorken flow \cite{hwa1974,bjorken1982}. They provide the possibility of non-boost-invariant temperature profiles \cite{csorgo2002}, and are naturally generalized to $3+1$ dimensions \cite{csorgo2003}. In this section, we will briefly review these solutions in $1+1$ dimensions. In this setup, one assumes that the fluid is expanding in time and only one spatial dimension, which, without loss of generality, can be taken to be the $z$-direction. The system is also assumed to possess translational invariance in the transverse plane, i.e. the $x$-$y$ plane. The latter assumption implies that the hydrodynamical fields, such as four-velocity $u^\mu(x)=\gamma(1,\mathbf{v})$ and entropy density $s(x)$, are independent of the transverse coordinates \cite{gubser2010}.\footnote{Here, $g^{\mu\nu}=\mbox{diag}(1,-1,-1,-1)$ and $u_{\mu}u^{\mu}=1$.}  Here, $\gamma=\left(1-v_{z}^{2}\right)^{-1/2}$ is the Lorentz factor. The equations of RIHD, consisting of conservation laws of the energy-momentum tensor of the fluid, $T^{\mu\nu}=(\epsilon+p)u^{\mu}u^{\nu}-pg^{\mu\nu}$ and entropy density current $su^{\mu}$, read
\begin{eqnarray}
\partial_\mu T^{\mu\nu}&=&0,\label{S1}\\
\partial_\mu \left(su^\mu \right )&=&0.\label{S2}
\end{eqnarray}
In $T^{\mu\nu}$, $\epsilon$ and $p$ are the energy density and pressure of the fluid, respectively.
In what follows, we will introduce the method used in \cite{csorgo2002}, where, in particular, general self-similar solutions for the above hydrodynamical fields are found.
\par
To this purpose, let us first consider an arbitrary continuity equation
\begin{equation}\label{S3}
\partial_\mu \left(fu^\mu \right)=0.
\end{equation}
Here, $f(t,z)$ is a conserved quantity such as the entropy density $s$. To determine the self-similar solution to (\ref{S3}), one introduces the scaling parameter $Z(t)$ and the scaling variable $\Theta(z,Z)$, such that
\begin{enumerate}
	\item[$i)$] if $f(t,z)$ is a solution to equation (\ref{S3}), then $f(t,z)\mathcal{F}_f(\Theta)$ is also a solution to the same equation. Here,  $\mathcal{F}_{f}$ is any differentiable function of the parameter $\Theta$, which will be defined below. In addition, if $f(t,z)$ is a positive quantity, then $\mathcal{F}_f$ must be positive as well,
	\item[$ii)$] the longitudinal velocity obeys a Hubble-like expansion law as $v_z = H(t) z$ with $H(t) = \frac{\dot{Z}}{Z}$.
\end{enumerate}
Assumption $i)$ requires
\begin{equation}\label{S4}
\tder{\Theta}=0.
\end{equation}
Here, $\tder\equiv u^{\mu}\partial_\mu$ is the conductive derivative.  Assumption $ii)$ can be used to solve (\ref{S4}) as
\begin{equation}\label{S5}
\Theta = \left(\frac{z}{Z}\right)^\alpha,
\end{equation}
where $\alpha$ is a parameter, which shall be fixed later. Having these in hand, the most general self-similar solution of (\ref{S3}) reads
\begin{equation}\label{S6}
f(t,z) = f_0 \left( \frac{Z_0}{\gamma Z}\right)\mathcal{F}_f(\Theta),
\end{equation}
with $Z_0 = Z(t_0)$ and $f_0=f(t_0,0)$. In (\ref{S6}), $\mathcal{F}_f$ is normalized as $\mathcal{F}_f(0)=1$.
\par
Let us consider again the RIHD equations (\ref{S1}) and (\ref{S2}).
These equations can be closed by incorporating a thermodynamic EoS, which is assumed to be
\begin{equation}\label{S7}
\epsilon=\kappa p,
\end{equation}
with $\kappa=\mbox{const}$.\footnote{For our purposes, it is enough that $\tder{\kappa}=0$.} Plugging (\ref{S7}) into the longitudinal component of (\ref{S1}),
\begin{equation}\label{S8}
\tder{\epsilon}+(\epsilon+p)\theta=0,
\end{equation}
and exploiting the entropy conservation of RIHD from (\ref{S2}), leads to the following continuity equation for the temperature $T$,
\begin{equation}\label{S9}
\partial_\mu(T^{\kappa}u^\mu)=0.
\end{equation}
Here, the thermodynamic relation $\epsilon+p=Ts$ is used.\footnote{The system is assumed to be baryon-free \cite{ollitrault2008}.}
According to (\ref{S6}), the solutions for $T$, $s$ and $p$ then reads \cite{csorgo2002}
\begin{eqnarray}
s(t,z)&=&s_0\left(\frac{Z_0}{Z\gamma}\right)\mathcal{S}(\Theta),\label{S10}\\
T(t,z)&=&T_0\left(\frac{Z_0}{Z\gamma}\right)^{1/\kappa}\mathcal{T}(\Theta),\label{S11}\\
p(t,z)&=&p_0\left(\frac{Z_0}{Z\gamma}\right)^{1+1/\kappa}\mathcal{T}(\Theta)\mathcal{S}(\Theta),\label{S12}
\end{eqnarray}
where $p_{0}=\frac{T_0 s_0}{1+\kappa}$. As concerns the power $\alpha$ in (\ref{S5}), we put (\ref{S12}) into the Euler equation
\begin{eqnarray}\label{S13}
\tder{u_{\mu}}=\frac{1}{\epsilon+p}\nabla_{\mu}p,
\end{eqnarray}
which arises from $\Delta_{\mu\nu}\partial_{\rho}T^{\rho\nu}=0$, with  $\nabla_{\mu}\equiv \Delta_{\mu\nu}\partial^{\nu}$
and  $\Delta_{\mu\nu}\equiv g_{\mu\nu}-u_{\mu}u_{\nu}$, and arrive at an expression for $\ddot{Z}$ in terms of $Z$, $\dot{Z}$ and $\Theta$. The requirement that $\ddot{Z}$ shall be finite leads to $\alpha=2$.\footnote{This result is in line with \cite{csorgo2002}, where $\alpha=2$ is \textit{a priori} assumed.} Moreover, the fact that $Z$ is $\Theta$-independent leads to $\ddot{Z}(\Theta,\cdots)=\ddot{Z}(0,\cdots)$. This, for its part, translates into a second-order equation for $\dot{Z}^2$ whose coefficients are only functions of $\Theta$. It thus has a solution of the form\footnote{The functional form of $\dot{Z}(\Theta)$ does not matter.}
\begin{equation}\label{S14}
\dot{Z}^2 = \dot{Z}^2(\Theta).
\end{equation}
Exploiting, at this stage, the $\Theta$ independence of $Z$ requires $\dot{Z}$ to be constant, and thus $\ddot{Z}=0$. This immediately results in vanishing of the proper acceleration, $\tder{u_\mu}=0$, and the emergence of the Hwa-Bjorken velocity profile $v_z = z/t$.
\par
Since $\tder{\gamma}=\frac{\partial\gamma}{\partial\tau}=0$, one is able to introduce new scaling functions
\begin{align}
\mathcal{U}(\Theta)&=\frac{\mathcal{S}(\Theta)}{\gamma^2},\label{S16}\\
\mathcal{V}(\Theta)&=\frac{\mathcal{T}(\Theta)}{\gamma^{2/\kappa}}.\label{S17}
\end{align}
Using $\tder{u_{\mu}}=0$, we arrive at the boost-invariance ($\eta$ independence) of $p$ and automatically at $\mathcal{U}\mathcal{V}=1$. If the process is isentropic \cite{ollitrault2008}, then $s$ and the number density $n$ share the same scaling function, and the ideal gas equation $p=nT$ holds.\footnote{The ideal gas equation which was used as an assumption in the original derivation of the solutions in \cite{csorgo2002}, seems not to be required for the case of baryon-free RIHD.} The latter can then be used to give the final results for $n,T$ and $p$,
\begin{eqnarray}
n&=&n_0\left(\frac{\tau_0}{\tau}\right)\mathcal{U}\left(\frac{\tanh^2\eta}{\dot{Z}_0^2}\right),\label{S18}\\
T&=&T_0\left(\frac{\tau_0}{\tau}\right)^{1/\kappa}\mathcal{U}^{-1}\left(\frac{\tanh^2\eta}{\dot{Z}_0^2}\right),\label{S19}\\
p&=&p_0\left(\frac{\tau_0}{\tau}\right)^{1+1/\kappa}.\label{S20}
\end{eqnarray}
Here, we have introduced  the Milne coordinates $z=\tau\sinh\eta$ and $t=\tau\cosh\eta$, with $\tau=\left(t^{2}-z^{2}\right)^{1/2}$ and $
\eta= \frac{1}{2}\ln\frac{t+z}{t-z}$ being the proper time and space-time rapidity, respectively.\footnote{Let us note that in these coordinates $\tder=u^{\mu}\partial_{\mu}$,  $\theta\equiv \partial_{\mu}u^{\mu}$ and $\nabla_{\mu}$ translate into $\tder=\frac{\partial}{\partial\tau}$,  $\theta=\frac{1}{\tau}$ and $\nabla_{\mu}=-\frac{1}{\tau}\left(\sinh\eta,0,0,\cosh\eta\right)\frac{\partial}{\partial\eta}$.} To derive (\ref{S18})-(\ref{S20}),  $Z(t)=\dot{Z}_{0}t$ is used.
The evolution of the energy density $\epsilon$ is determined by plugging (\ref{S20}) into (\ref{S8}). It is given by
\begin{eqnarray}\label{S21}
\epsilon=\epsilon_0\left(\frac{\tau_0}{\tau}\right)^{1+1/\kappa},
\end{eqnarray}
with $\epsilon_0=\kappa p_{0}$. Let us note that although the velocity and pressure profile are the same as the Hwa-Bjorken solution, but the temperature, number and entropy densities have an arbitrary rapidity dependence through the factors ${\cal{U}}\left(\frac{\tanh^2\eta}{\dot{Z}_0^2}\right)$. It is also worth to mention that although in deriving (\ref{S9}) $\kappa$ was only assumed to have vanishing conductive derivative, the treatment of the Euler equation was based on $\kappa$ being a constant.
\par
Transverse MHD is previously studied in \cite{rischke2015, rischke2016}. It is found that the Hwa-Bjorken solution for the energy density (and temperature profile) is also valid for $1+1$ dimensional ideal MHD.\footnote{In ideal MHD, apart from hydrodynamical dissipative effects, the resistivity of the medium is assumed to vanish.} This is not surprising since ideal MHD has no extra energy dissipation channel in addition to RIHD, and the energy equation (\ref{S8}) still holds. Therefore any solution of the RIHD energy density holds in ideal MHD as well. It thus seems that self-similar solutions of  thermodynamic fields in RIHD are automatically generalized to the case of ideal MHD. However, the boost-invariance needs extra care. As it is shown in the following sections, in the transverse MHD setup \cite{rischke2015} electrical charge density vanishes. The inspection of the equation of motion for the fluid parcels shows that proper acceleration gains therefore no contribution from electromagnetic fields. Since the equation of motion is linear, one can take the proper acceleration to remain zero by superposition. As we will show, any additive term in the Euler equation becomes boost-invariant when proper acceleration $\tder{u_{\mu}}$ vanishes. The results of self-similar solutions of thermodynamic fields in RIHD are thus generalized to the case of ideal MHD.
\par
However, in the treatment of MHD one is not only concerned about the evolution of hydrodynamical fields, but also wants to know how the electromagnetic fields evolve as observed in the LRF of the fluid. As it turns out, the frozen flux theorem of ideal MHD \cite{rischke2015} is translated into another continuity equation for the magnitude of the local magnetic field, $B$,
\begin{eqnarray}\label{S22}
\partial_{\mu}(Bu^{\mu})=0.
\end{eqnarray}
The most general self-similar solution of $B$ will be therefore given by
\begin{eqnarray}\label{S23}
B=B_{0}\left(\frac{\tau_{0}}{\tau}\right){\cal{B}}\left(\frac{\tanh^{2}\eta}{\dot{Z}_{0}^{2}}\right).
\end{eqnarray}
Here, as in the case of pressure in the RIHD case, the $\eta$-dependent scaling factor ${\cal{B}}=1$, because, by inspecting the Euler equation, it turns out that an additional term $B^2$ appears on the right hand side (r.h.s.) of the Euler equation (\ref{S13}) (see also Sec. 3). The fact that $\tder{u_{\mu}}=0$ thus leads to the boost-invariance ($\eta$ independence) of $B$, or equivalently to ${\cal{B}}=1$.
\par
In non-ideal transverse MHD, when one relaxes the assumption of infinite conductivity, things becomes more complicated. The frozen flux (\ref{S22}) and the continuity equation for the temperature (\ref{S9}) are then violated, and the evolution of the electric field becomes also important. In this case, we have found it useful to introduce a method to solve the non-ideal MHD equations, which for further convenience will be referred to as \textit{the method of non-conserved charges}. Here, one basically considers a non-conserved charge $f(t,z)$, which satisfies
\begin{equation}\label{S24}
\partial_\mu\left(fu^\mu\right) = f\tder{\lambda},
\end{equation}
with $\lambda=\lambda(t,z)$ a differentiable function of space and time. From (\ref{S24}), one finds
\begin{equation}\label{S25}
\partial_\mu\left(fu^\mu \exp{(-\lambda)}\right) = 0,
\end{equation}
leading to
\begin{equation}\label{S26}
f(t,z)=f_0 \left(\frac{Z_0}{\gamma Z}\right)\exp{(\lambda-\lambda_0)}\mathbb{F}_f(\Theta),
\end{equation}
with $\lambda=\lambda(t,z)$ and $\lambda_0 = \lambda(t_0,0)$. One can add any function $g(t,z)$ to $\lambda$ as long as $\tder{g}=0$. Specifically, any differentiable function of $\Theta$ can be added to $\lambda$. The resulting factor $\exp{(g-g_0)}$ can however be absorbed into $\mathbb{F}_f$, as a purely $\eta$-dependent function. In a uniformly expanding fluid, where $v_z$ satisfies the Hwa-Bjorken profile $v_{z}=\frac{z}{t}$, the final result of the non-conserved charge $f$ can thus be given in terms of $\tau$ and $\eta$ as
\begin{eqnarray}\label{S27}
\hspace{-1cm}f(\tau,\eta)=f_{0}\left(\frac{\tau_0}{\tau}\right)\exp\left(\lambda-\lambda_0\right)\mathbb{U}\left(\frac{\tanh^{2}\eta}{\dot{Z}_{0}^{2}}\right),
\end{eqnarray}
where $\lambda=\lambda(\tau)$ and $\lambda_0=\lambda(\tau_0)$. Without loss of generality, we set $\lambda_{0}=0$ in the rest of this work.
In the next sections, we will apply this method to non-ideal MHD, and, in particular,  find a master equation that governs the deviation of an electromagnetized non-ideal fluid from the frozen flux theorem. The latter leads to the solutions of the equations of transverse MHD in some specific cases. We will show that the $\eta$ dependence of the relative angle of the $B$ field with a certain axis in the LRF of the fluid may distinguish between various solutions of this master equation.
%%%%%%%%%%%%%%%%%%%%%%%%%%
\section{Relativistic Magnetohydrodynamics}\label{sec3}
\setcounter{equation}{0}
%%%%%%%%%%%%%%%%%%%%%%%%%%
In this section, we will first focus on transverse MHD in $1+1$ dimensions, and will introduce the main definitions and a number of useful relations (Sec. \ref{sec3A}). To be brief, we will only consider the case of non-ideal magnetized fluid with finite magnetization $M$, electric polarization $P$ and electric conductivity $\sigma$. Taking the limit $\sigma\to\infty$ as well as $M,P\to 0$, the case of ideal MHD can be retrieved. We will compare the results of ideal and non-ideal fluid whenever necessary. Apart from energy and Euler equations, we will consider the homogeneous and non-homogeneous Maxwell equations.  Combining these equations, we will derive in Sec. \ref{sec3C} the aforementioned master equation, whose solutions will be explored in Sec. \ref{sec4}. The aim is to use the method of non-conserved charges in order to determine the space-time evolution of thermodynamic quantities $n,T,p,\epsilon$ as well as those of electric and magnetic fields $E^{\mu}$ and $B^{\mu}$. Formal self-similar solutions to these fields are presented in \ref{sec3B}.
%%%%%%%%%%%%%%%%%%%%%%%%%%
\subsection{Transverse MHD: Definitions and useful relations}\label{sec3A}
%%%%%%%%%%%%%%%%%%%%%%%%%%
A locally equilibrated relativistic fluid in $1+1$ dimensions is characterized by the four-velocity $u_{\mu}=\gamma(1,0,0,v_{z})$, which is defined by the variation of the four coordinate $x^{\mu}=(t,\mathbf{x})$ with respect to proper time $\tau=(t^{2}-z^{2})^{1/2}$ and satisfies $u_{\mu}u^{\mu}=1$. Continuity equations
\begin{align}\label{A1}
\partial_{\mu}(nu^{\mu})=0,\quad
\partial_{\mu}T^{\mu\nu}=0,\quad \partial_{\mu}J^{\mu}=0,
\end{align}
then govern the dynamics of the fluid. Here, $n$ is the baryonic number density and $T^{\mu\nu}$ and $J^{\mu}$ are the total energy momentum tensor and electric current, respectively.
\par
In the presence of electromagnetic fields, $T^{\mu\nu}$ is given by a combination of the fluid and electromagnetic energy-momentum tensor, $T_{F}^{\mu\nu}$ and $T^{\mu\nu}_{EM}$, as
\begin{eqnarray}\label{A2}
T^{\mu\nu}=T_{F}^{\mu\nu}+T^{\mu\nu}_{EM},
\end{eqnarray}
with\footnote{Apart from electric conductivity, other dissipative effects, such as shear and bulk viscosities, will not be considered in this paper.}
\begin{eqnarray}\label{A3}
T_{F}^{\mu\nu}&=&(\epsilon+p) u^{\mu}u^{\nu}-p g^{\mu\nu}
\nonumber\\
&&-\frac{1}{2}(M^{\mu\lambda}F_{\lambda}^{~\nu}+M^{\nu\lambda}F_{\lambda}^{~\mu}),
\end{eqnarray}
and
\begin{eqnarray}\label{A4}
T^{\mu\nu}_{EM}=-F^{\mu\lambda}F^{\nu}_{~\lambda}+\frac{1}{4}F^{\rho\sigma}F_{\rho\sigma}.
\end{eqnarray}
The antisymmetric field strength and polarization tensors, $F^{\mu\nu}$ and $M^{\mu\nu}$ are defined by
\begin{eqnarray}\label{A5}
\hspace{-1cm}F^{\mu\nu}&=& E^{\mu}u^{\nu}-E^{\nu}u^{\mu}-\epsilon^{\mu\nu\alpha\beta}B_{\alpha}u_{\beta},\nonumber\\
\hspace{-1cm}M^{\mu\nu}&=&-\chi_{e}(E^{\mu}u^{\nu}-E^{\nu}u^{\mu})-\chi_m\epsilon^{\mu\nu\alpha\beta}B_{\alpha}u_{\beta},
\end{eqnarray}
where $\epsilon^{\mu\nu\alpha\beta}$ is the totaly antisymmetric Levi-Civita symbol,\footnote{Here, $\epsilon^{0123}=-\epsilon_{0123}=1$.} and the four-vector of electric and magnetic fields are given by $E^{\mu}\equiv F^{\mu\nu}u_{\nu}$ and $B^{\mu}\equiv \frac{1}{2}\epsilon^{\mu\nu\alpha\beta}F_{\nu\alpha}u_{\beta}$. They satisfy $E^{\mu}E_{\mu}=-E^{2}$ and $B^{\mu}B_{\mu}=-B^{2}$. In the LRF of the fluid, with $u^{\mu}=(1,\mathbf{0})$, we have $E^{\mu}=(0,\mathbf{E})$ and $B^{\mu}=(0,\mathbf{B})$. Moreover, using the definitions of $E^{\mu}$ and  $B^{\mu}$ in terms of $F^{\mu\nu}$, we arrive at $u_{\mu}E^{\mu}=0$ and $u_{\mu}B^{\mu}=0$. Combining  these relations with $\mathbf{v}\cdot \mathbf{E}=0$ as well as $\mathbf{v}\cdot\mathbf{B}=0$, which are valid in $1+1$ dimensional transverse MHD, we have, in particular, $E^{\mu}=(0,E_x,E_y,0)$ as well as $B^{\mu}=(0,B_x,B_{y},0)$. For later convenience, we will parameterize $E^{\mu}$ and $B^{\mu}$ in terms of the magnitudes of the fields, $E$ and $B$, as well as the relative angles of $\mathbf{E}$ and $\mathbf{B}$ fields with respect to the $x$-axis in the LRF of the fluid, $\zeta$ and $\phi$
\begin{eqnarray}\label{A6}
E^{\mu}&=&(0,E\cos\zeta,E\sin\zeta,0),\nonumber\\
B^{\mu}&=&(0,B\cos\phi,B\sin\phi,0).
\end{eqnarray}
The antisymmetric polarization tensor $M^{\mu\nu}$ in (\ref{A4}) describes the response of the system to an applied electromagnetic field. Assuming a linear response from the medium, the electric and magnetic susceptibilities $\chi_{e}$ and $\chi_{m}$ are defined by $\chi_{e}\equiv P/E$ and $\chi_{m}\equiv M/B$, where $P$ and $M$ are given by $P^{2}=-P^{\mu}P_{\mu}$ and $M^{2}=-M^{\mu}M_{\mu}$, with the electric polarization $P^{\mu}\equiv -M^{\mu\nu}u_{\nu}$ and magnetization $M^{\mu}\equiv \frac{1}{2}\epsilon^{\mu\nu\alpha\beta}M_{\nu\alpha}u_{\beta}$. In this paper, $\partial_{\mu}\chi_{e}=0$ and $\partial_{\mu}\chi_{m}=0$ as well as $\chi_{e}\neq -1$ are assumed.
\par
The electromagnetic field strength tensor satisfies the homogeneous Maxwell equation
\begin{eqnarray}\label{A7}
\partial_{\mu}\tilde{F}^{\mu\nu}=0,
\end{eqnarray}
with
$\tilde{F}^{\mu\nu}\equiv \frac{1}{2}\epsilon^{\mu\nu\alpha\beta}F_{\alpha\beta}$, or equivalently,
\begin{eqnarray}\label{A8}
\hspace{-0.8cm}\tilde{F}^{\mu\nu}&=& B^{\mu}u^{\nu}-B^{\nu}u^{\mu}+\epsilon^{\mu\nu\alpha\beta}E_{\alpha}u_{\beta},
\end{eqnarray}
and the inhomogeneous Maxwell equation
\begin{eqnarray}\label{A9}
\partial_{\mu}F^{\mu\nu}=J^{\nu},
\end{eqnarray}
with the electromagnetic current
\begin{eqnarray}\label{A10}
J^{\mu}=\rho_{e}u^{\mu}+\partial_{\rho}M^{\rho\mu}+\sigma E^{\mu}.
\end{eqnarray}
Here, $\rho_{e}$ is the electric charge density, and $\partial_{\rho}M^{\rho\mu}$ is the magnetization current. Differentiating  (\ref{A9}) with respect to $x^{\nu}$ leads to the third continuity equation $\partial_{\nu}J^{\nu}=0$ in (\ref{A1}). Contracting further  (\ref{A9}) with $u_\nu$ leads to $\rho_e=0$. Exploring the equation of motion of the fluid parcel shows that the proper acceleration, $\tder{u_{\mu}}$, vanishes once $\rho_{e}=0$, and, according to the arguments presented in Sec. \ref{sec2}, this leads to the Hwa-Bjorken velocity profile $v_{z}=z/t$. Setting $\tder{u_{\mu}}=0$ on the left hand side (l.h.s.) of the Euler equation,
\begin{widetext}
\begin{eqnarray}\label{A11}
\tder{u_{\mu}}=\frac{1}{\epsilon+p+(1+\chi_{e})E^{2}+(1-\chi_{m})B^{2}}\nabla_{\mu}\left(p-\chi_{m}B^{2}
+\frac{1}{2}\left(E^{2}+B^{2}\right)\right),
\end{eqnarray}
\end{widetext}
we arrive at the boost-invariance ($\eta$ independence) of $p$ as well as of $E$ and $B$. Let us note that (\ref{A11}) arises from $\Delta_{\mu\nu}\partial_{\rho}T^{\rho\nu}=0$, with $T^{\mu\nu}$ defined in (\ref{A2}).
\par
In what follows, we will derive a number of useful relations, which will help us to determine the space and time evolution of $n,T,p,\epsilon$ as well as $E$ and $B$.
Let us first consider the homogeneous Maxwell equation (\ref{A7}). Plugging the definition of $\tilde{F}^{\mu\nu}$ from (\ref{A8}) into this equation, and using the fact that in a non-accelerating expansion
\begin{eqnarray}\label{A12}
u^{\mu}=\left(\cosh\eta,0,0,\sinh\eta\right),
\end{eqnarray}
we arrive for $\nu=1$ and $\nu=2$ at
\begin{eqnarray}\label{A13}
\hspace{-1cm}&\partial_{\mu}(Bu^{\mu})\cos\phi-B\sin\phi\frac{\partial\phi}{\partial\tau}-\frac{E}{\tau}\cos\zeta\frac{\partial\zeta}{\partial\eta}=0,\nonumber\\
\hspace{-1cm}&\partial_{\mu}(Bu^{\mu})\sin\phi+B\cos\phi\frac{\partial\phi}{\partial\tau}-\frac{E}{\tau}\sin\zeta\frac{\partial\zeta}{\partial\eta}=0,
\end{eqnarray}
respectively. Here, the parametrization (\ref{A6}) for the electromagnetic fields and
\begin{eqnarray}\label{A14}
\frac{\partial}{\partial t}&=&+\cosh\eta\frac{\partial}{\partial\tau}-\frac{1}{\tau}\sinh\eta\frac{\partial}{\partial\eta},\nonumber\\
\frac{\partial}{\partial z}&=&-\sinh\eta\frac{\partial}{\partial\tau}+\frac{1}{\tau}\cosh\eta\frac{\partial}{\partial\eta},
\end{eqnarray}
are used. Combining the relations arising in (\ref{A13}), we arrive at
\begin{eqnarray}\label{A15}
\partial_{\mu}(Bu^{\mu})-\frac{E}{\tau}\cos\delta\frac{\partial\zeta}{\partial\eta}&=&0,\nonumber\\
B\frac{\partial\phi}{\partial\tau}+\frac{E}{\tau}\sin\delta\frac{\partial\zeta}{\partial\eta}&=&0,
\end{eqnarray}
where $\delta\equiv \phi-\zeta$ is the relative angles between $\mathbf{E}$ and $\mathbf{B}$. As concerns the inhomogeneous Maxwell equation, plugging $F^{\mu\nu}$ from (\ref{A5}) into the l.h.s. of (\ref{A9}), we arrive for $J^{\mu}$ from (\ref{A10}) first at
\begin{eqnarray}\label{A16}
(1+\chi_{e})\partial_{\nu}(E^{\mu}u^{\nu})-(1-\chi_{m})\epsilon^{0\mu\nu 3}\frac{1}{\tau}\frac{\partial B_{\nu}}{\partial\eta}+\sigma E^{\mu}=0.\hspace{-0.3cm}\nonumber\\
\end{eqnarray}
Here, we have, in particular, used $\partial\cdot E=0$, $\partial\cdot B=0$ as well as $E\cdot \partial=0$, and $B\cdot \partial=0$, which are valid in $1+1$ dimensional transverse MHD.
For $\mu=1$ and $\mu=2$, (\ref{A16}) then yields
%\begin{widetext}
\begin{eqnarray}\label{A17}
(1+\chi_{e})\partial_{\mu}(Eu^{\mu})\cos\zeta-(1+\chi_{e})E\sin\zeta\frac{\partial\zeta}{\partial\tau}&&\nonumber\\
+(1-\chi_{m})\frac{B}{\tau}\cos\phi\frac{\partial\phi}{\partial\eta}+\sigma E\cos\zeta&=&0,\nonumber\\
(1+\chi_{e})\partial_{\mu}(Eu^{\mu})\sin\zeta+(1+\chi_{e})E\cos\zeta\frac{\partial\zeta}{\partial\tau}&&\nonumber\\
+(1-\chi_{m})\frac{B}{\tau}\sin\phi\frac{\partial\phi}{\partial\eta}+\sigma E\sin\zeta&=&0.\nonumber\\
\end{eqnarray}
%\end{widetext}
Combining these two equations results in
\begin{eqnarray}\label{A18}
(1+\chi_{e})E\frac{\partial\zeta}{\partial\tau}+(1-\chi_{m})\frac{B}{\tau}\sin\delta\frac{\partial\phi}{\partial\eta}&=&0,\nonumber\\
(1+\chi_{e})\partial_{\mu}(Eu^{\mu})+(1-\chi_{m})\frac{B}{\tau}\cos\delta\frac{\partial\phi}{\partial\eta}+\sigma E&=&0.
\nonumber\\
\end{eqnarray}
Using, at this stage, the previously derived boost-invariance of $p,E$ and $B$ in transverse MHD in combination with  $\Delta_{\mu\nu}\partial_{\rho}T^{\rho\nu}_{EM}=
\Delta_{\mu\nu}J_{\rho}F^{\rho\nu}$, we also obtain
\begin{eqnarray}\label{A19}
\hspace{-0.5cm}[\chi_{e}\partial_{\mu}(Eu^{\mu})+\sigma E]\sin\delta=\chi_{e}E\cos\delta\frac{\partial\zeta}{\partial\tau}.
\end{eqnarray}
Another useful relation, which will be used later to determine the evolution of thermodynamic quantities $T,p$ and $\epsilon$ arises from $u_{\nu}\partial_{\mu}T_{F}^{\mu\nu}=-u_{\nu}J_{\mu}F^{\mu\nu}$, and reads
\begin{eqnarray}\label{A20}
\lefteqn{\hspace{-1.5cm}
\tder{\left(\epsilon+\frac{1}{2}\chi_{e}E^{2}\right)}+\theta\left(\epsilon+p-\chi_{m}B^{2}\right)
}\nonumber\\
&=&\sigma E^{2}-\chi_{m}\frac{EB}{\tau}\cos\delta\frac{\partial\phi}{\partial\eta}.
\end{eqnarray}
Here,
\begin{eqnarray}\label{A21}
\epsilon^{0\mu\nu 3}E_{\mu}\frac{\partial B_{\nu}}{\partial\eta}=EB\cos\delta\frac{\partial\phi}{\partial\eta},
\end{eqnarray}
is used.
The full energy equation
\begin{eqnarray}\label{A22}
\lefteqn{\hspace{-0.5cm}
\tder{\bigg[\epsilon+\left(\frac{1}{2}+\chi_{e}\right)E^{2}+\frac{1}{2}B^{2}\bigg]}
}\nonumber\\
&&+\theta\big[\epsilon+p+(1+\chi_{e})E^{2}+(1-\chi_{m})B^{2}\big]=0,\nonumber\\
\end{eqnarray}
is derived from $u_{\nu}\partial_{\mu}T^{\mu\nu}=0$.
In the next section, we will use the relations (\ref{A15}), (\ref{A18}) and (\ref{A19}) to derive a differential equation, whose solution yields the space and time dependence of the $\mathbf{E}$ and $\mathbf{B}$ vectors. In particular, the evolution of the magnitude of these fields, $E=|\mathbf{E}|$ and $B=|\mathbf{B}|$ as well as their relative angles $\zeta$ and $\phi$ with respect to $x$-axis in the LRF of the fluid will be determined as functions of independent coordinates $\tau$ and $\eta$. Moreover, the method of self-similar solutions for non-conserved charges, introduced in Sec. \ref{sec2}, will be used to determine the space-time evolution of thermodynamic quantities $n,T, p$ and $\epsilon$.
%%%%%%%%%%%%%%%%%%%%%%%%%%
\subsection{Formal self-similar solutions for electromagnetic and thermodynamic quantities in non-ideal transverse MHD}\label{sec3B}
%%%%%%%%%%%%%%%%%%%%%%%%%%
In non-ideal transverse MHD, the dynamics of the electromagnetic fields $B$ and $E$ as well as the thermodynamic quantities $n,T,p$ and $\epsilon$ are governed by following homogeneous and inhomogeneous differential equations:
\begin{eqnarray}
\partial_{\mu}(nu^{\mu})&=&0,\label{A23}\\
\partial_{\mu}(T^{\kappa}u^{\mu})&=&T^{\kappa}\tder{\cal{L}},\label{A24}\\
\partial_{\mu}(Bu^{\mu})&=&B\tder{\cal{M}},\label{A25}\\
\partial_{\mu}(Eu^{\mu})&=&E\tder{\cal{N}},\label{A26}
\end{eqnarray}
where functions ${\cal{L}},{\cal{M}}$ and ${\cal{N}}$ are to be determined. Here, the baryonic current $\partial_{\mu}(nu^{\mu})$ is assumed to be conserved. This leads to (\ref{A23}). Whereas the last two equations (\ref{A25}) and (\ref{A26}) are only assumed to be valid at this stage, the second equation (\ref{A24}) arises by assuming the ideal gas equation $p=nT$, as in the previous Sec. \ref{sec2}, and by plugging the EoS (\ref{S7}), with $\kappa$ satisfying $\tder{\kappa}=0$, into (\ref{A20}). Using (\ref{A23}), we thus obtain (\ref{A24}) with
\begin{eqnarray}\label{A27}
\lefteqn{
\tder{{\cal{L}}}=\frac{1}{p}}\nonumber\\
&&\times\left(\sigma E^{2}-\chi_{e}E\frac{\partial E}{\partial\tau}+\chi_{m}B^{2}\theta-\chi_{m}\frac{EB}{\tau}\cos\delta\frac{\partial\phi}{\partial\eta}\right).\nonumber\\
 \end{eqnarray}
As expected, in ideal MHD, with $\sigma E^{2}\to 0$ and $\chi_{e}=\chi_{m}=0$, we have $\tder{{\cal{L}}}=0$. In this case, the self-similar solution of the resulting equation $\partial_{\mu}(T^{\kappa}u^{\mu})=0$ will be given by (\ref{S19}).
\par
Following the method presented in the previous section, the self-similar solution of $n$ reads
\begin{eqnarray}\label{A28}
n(\tau,\eta)=n_{0}\left(\frac{\tau_{0}}{\tau}\right){\cal{U}}\left(\frac{\tanh^{2}\eta}{\dot{Z}_{0}^{2}}\right),
\end{eqnarray}
[see (\ref{S18})].  Here, ${\cal{U}}$ is an arbitrary $\eta$-dependent scaling factor.  Using further the method of non-conserved charges from Sec. \ref{sec2}, the formal solution of (\ref{A24}) reads
\begin{eqnarray}\label{A29}
\hspace{-1cm}T(\tau,\eta)=T_{0}\left(\frac{\tau_{0}}{\tau}\right)^{1/\kappa}e^{\frac{{\cal{L}}}{\kappa}}\mathbb{V}\left(\frac{\tanh^{2}\eta}{\dot{Z}_{0}^{2}}\right),
\end{eqnarray}
[see (\ref{S27})]. Here, $\mathbb{V}={\cal{U}}^{-1}$ guarantees the boost-invariance of $p$, whose formal solution can be derived from the ideal gas equation $p=nT$,
\begin{eqnarray}\label{A30}
p=p_{0}\left(\frac{\tau_{0}}{\tau}\right)^{1+\frac{1}{\kappa}}e^{\frac{{\cal{L}}}{\kappa}}.
\end{eqnarray}
Here, $p_{0}=n_{0}T_{0}$. Using further the EoS $\epsilon=\kappa p$, the formal solution of the energy density is given by
\begin{eqnarray}\label{A31}
\epsilon=\epsilon_{0}\left(\frac{\tau_{0}}{\tau}\right)^{1+\frac{1}{\kappa}}e^{\frac{{\cal{L}}}{\kappa}},
\end{eqnarray}
with $\epsilon_{0}=\kappa p_{0}$.
To determine ${\cal{L}}$ explicitly from (\ref{A27}) the space-time evolutions of $B$ and $E$ are first to be determined. Using the boost-invariance of $E$ and $B$, which arises from the Euler equation (\ref{A11}) in a uniformly expanding fluid with $\tder{u_{\mu}}=0$, the formal solutions of $B$ and $E$ are given by
\begin{eqnarray}
B(\tau)&=&B_{0}\left(\frac{\tau_{0}}{\tau}\right)e^{{\cal{M}}}, \label{A32}\\
E(\tau)&=&E_{0}\left(\frac{\tau_{0}}{\tau}\right)e^{{\cal{N}}}.\label{A33}
\end{eqnarray}
Here, ${\cal{M}}$ and ${\cal{N}}$ are functions of $\tau$, and ${\cal{M}}_{0}$ and ${\cal{N}}_{0}$ are chosen to be ${\cal{M}}_{0}={\cal{M}}(\tau_0)=0$ and ${\cal{N}}_{0}={\cal{N}}(\tau_0)=0$.
Plugging finally (\ref{A30}), (\ref{A32}) and (\ref{A33}) into (\ref{A27}), we arrive at
\begin{widetext}
\begin{eqnarray}\label{A34}
e^{\frac{{\cal{L}}}{\kappa}}&=&1+
\frac{\sigma E_{0}^{2}}{\epsilon_0}\int_{\tau_0}^{\tau}d\tau'\left(\frac{\tau_0}{\tau'}\right)^{1-\frac{1}{\kappa}}e^{2{\cal{N}}}\nonumber\\
&&+\frac{\chi_e E_{0}^{2}}{\epsilon_0\tau_0}\int_{\tau_0}^{\tau}d\tau'\left(\frac{\tau_0}{\tau'}\right)^{2-\frac{1}{\kappa}}e^{2{\cal{N}}}
-\frac{\chi_{e}E_{0}^{2}}{\epsilon_{0}}\int_{\tau_0}^{\tau}d\tau'\left(\frac{\tau_0}{\tau'}\right)^{1-\frac{1}{\kappa}}e^{2{\cal{N}}}\frac{d{\cal{N}}}{d\tau'}\nonumber\\
&&
+\frac{\chi_m B_{0}^{2}}{\epsilon_0\tau_0}\int_{\tau_0}^{\tau}d\tau'\left(\frac{\tau_0}{\tau'}\right)^{2-\frac{1}{\kappa}}e^{2{\cal{M}}}
-\frac{\chi_{m}E_0B_{0}}{\epsilon_{0}\tau_0}\int_{\tau_0}^{\tau}d\tau'\left(\frac{\tau_0}{\tau'}\right)^{2-\frac{1}{\kappa}}e^{{\cal{M}}+{\cal{N}}}\cos\delta\frac{\partial\phi}{\partial\eta}.
\end{eqnarray}
\end{widetext}
Here, $\delta\equiv \phi-\zeta$. To have the full $\tau$ dependence of $T,p,\epsilon$ as well as $B$ and $E$, the $\tau$ dependence of ${\cal{M}}$ and ${\cal{N}}$ as well as the $\eta$ dependence of $\phi$ and $\delta$ are to be determined. This will be done in the following section.
%%%%%%%%%%%%%%%%%%%%%%%%%%
\subsection{Master equation for ${\cal{M}}$}\label{sec3C}
%%%%%%%%%%%%%%%%%%%%%%%%%%
Let us start by considering (\ref{A15}), (\ref{A18}) and (\ref{A19}). These are a set of five constituent equations, whose solutions lead to $(\tau,\eta)$ dependence of ${\cal{M}}, {\cal{N}}$ as well as $\phi$ and $\zeta$.
To arrive at these solutions, we will go through following steps:
\par\vspace{0.3cm}\par
$i)$ We will first show that a combination of these equations leads automatically to $\sin\delta=0$. Mathematically, $\sin\delta=0$ leads to  $\delta=\phi-\zeta=n\pi$ with $n=0,1, 2,\cdots$. Physically, this would mean that in a uniformly expanding fluid, where a transverse MHD setup is applicable, the electric and magnetic vectors, $\mathbf{E}$ and $\mathbf{B}$ are either parallel or anti-parallel with respect to each other.
Moreover, since the relative angle, $\delta$ of these fields remains constant in $\tau$ and $\eta$, if at $\tau_{0}$ they are parallel/anti-parallel, they remain so at any later time $\tau>\tau_0$ and for any $\eta=-\infty,\cdots,\infty$. Let us notice that the fact that electric and magnetic fields are either parallel or anti-parallel leads to vanishing  local Poynting vector $\mathbf{S}=\mathbf{E}\times \mathbf{B}$, and consequently to vanishing electromagnetic energy flow between fluid parcels.
\par
$ii)$ By solving these equations, we will, in particular, show that $\phi$ and $\eta$ evolves as
\begin{eqnarray}\label{A35}
\phi(\eta)&=&\omega_{0}\eta+\phi_{0},\nonumber\\
\zeta(\eta)&=&\omega_{0}\eta+\zeta_{0},
\end{eqnarray}
with $\phi(\eta)-\zeta(\eta)=\phi_0-\zeta_0=\delta=n\pi$. Here, $\omega_{0}=\frac{\partial\phi}{\partial\eta}=\frac{\partial\zeta}{\partial\eta}=$ const.  A non-vanishing $\omega_{0}$ implies a rotation of $\mathbf{B}$ and $\mathbf{E}$ vectors around an axis parallel to the $z$-axis. It can also be regarded as the source for non-boost-invariance ($\eta$ dependence) of rotating solutions in non-ideal transverse MHD.
\par
$iii)$ Finally, by combining these equations, we will show that ${\cal{M}}$ either satisfies
\begin{eqnarray}\label{A36}
\frac{d{\cal{M}}}{du}=0,
\end{eqnarray}
where $u\equiv \ln\left(\frac{\tau}{\tau_0}\right)$, or the following second-order non-linear differential equation:
\begin{eqnarray}\label{A37}
\frac{d^{2}{\cal{M}}}{du^{2}}+\frac{d{\cal{M}}}{du}\left(\frac{d{\cal{M}}}{du}+\frac{\sigma\tau_{0}e^{u}}{1+\chi_{e}}\right)+\omega_{0}^{2}\frac{(1-\chi_{m})}{1+\chi_{e}}=0.\nonumber\\
\end{eqnarray}
Here, $\omega_{0}=\frac{\partial\phi}{\partial\eta}=\frac{\partial\zeta}{\partial\eta}$, being part of the initial condition, remains constant for all $\tau$ and $\eta$. We will show that (\ref{A36}) corresponds to $\omega_{0}=0$, which leads, using (\ref{A35}), to constant $\phi$ and $\zeta$. Physically, this corresponds to non-rotating vectors $\mathbf{E}$ and ${\mathbf{B}}$. Moreover, for ${\cal{M}}$ satisfying (\ref{A36}), we have ${\cal{M}}=0$. Using (\ref{A25}), this leads to frozen flux relation $\partial_{\mu}(Bu^{\mu})=0$, even in the non-ideal MHD with non-vanishing magnetization and electric polarization. In addition, any solution of (\ref{A37}) leads to a deviation from frozen flux theorem in such a medium. Since for the derivation of (\ref{A37}), $\omega_{0}$ is assumed to be non-zero, these solutions correspond to rotating $\mathbf{E}$ and ${\mathbf{B}}$ fields.
\par
Let us finally notice that whenever ${\cal{M}}(\tau)$ and $\phi(\tau,\eta)$ are computed, it is then easy to determine ${\cal{N}}(\tau)$ from the second equation in (\ref{A18}) in combination with the Ansatz (\ref{A26}). The $\tau$ dependence of $T$ arises then from (\ref{A34}) in combination with the formal self-similar solution (\ref{A29}) of $T$.
\par\vspace{0.3cm}\par\noindent
\textbf{Proofs:}\par
$i)$ In order to show that $\sin\delta=0$, let us consider the first equation in (\ref{A18}). Plugging
$$\chi_{e} E\frac{\partial\zeta}{\partial\tau}=-E\frac{\partial\zeta}{\partial\tau}-(1-\chi_m)\frac{B}{\tau}\sin\delta\frac{\partial\phi}{\partial\eta},$$
from this equation into the r.h.s. of (\ref{A19}), we arrive, in particular, at
\begin{eqnarray}\label{A38}
\lefteqn{\hspace{-0.9cm}
[\chi_{e}\partial_{\mu}(Eu^{\mu})+\sigma E]\sin\delta
}\nonumber\\
&&\hspace{-0.9cm}=-E\cos\delta\frac{\partial\zeta}{\partial\tau}-(1-\chi_{m})\sin\delta\cos\delta\frac{B}{\tau}\frac{\partial\phi}{\partial\tau}.
\end{eqnarray}
From the second equation in (\ref{A18}), we then have
\begin{eqnarray}\label{A39}
\lefteqn{\hspace{-1.2cm}
\chi_{e}\partial_{\mu}(Eu^{\mu})+\sigma E=-\partial_{\mu}(Eu^{\mu})}\nonumber\\
&&-(1-\chi_m)\cos\delta\frac{B}{\tau}\frac{\partial\phi}{\partial\eta}.
\end{eqnarray}
Plugging (\ref{A38}) into the l.h.s. of (\ref{A39}) results in
\begin{eqnarray}\label{A40}
E\cos\delta\frac{\partial\zeta}{\partial\tau}=\partial_{\mu}(Eu^{\mu})\sin\delta,
\end{eqnarray}
which together with (\ref{A19}) leads to
\begin{eqnarray}\label{A41}
\sigma E\sin\delta=0.
\end{eqnarray}
In non-ideal transverse MHD, where $\sigma E\neq 0$, (\ref{A41}) leads to $\sin\delta=0$, and consequently to $\delta=n\pi$ with $n=0,1,2,\cdots$, and $\ell\equiv \cos\delta=\pm 1$. Here, the plus and minus signs correspond to parallel and anti-parallel orientation of $\mathbf{E}$ and $\mathbf{B}$ fields with respect to each other.
\par
$ii)$ Let us now reconsider the relations from (\ref{A15}), (\ref{A18}) and (\ref{A19}) with $\sin\delta=0$ and $\cos\delta=\pm 1$. In this case, (\ref{A19}) leads to
\begin{eqnarray}\label{A42}
\frac{\partial \zeta(\tau,\eta)}{\partial \tau}=0, \qquad \forall \tau,\eta.
\end{eqnarray}
This is also compatible with the first equation of (\ref{A18}). From the second equation of (\ref{A15}), we also obtain
\begin{eqnarray}\label{A43}
\frac{\partial\phi(\tau,\eta)}{\partial\tau}=0, \qquad \forall \tau,\eta.
\end{eqnarray}
Introducing at this stage $u=\ln\left(\frac{\tau}{\tau_0}\right)$, the first equation in (\ref{A15}) leads to
\begin{eqnarray}\label{A44}
\frac{\partial\zeta(\tau,\eta)}{\partial\eta}=\ell\frac{B(\tau)}{E(\tau)}\frac{d{\cal{M}}(u)}{du}.
\end{eqnarray}
Here, (\ref{A25}) is used. Let us note that since $\delta=\phi-\zeta=\mbox{const}.$, we also have
\begin{eqnarray}\label{A45}
\frac{\partial\zeta(\tau,\eta)}{\partial\eta}=\frac{\partial\phi(\tau,\eta)}{\partial\eta}.
\end{eqnarray}
Bearing in mind that the electromagnetic fields, $E$ and $B$, are boost-invariant ($\eta$-independent), and that ${\cal{M}}$ depends only on $\tau$, the r.h.s of  (\ref{A44}) turns out to be independent of $\eta$. We thus have
\begin{eqnarray}\label{A46}
\frac{\partial^{2}\zeta(\tau,\eta)}{\partial\eta^{2}}=0, \qquad \forall \tau,\eta,
\end{eqnarray}
and upon using (\ref{A45}),
\begin{eqnarray}\label{A47}
\frac{\partial^{2}\phi(\tau,\eta)}{\partial\eta^{2}}=0, \qquad \forall \tau,\eta.
\end{eqnarray}
The last two relations together with (\ref{A45}) lead first to
\begin{eqnarray}\label{A48}
\phi(\tau,\eta)&=&\omega(\tau)\eta+\phi_{0}(\tau),\nonumber\\
\zeta(\tau,\eta)&=&\omega(\tau)\eta+\zeta_{0}(\tau).
\end{eqnarray}
Using then
\begin{eqnarray}\label{A49}
\frac{\partial}{\partial\tau}\left(\frac{\partial\phi}{\partial\eta}\right)=\frac{\partial}{\partial\tau}\left(\frac{\partial\zeta}{\partial\eta}\right)=0,
\end{eqnarray}
arising from (\ref{A42}) and (\ref{A43}), we obtain
\begin{eqnarray}\label{A50}
\frac{\partial\zeta}{\partial\eta}=\frac{\partial\phi}{\partial\eta}\equiv \omega_{0}=\mbox{const},
\end{eqnarray}
as well as
\begin{eqnarray}\label{A51}
\phi_{0}(\tau)&\equiv& \phi_{0}=\mbox{const.},\nonumber\\
\zeta_{0}(\tau)&\equiv& \zeta_{0}~=\mbox{const.}
\end{eqnarray}
Plugging the above results into (\ref{A48}), we arrive at (\ref{A35}), as claimed.
\par
$iii)$ To derive the differential equation (\ref{A37}) for ${\cal{M}}$, we use (\ref{A35}) together with
(\ref{A44}), and arrive at
\begin{eqnarray}\label{A52}
\frac{d{\cal{M}}}{du}=\ell\omega_{0}\frac{E}{B},
\end{eqnarray}
which, upon using (\ref{A25}) and (\ref{A26}), yields
\begin{eqnarray}\label{A53}
\frac{d{\cal{M}}}{du}=\ell\omega_{0}\beta_{0}e^{{\cal{N}}-{\cal{M}}},
\end{eqnarray}
with $\beta_{0}\equiv\frac{E_{0}}{B_{0}}$. For $\omega_{0}=0$, (\ref{A53}) leads to (\ref{A36}). For $\omega_{0}\neq 0$, we use the second equation of (\ref{A18}), together with (\ref{A26}) and (\ref{A53}), and  arrive, for $\chi_{e}\neq -1$ and $\frac{d{\cal{M}}}{du}\neq 0$, at the differential equation (\ref{A37}).
\par
As aforementioned, the solutions of (\ref{A36}) and (\ref{A37}) correspond to non-rotating and rotating electromagnetic fields, respectively. In the next section, we will present exact and approximate solutions to these equations.
%%%%%%%%%%%%%%%%%%%%%%%%%%
\section{Analytical solutions of (\ref{A36}) and (\ref{A37})}\label{sec4}
\setcounter{equation}{0}
%%%%%%%%%%%%%%%%%%%%%%%%%%
\subsection{Non-Rotating electric and magnetic fields}\label{sec4A}
%%%%%%%%%%%%%%%%%%%%%%%%%%
%%%%%%%%%%%%%%%%
As we have argued in the previous section, in non-ideal transverse MHD with non-vanishing electric field, the case $\frac{d{\cal{M}}}{du}=0$ corresponds to $\omega_{0}=0$. This implies constant angles of $\mathbf{B}$ and $\mathbf{E}$ fields with respect to a certain $x$-axis in the LRF of the fluid, i.e.,
\begin{eqnarray}\label{D1}
\phi(\tau,\eta)=\phi_{0}=\mbox{const.}\quad\mbox{and}\quad\zeta(\tau,\eta)=\zeta_{0}=\mbox{const.,}\nonumber\\
\end{eqnarray}
[see (\ref{A35}]. To determine the magnitude of the electric and magnetic fields, let us consider $\frac{d{\cal{M}}}{du}=0$, or equivalently, $\tder{{\cal{M}}}=0$. Using (\ref{A25}) and ${\cal{M}}_{0}=0$, we have
\begin{eqnarray}\label{D2}
{\cal{M}}=0.
\end{eqnarray}
Plugging this relation into (\ref{A25}), it turns out that $B$ satisfies (\ref{S22}), as in the ideal case. In other words, the fluxes are, as in the case of ideal MHD, frozen. Bearing in mind that $B=|\mathbf{B}|$ is $\eta$-independent, the most general self-similar solution of $B$ reads, \begin{eqnarray}\label{D3}
B(\tau)=B_{0}\left(\frac{\tau_{0}}{\tau}\right),
\end{eqnarray}
[see (\ref{A32})]. Using, at this stage, the second relation in (\ref{A18}) with $\frac{\partial\phi}{\partial\eta}=0$, we arrive at
\begin{eqnarray}\label{D4}
\partial_{\mu}(Eu^{\mu})=-E\frac{\sigma}{1+\chi_{e}},
\end{eqnarray}
which, upon comparing with (\ref{A26}), leads to
\begin{eqnarray}\label{D5}
{\cal{N}}=-\frac{\sigma(\tau-\tau_{0})}{1+\chi_{e}}.
\end{eqnarray}
Hence, according to (\ref{A33}), $E(\tau)$ evolves as
\begin{eqnarray}\label{D6}
E(\tau)=E_{0}\left(\frac{\tau_{0}}{\tau}\right)e^{-\frac{\sigma(\tau-\tau_{0})}{1+\chi_{e}}}.
\end{eqnarray}
The ideal transverse MHD limit $E\to 0$ is thus recovered for $\frac{\sigma}{1+\chi_{e}}\ll \frac{1}{\tau-\tau_{0}}$. Let us notice that in the ideal MHD, where $E$ is assumed to vanish, $\frac{d{\cal{M}}}{du}=0$ leads also to (\ref{D3}). Combining now (\ref{D3}) and (\ref{D6}), we also arrive at
\begin{eqnarray}\label{D7}
\frac{E}{B}=\beta_{0}e^{-\frac{\sigma(\tau-\tau_{0})}{1+\chi_{e}}},
\end{eqnarray}
with $\beta_{0}=\frac{E_0}{B_{0}}$.
\begin{figure}[hbt]
\begin{center}
\includegraphics[width=7.cm,height=7.5cm]{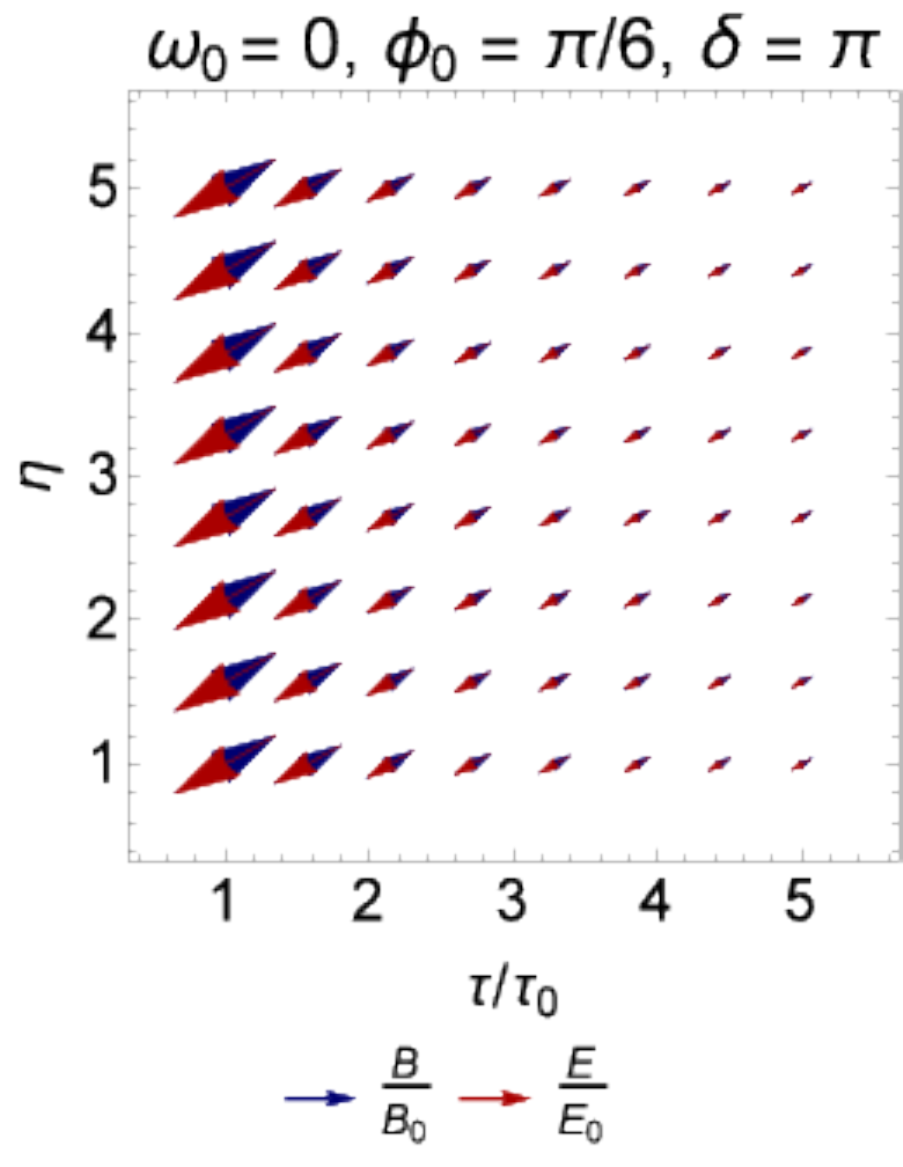}
\end{center}
\caption{(color online). Non-rotating $\frac{\mathbf{B}}{B_{0}}$ and $\frac{\mathbf{E}}{E_{0}}$ from (\ref{D8}) are plotted in an $\eta$ vs. $\tau/\tau_{0}$ plane. Blue (Red) arrows correspond to the magnetic (electric) vector field. Here, $\phi_{0}=\pi/6$ and $\delta=\pi$, i.e. $\mathbf{B}$ and $\mathbf{E}$ are anti-parallel. For $\mathbf{E}$,  $\sigma=4$ MeVc and $\tau_{0}=0.5$ fm/c. Whereas, the magnetic flux is frozen, $\partial_{\mu}(Bu^{\mu})=0$, the electric field satisfies (\ref{D4}). }\label{fig1}
\end{figure}
\par
In Fig. \ref{fig1}, the spatial components of non-rotating $B^{\mu}=(0,\mathbf{B})$ and $E^{\mu}=(0,\mathbf{E})$ fields with
\begin{eqnarray}\label{D8}
\frac{\mathbf{B}}{B_{0}}&=&\left(\frac{\tau_{0}}{\tau}\right)\left(\cos\phi_{0}, \sin\phi_{0},0\right),\nonumber\\
\frac{\mathbf{E}}{E_{0}}&=&\left(\frac{\tau_{0}}{\tau}\right)e^{-\frac{\sigma(\tau-\tau_{0})}{1+\chi_{e}}}\left(\cos\zeta_{0}, \sin\zeta_{0},0\right),
\end{eqnarray}
are plotted in an $\eta$ vs. $\tau/\tau_{0}$ plane. Here, $\phi_{0}=\frac{\pi}{6}$ and $\zeta_{0}=\frac{7\pi}{6}$. Moreover, $\tau_{0}=0.5$ fm/c, $\sigma=4$ MeVc and $\chi_{e}=0$ are assumed. In this case, $\delta=\phi_{0}-\zeta_{0}=\pi$. This corresponds to anti-parallel $\mathbf{B}$ (blue arrows) and $\mathbf{E}$ (red arrows) vectors. As it is shown, whereas  $B=|\mathbf{B}|$ and $E=|\mathbf{E}|$ decrease with increasing $\tau$, the orientations of magnetic and electric fields remain constant in the $\eta$ direction.
\par
Plugging, at this stage, ${\cal{M}}$ and ${\cal{N}}$ from (\ref{D2}) and (\ref{D5}) into (\ref{A34}), and performing the integration over $\tau'$ by making use of
\begin{widetext}
\begin{eqnarray*}
\int_{\tau_{0}}^{\tau}d\tau'~\left(\frac{\tau_{0}}{\tau'}\right)^{n-\frac{1}{\kappa}}e^{-\frac{m\sigma(\tau'-\tau_{0})}{1+\chi_{e}}}
=\tau_{0}\left(\frac{m\sigma\tau_{0}}{1+\chi_{e}}\right)^{n-\frac{1}{\kappa}-1}e^{\frac{m\sigma\tau_{0}}{1+\chi_{e}}}\left\{
\Gamma\left(\frac{1}{\kappa}-n+1,\frac{m\sigma\tau_{0}}{1+\chi_{e}}\right)-\Gamma\left(
\frac{1}{\kappa}-n+1,\frac{m\sigma\tau}{1+\chi_{e}}
\right)\right\},
\end{eqnarray*}
\end{widetext}
which arises from
\begin{eqnarray}\label{D9}
\int_{\tau_{1}}^{\tau_{2}}dt~t^{a-1}e^{-t}=\Gamma(a,\tau_{1})-\Gamma(a,\tau_{2}),
\end{eqnarray}
we obtain
%\begin{widetext}
\begin{eqnarray}\label{D10}
e^{\frac{{\cal{L}}}{\kappa}}&=&1+\frac{(1+2\chi_{e})E_{0}^{2}}{2\epsilon_{0}}\left(\frac{2\sigma\tau_0}{1+\chi_e}\right)^{1-c_s^{2}}e^{\frac{2\sigma\tau_0}{1+\chi_e}}
\nonumber\\
&&\times \bigg\{\Gamma\left(c_s^{2},\frac{2\sigma\tau_0}{1+\chi_e}\right)-\Gamma\left(c_s^{2},\frac{2\sigma\tau}{1+\chi_e}\right)\bigg\}\nonumber\\
&&+\frac{\chi_{e}E_{0}^{2}}{\epsilon_0}\left(\frac{2\sigma\tau_0}{1+\chi_e}\right)^{1-c_s^{2}}e^{\frac{2\sigma\tau_0}{1+\chi_e}}\nonumber\\
&&\times\bigg\{\Gamma\left(c_s^{2}-1,\frac{2\sigma\tau_0}{1+\chi_e}\right)-\Gamma\left(c_s^{2}-1,\frac{2\sigma\tau}{1+\chi_e}\right)\bigg\}\nonumber\\
&&-\frac{\chi_m B_0^{2}}{\epsilon_0(1-c_{s}^{2})}\left(\left(\frac{\tau_0}{\tau}\right)^{1-c_s^2}-1\right).
\end{eqnarray}
%\end{widetext}
Here, $c_{s}^{2}\equiv \kappa^{-1}$ is the sound velocity and $\epsilon_{0}=\kappa p_{0}$.
In Sec. \ref{sec5}, we will use (\ref{D10}) to demonstrate the evolution of thermodynamic fields $T,p$ and $\epsilon$, whose self-similar solutions are presented in (\ref{A29}), (\ref{A30}) and (\ref{A31}). In particular, we will compare the evolution of these fields for non-rotating and rotating electromagnetic fields, characterized by  $\omega_{0}=0$ and $\omega_{0}\neq 0$, respectively. The latter case will be discussed as a next step.
%%%%%%%%%%%%%%%%%%%%%%%%%%
\subsection{Rotating electric and magnetic fields; Approximate analytical solutions}\label{sec4B}
%%%%%%%%%%%%%%%%%%%%%%%%%%
In this section, we will present approximate analytical solutions to (\ref{A37}). We will consider two different cases
\begin{eqnarray}
\begin{array}{llcl}
\mbox{Case 1:}&\frac{E}{B}\sim\frac{E_0}{B_0}\left(\frac{\tau_{0}}{\tau}\right)^{n}&&\mbox{for $n\neq 0$ and $n=0$},\nonumber\\
\mbox{Case 2:}&\mbox{Small $\omega_{0}$}&&\mbox{with ${\cal{M}}(\tau)\sim \omega_{0} f(\tau)$}.
\end{array}
\end{eqnarray}
In both cases, ${\mathbf{B}}$ and ${\mathbf{E}}$ are either parallel ($\ell=+1$) or anti-parallel ($\ell=-1$), and rotate gradually with increasing $\eta$. The angular velocity is given by $\omega_{0}=\mbox{const}$. The $\tau$ dependence of the magnitudes of the electromagnetic fields turns out to be given by any non-vanishing solution of (\ref{A37}), which, in particular, represents a deviation from frozen flux theorem. Apart from the evolution of $\mathbf{B}$ and $\mathbf{E}$, we are interested in the evolution of $T,p$ and $\epsilon$. To this purpose, we insert the corresponding ${\cal{M}}$ and ${\cal{N}}$ from these two cases into (\ref{A34}), and arrive at $\exp\left(\frac{{\cal{L}}}{\kappa}\right)$, which, upon insertion into (\ref{A29}), (\ref{A30}) and (\ref{A31}) leads to the evolution of $T,p$ and $\epsilon$, respectively.
%%%%%%%%%%%%%%%%%%%%%%%%%%
\subsubsection{Case 1: $\frac{E}{B}\sim\frac{E_0}{B_0}\left(\frac{\tau_{0}}{\tau}\right)^{n}$ for $n\neq 0$ and $n=0$}\label{sec4B1}
%%%%%%%%%%%%%%%%%%%%%%%%%%
\begin{figure*}[hbt]
\includegraphics[width=5.35cm,height=5.35cm]{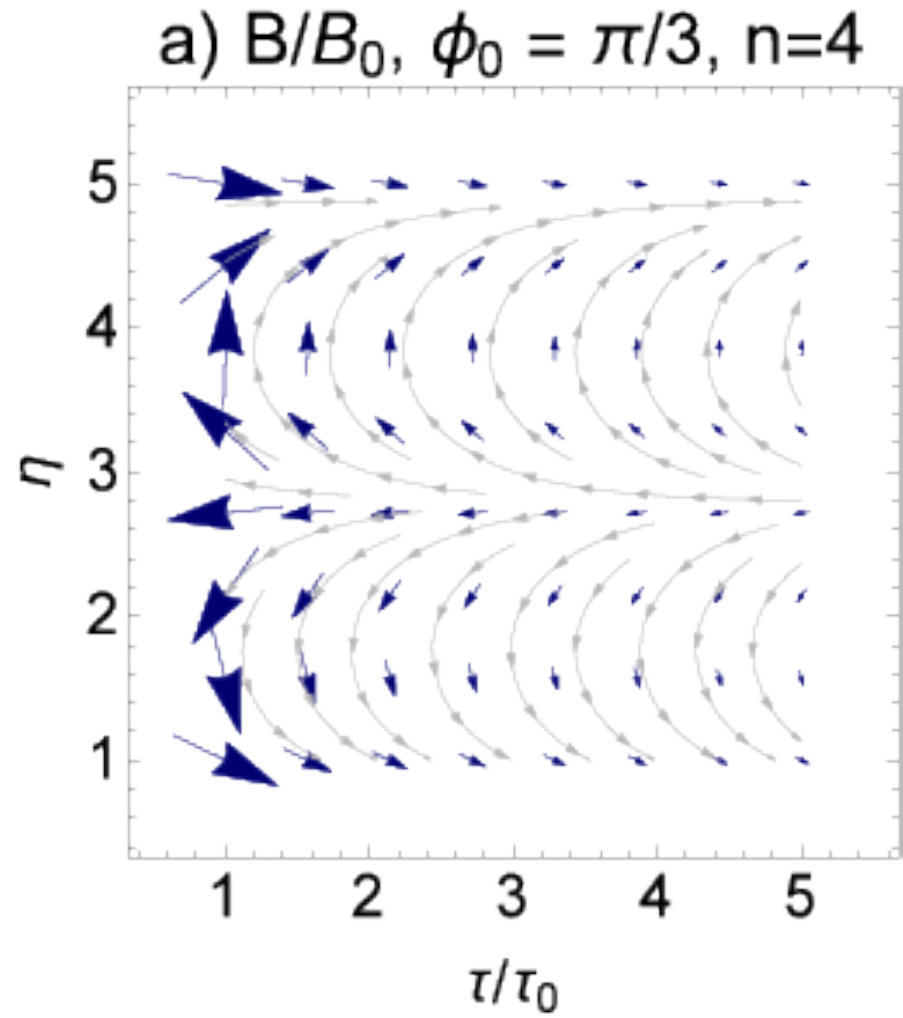}\hspace{0.35cm}
\includegraphics[width=5.35cm,height=5.35cm]{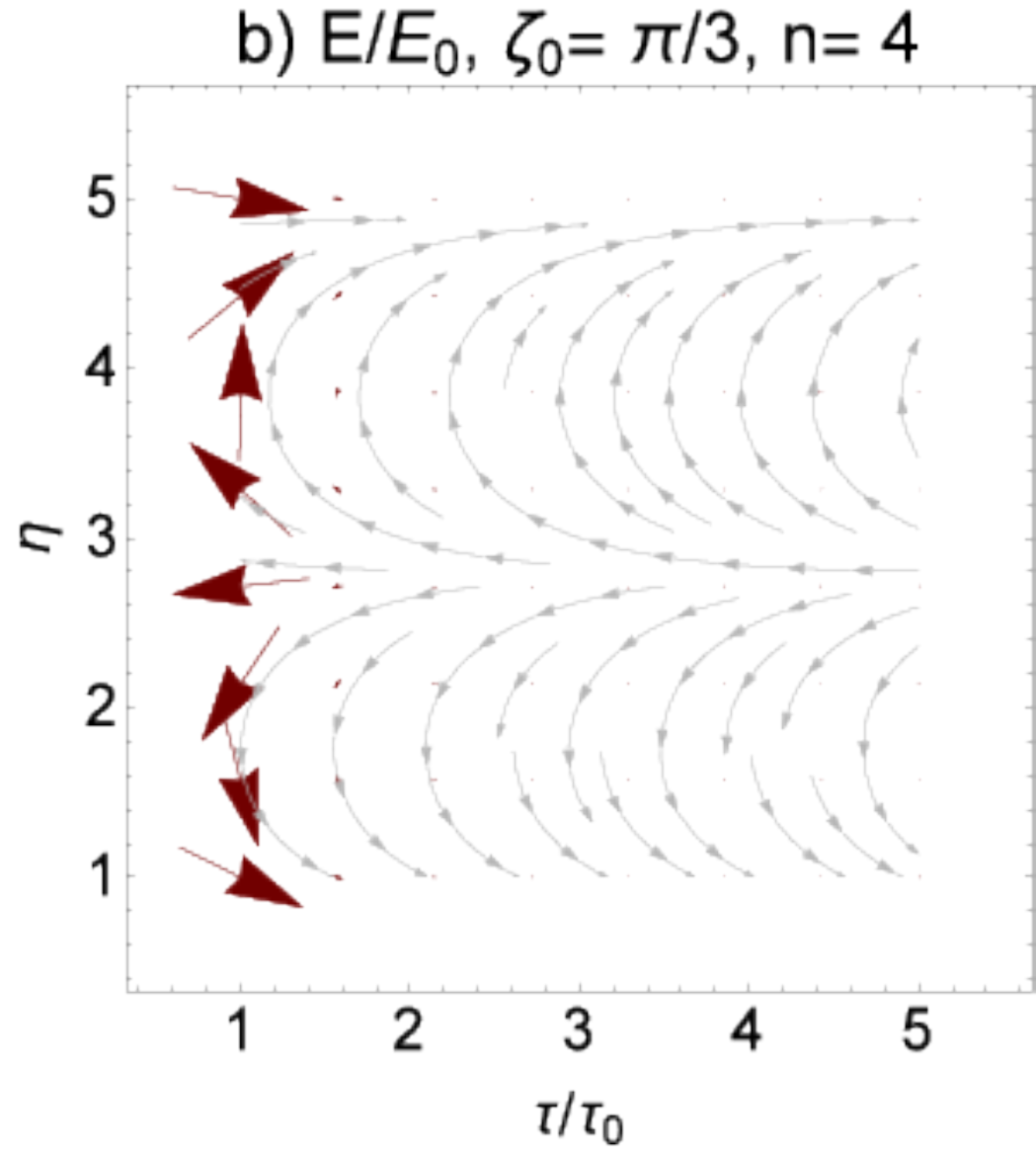}\hspace{0.35cm}
\includegraphics[width=5.35cm,height=5.35cm]{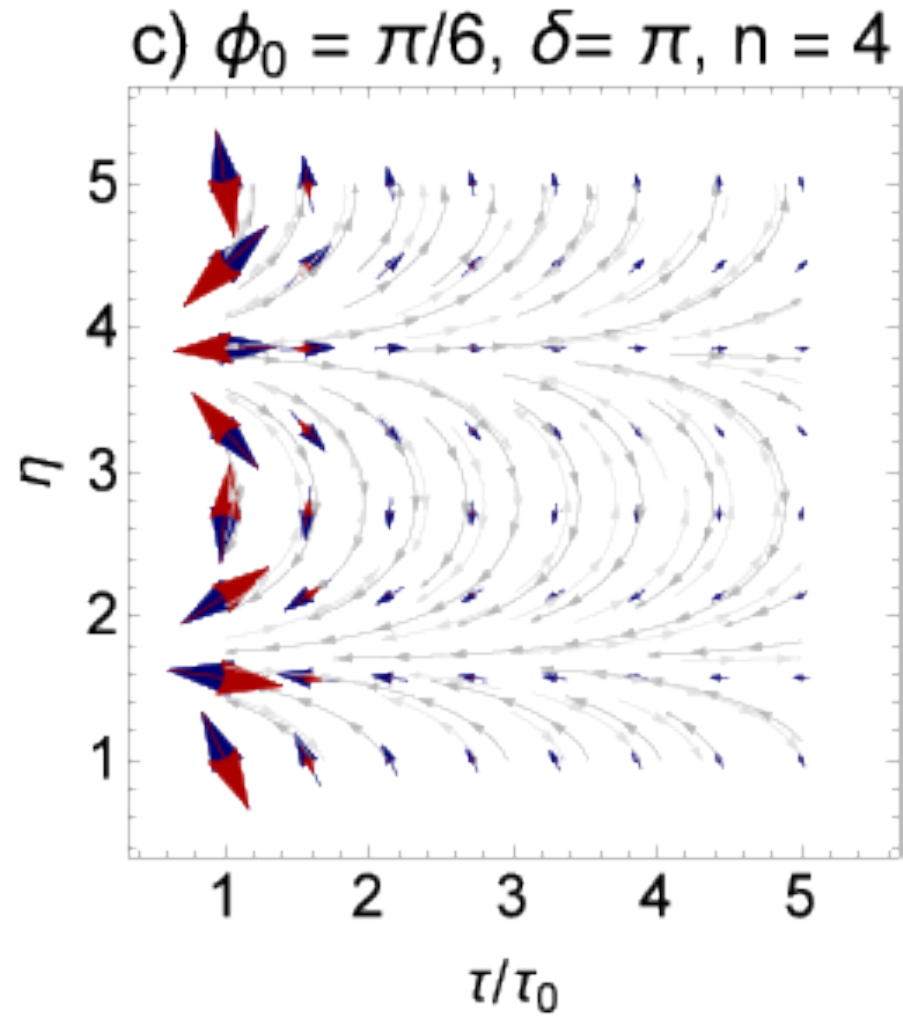}
\caption{(color online). $\mathbf{B}/B_{0}$ (panel a) and $\mathbf{E}/E_{0}$ (panel b) from (\ref{D21}) are plotted  for the  set of free parameters $\{n,\sigma, \beta_{0}, \phi_{0},\chi_e, \chi_m,\ell\}= \{4,400,1,\frac{\pi}{3},0,0,+1\}$ in an $\eta$ vs. $\tau/\tau_{0}$ plane. The magnetic and electric vectors,
indicated by blue (panel a) and red (panel b) vectors, are parallel. This implies a clockwise rotation of $\mathbf{B}$ and $\mathbf{E}$ vectors while $\eta$ increases. This is demonstrated with gray stream lines. The angular velocity $ \tilde{\omega}_{0}$ is given in (\ref{D18}). (c) $\mathbf{B}/B_{0}$ (blue arrows) and $\mathbf{E}/E_{0}$ (red arrows) from (\ref{D21}) are plotted  for the set of free parameters $\{n,\sigma, \beta_{0}, \phi_{0},\chi_e, \chi_m,\ell\}= \{4,400,1,\frac{\pi}{6},0,0,-1\}$ in an $\eta$ vs. $\tau/\tau_{0}$ plane. In this case, $\mathbf{B}$ and $\mathbf{E}$ are anti-parallel, and rotate counterclockwise while $\eta$ increases. The electric field decreases much faster than the magnetic field with increasing $\tau/\tau_0$. }\label{fig2}
\end{figure*}
%%%%%%%%%%%%%%%%%%%%%%%%
\begin{figure*}[hbt]
\includegraphics[width=6.5cm,height=7cm]{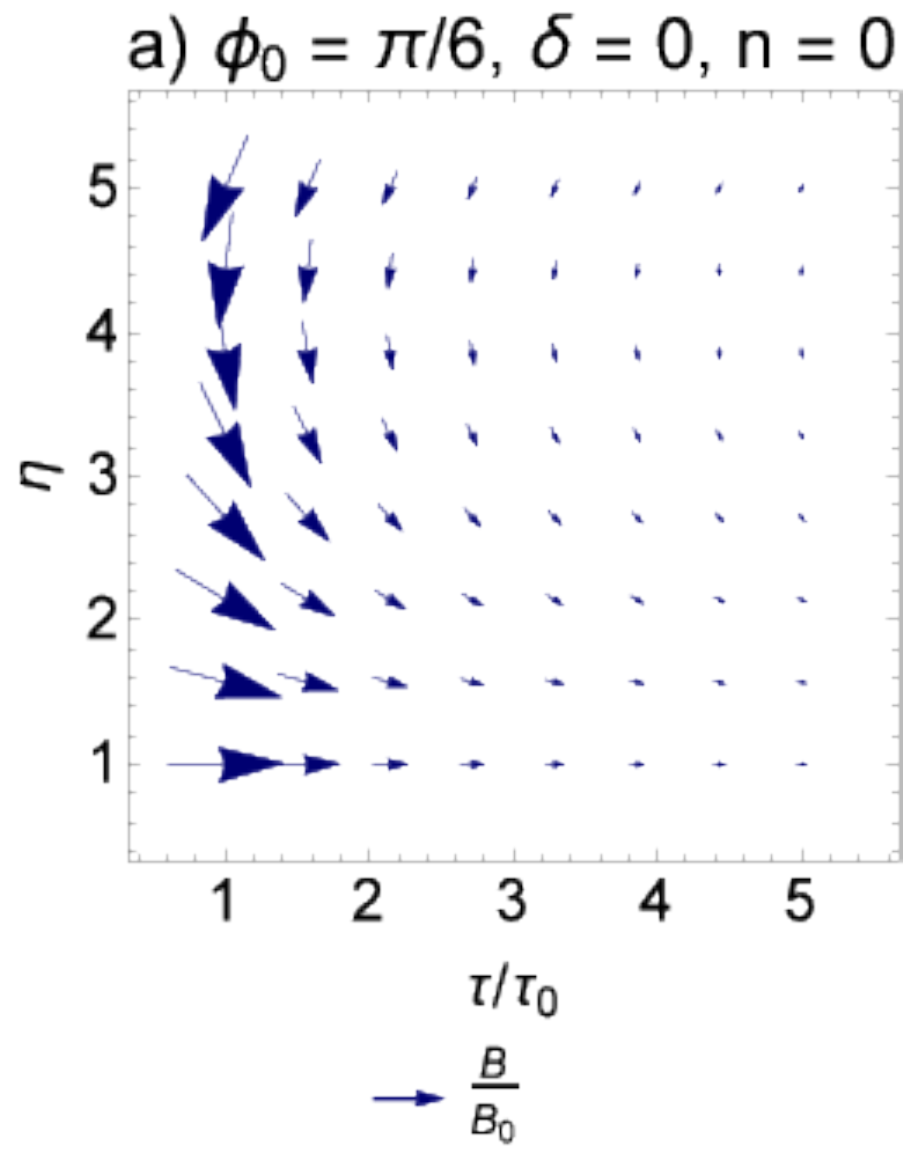}\hspace{1cm}
\includegraphics[width=6.5cm,height=7cm]{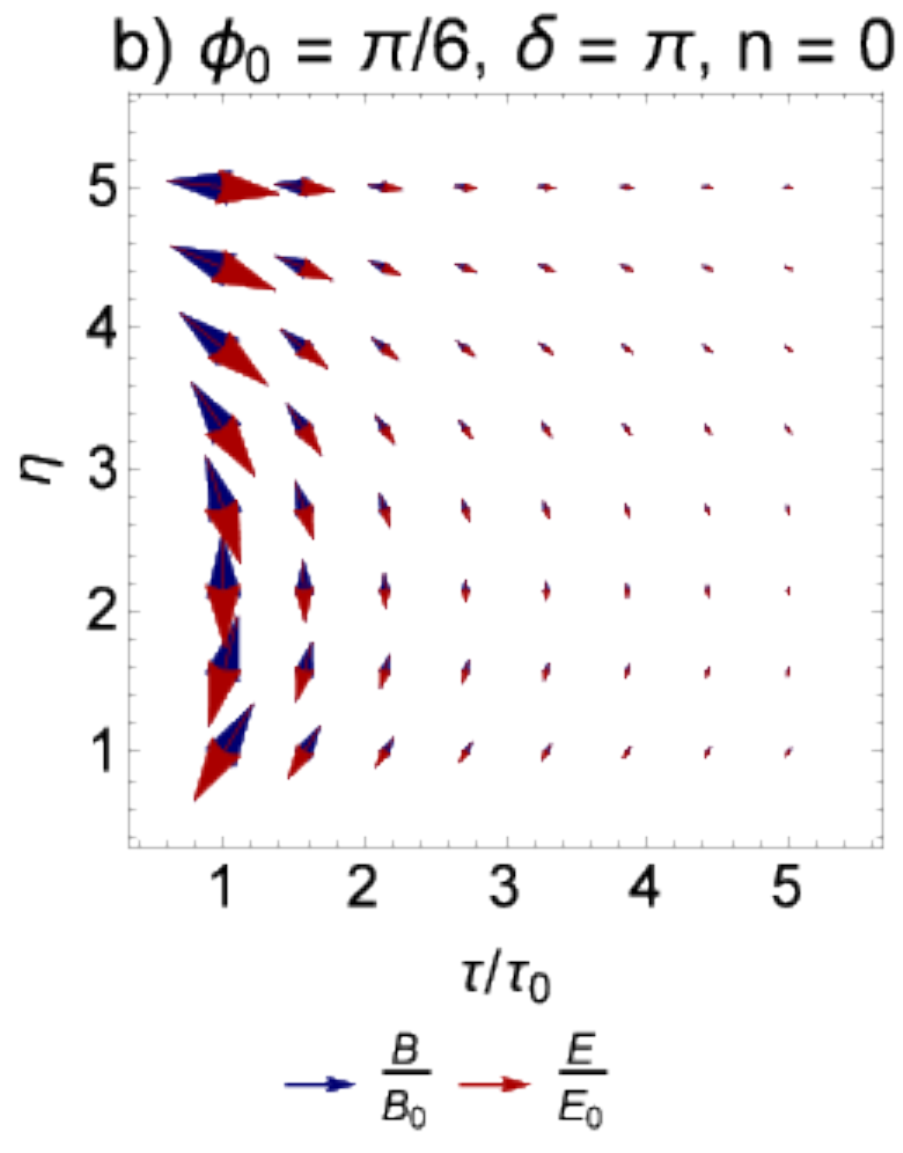}
\caption{(color online). (a) $\mathbf{B}/B_{0}$ from (\ref{D22}) is plotted for the set of free parameters $\{\sigma, \beta_{0}, \phi_{0},\chi_e, \chi_m,\ell\}= \{400,1,\frac{\pi}{6},0,0,+1\}$ in an $\eta$ vs. $\tau/\tau_{0}$ plane. The electric field vectors are parallel to the magnetic field vectors (denoted by blue arrows), and are not demonstrated in this plot. A clockwise rotation is set up with increasing $\eta$. The angular velocity is given by (\ref{D20}). (b) $\mathbf{B}/B_{0}$ (blue arrows) and $\mathbf{E}/E_{0}$ (red arrows) from (\ref{D22}) are plotted for the set of free parameters $\{\sigma, \beta_{0}, \phi_{0},\chi_e, \chi_m,\ell\}= \{400,1,\frac{\pi}{6},0,0,-1\}$ in an $\eta$ vs. $\tau/\tau_{0}$ plane. The rotation turns out to be counterclockwise while $\eta$ increases.  }\label{fig3}
\end{figure*}
\begin{figure*}[hbt]
\includegraphics[width=6.5cm,height=7cm]{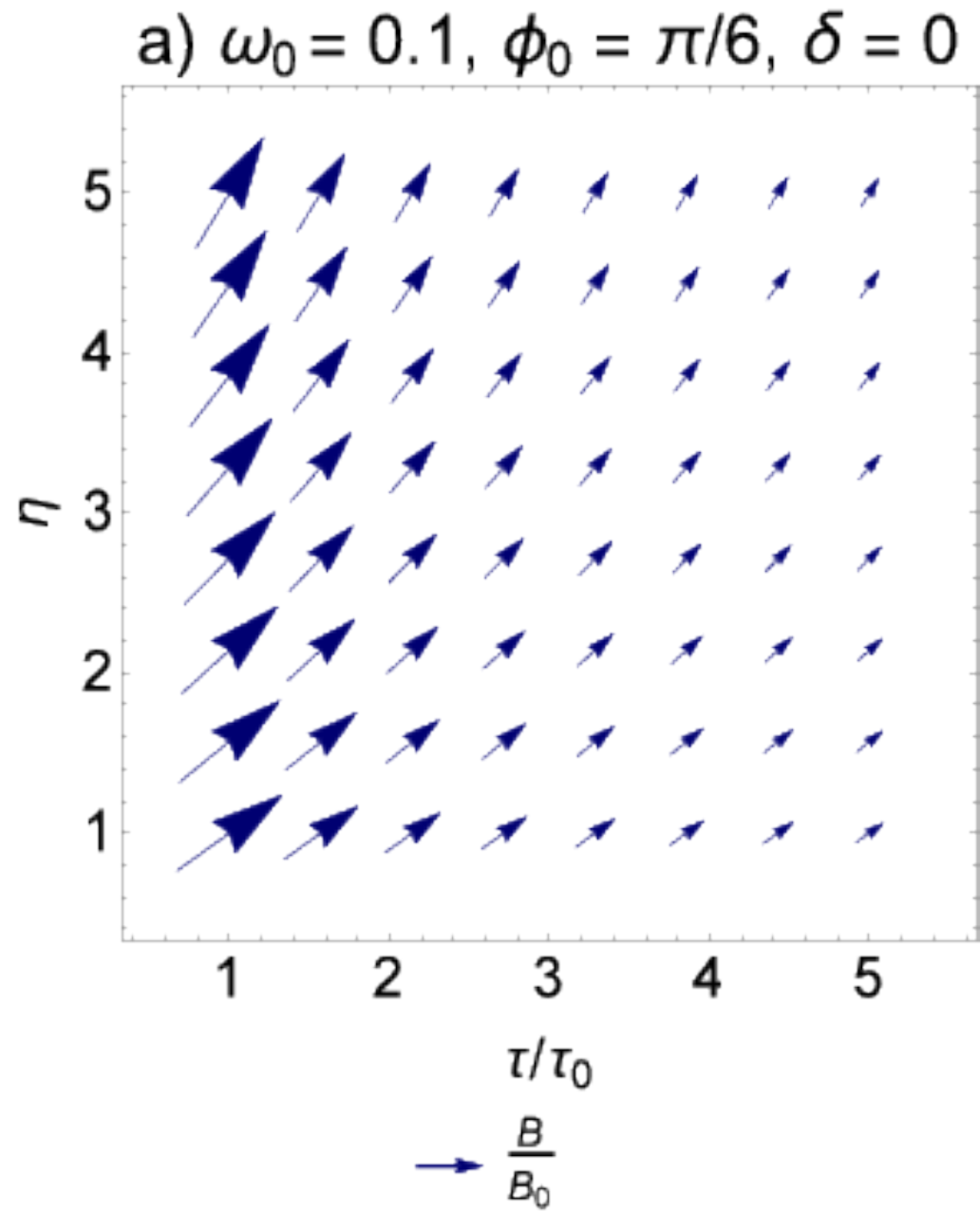}\hspace{1cm}
\includegraphics[width=6.5cm,height=7cm]{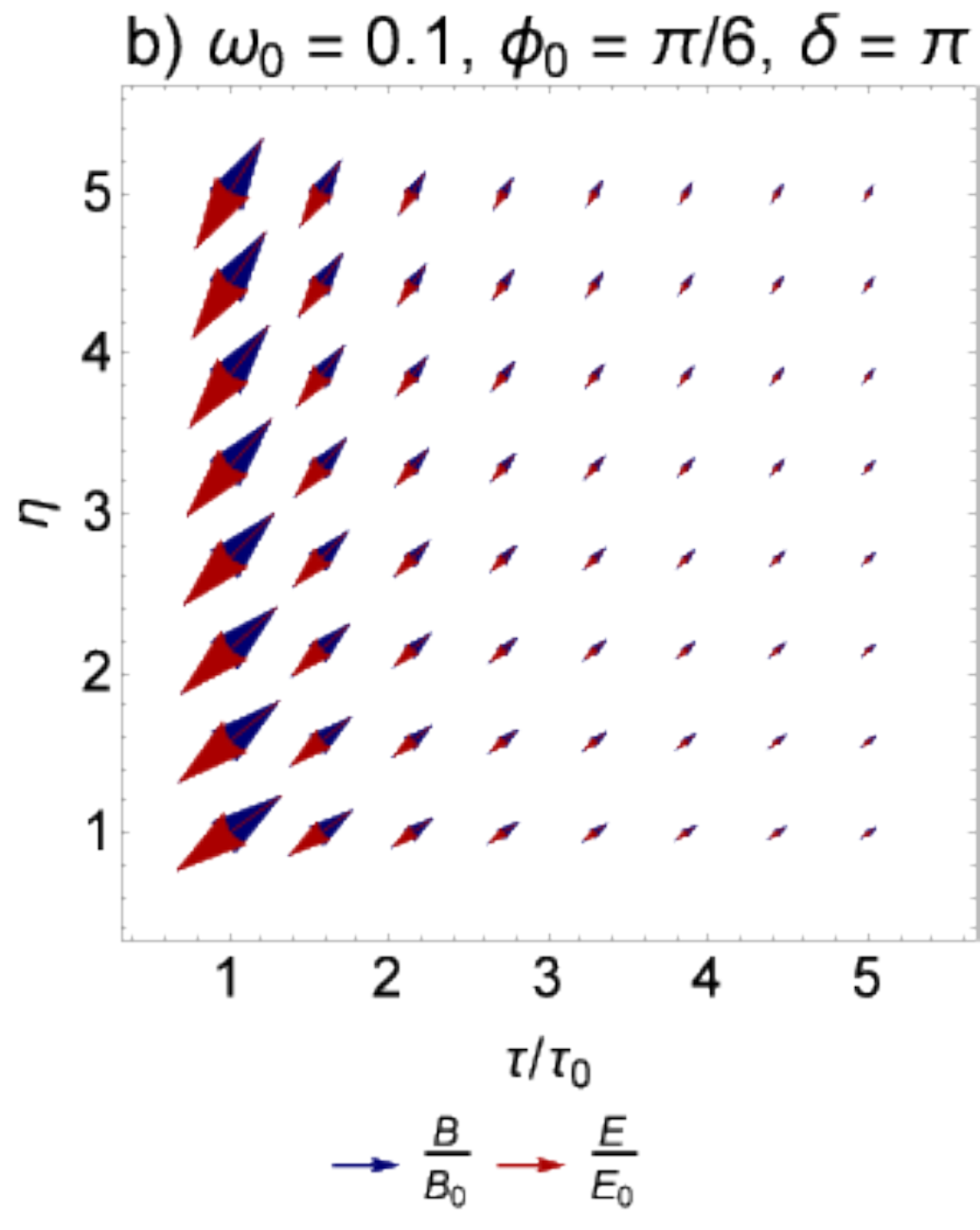}
\caption{(color online). (a) $\mathbf{B}/B_{0}$ from (\ref{D36}) is plotted for the set of free parameters $\{\sigma,\omega_{0},\beta_{0},\phi_{0},\chi_{e},\chi_{m},\ell\}=\{40,0.1,0.1,\frac{\pi}{6},0,0,+1\}$ in an $\eta$ vs. $\tau/\tau_{0}$ plane. The electric field vectors are parallel to the magnetic field vectors (denoted by blue arrows), and are not demonstrated in this plot. A slow rotation sets up with increasing $\eta$. (b) $\mathbf{B}/B_{0}$ (blue arrows) and $\mathbf{E}/E_{0}$ (red arrows) from (\ref{D36}) are plotted for the set of free parameters $\{\sigma,\omega_{0},\beta_{0},\phi_{0},\chi_{e},\chi_{m},\ell\}=\{40,0.1,0.1,\frac{\pi}{6},0,0,-1\}$ in an $\eta$ vs. $\tau/\tau_{0}$ plane.  A slow rotation sets up with increasing $\eta$. }\label{fig4}
\end{figure*}
\par\noindent
Plugging $\frac{E}{B}=\frac{E_0}{B_0}\left(\frac{\tau_{0}}{\tau}\right)^{n}$ into the r.h.s. of (\ref{A52}), we arrive first at
\begin{eqnarray}\label{D12}
{\cal{M}}(u)=-\frac{\ell\omega_{0}\beta_{0}}{n}\left(e^{-nu}-1\right),
\end{eqnarray}
with $u=\ln\left(\frac{\tau}{\tau_{0}}\right)$. Then, using the formal solution of $B(\tau)$ from (\ref{A32}), arising from the method of non-conserved charges introduced in Sec. \ref{sec2}, the most general self-similar solution for $B=|\mathbf{B}|$ reads
\begin{eqnarray}\label{D13}
B(\tau)=B_{0}\left(\frac{\tau_0}{\tau}\right)e^{-\mathfrak{b}_{n}(\tau,\omega_{0})},
\end{eqnarray}
with
\begin{eqnarray}\label{D14}
\mathfrak{b}_{n}(\tau,\omega_{0})\equiv \frac{\ell\omega_{0}\beta_{0}}{n}\left(\left(\frac{\tau_0}{\tau}\right)^{n}-1\right).
\end{eqnarray}
To determine ${\cal{N}}$, we insert (\ref{D14}) into the r.h.s. of (\ref{A53}). We obtain
\begin{eqnarray}\label{D15}
{\cal{N}}(u)&=&{\cal{M}}(u)-nu\nonumber\\
&=&-\frac{\ell\omega_{0}\beta_{0}}{n}\left(e^{-nu}-1\right)-nu,
\end{eqnarray}
which leads, upon using (\ref{A33}), to
\begin{eqnarray}\label{D16}
E=E_{0}\left(\frac{\tau_0}{\tau}\right)^{1+n}e^{-\mathfrak{b}_{n}(\tau,\omega_{0})},
\end{eqnarray}
with $\mathfrak{b}_{n}(\tau,\omega_{0})$ from (\ref{D14}).
As concerns the evolution of $\phi$ and $\zeta$, the relative angles of $\mathbf{B}$ and $\mathbf{E}$ with respect to $x$-axis in the LRF of the fluid, they are, as before, given by (\ref{A35}), where the constant angular velocity of these fields $\omega_{0}$, can be fixed from the master equation (\ref{A37}) evaluated at $u=0$ (or equivalently $\tau=\tau_{0}$).
To this purpose, we use (\ref{A52}), which for ${\cal{M}}$ from (\ref{D12}) yields
\begin{eqnarray}\label{D17}
\frac{d{\cal{M}}}{du}\bigg|_{u=0}&=&\ell\omega_{0}\beta_{0},\nonumber\\
\frac{d^{2}{\cal{M}}}{du^{2}}\bigg|_{u=0}&=&-n\ell\omega_{0}\beta_{0}.
\end{eqnarray}
Plugging these expressions into (\ref{A37}), and setting $u=0$, we obtain
\begin{eqnarray}\label{D18}
\tilde{\omega}_{0}=\frac{\ell\beta_{0}[n(1+\chi_e)-\sigma\tau_0]}{1-\chi_m+\beta_{0}^{2}(1+\chi_e)}.
\end{eqnarray}
We are, in particular, interested in the evolution of $\mathbf{B}$ and $\mathbf{E}$ in the limit $n\to 0$. Using (\ref{D13}) and (\ref{D16}), and taking the limit $n\to 0$, we arrive at the power-law solutions
\begin{eqnarray}\label{D19}
B=B_{0}\left(\frac{\tau_0}{\tau}\right)^{a},\nonumber\\
E=E_{0}\left(\frac{\tau_0}{\tau}\right)^{a},
\end{eqnarray}
with $a\equiv 1-\ell\omega_{0}\beta_{0}$ and
\begin{eqnarray}\label{D20}
\omega_{0}=-\frac{\ell\beta_{0}\sigma\tau_{0}}{1-\chi_{m}+\beta_{0}^{2}(1+\chi_{e})},
\end{eqnarray}
which arises from (\ref{D18}) by taking the limit $n\to 0$. Let us notice, at this stage, that the power-law solutions (\ref{D19}) for the $B$ fields is similar to the power-la decay Ansatz which was previously introduced in \cite{rischke2015}. In contrast to our method, the authors took the Ansatz $B(\tau)\sim \tau^{-a}$, with $a$ being an arbitrary constant free parameter, as the starting point of their analysis, without bringing the power $a$ into relation with $\omega_{0}$ and $\beta_{0}$. Let us note that according to (\ref{D20}), two cases of $a>1$ and $a<1$, discussed in \cite{rischke2015, rischke2016}, are controlled by
\begin{eqnarray*}
&&\chi_{m}<1+\beta_{0}^{2}(1+\chi_{e}),\\
&&\chi_{m}>1+\beta_{0}^{2}(1+\chi_{e}),\\
\end{eqnarray*}
leading to $\ell\omega_{0}<0$ and $\ell\omega_{0}>0$, respectively.
\par
In Fig. \ref{fig2}, we have demonstrated the spatial components of rotating magnetic and electric fields $B^{\mu}=(0,\mathbf{B})$ and $E^{\mu}=(0,\mathbf{E})$ with
\begin{eqnarray}\label{D21}
\hspace{0cm}\frac{\mathbf{B}}{B_{0}}&=&\left(\frac{\tau_{0}}{\tau}\right)e^{-\mathfrak{b}_{n}(\tau, \tilde{\omega}_{0})}\left(\cos\phi(\eta),\sin\phi(\eta),0\right),\nonumber\\
\hspace{0cm}\frac{\mathbf{E}}{E_{0}}&=&\left(\frac{\tau_{0}}{\tau}\right)^{1+n}e^{-\mathfrak{b}_{n}(\tau, \tilde{\omega}_{0})}\left(\cos\zeta(\eta),\sin\zeta(\eta),0\right).\nonumber\\
\end{eqnarray}
Here, $\mathfrak{b}_{n}(\tau, \tilde{\omega}_{0})$ is defined in (\ref{D14}) and  $ \tilde{\omega}_{0}$ in (\ref{D18}). The angles $\phi(\eta)$ and $\zeta(\eta)$ are given in (\ref{A35}) with $\omega_{0}$ replaced by $ \tilde{\omega}_{0}$.
In Figs. \ref{fig2}(a) and \ref{fig2}(b), the vectors corresponding to $\mathbf{B}/B_{0}$ [blue arrows in Fig. \ref{fig2}(a)] and $\frac{\mathbf{E}}{E_{0}}$ [red arrows in Fig. \ref{fig2}(b)] are plotted for the set of free parameters
$\{n,\sigma, \beta_{0}, \phi_{0},\chi_e, \chi_m,\ell\}= \{4,400,1,\frac{\pi}{3},0,0,+1\}$ in an $\eta$ vs. $\tau/\tau_{0}$ plane. The gray stream lines are plotted to demonstrate the rotation of $\mathbf{B}$ and $\mathbf{E}$ vectors, which are, in this case, parallel to each other ($\ell=+1$).\footnote{$\ell=+1$ corresponds to $\delta=\phi_{0}-\zeta_{0}=2n\pi$ with $n=0,1,2,\cdots$.} Here, the rotation turns out to be clockwise while $\eta$ increases. The magnitudes of the fields, $B$ and $E$ decrease with increasing $\tau$. Because of an additional power $n$ of $\frac{\tau_{0}}{\tau}$, $E$ decays much faster than $B$ with increasing $\tau$ [see (\ref{D21})].
\par
In Fig. \ref{fig2}(c),  $\mathbf{B}$ (blue arrows) and $\mathbf{E}$ (red arrows) are anti-parallel ($\ell=-1$).\footnote{$\ell=-1$ corresponds to $\delta=\phi_{0}-\zeta_{0}=(2n+1)\pi$ with $n=0,1,2,\cdots$.} In this case, a counterclockwise rotation occurs with increasing $\eta$. Here, the set of free parameters is chosen to be $\{n,\sigma, \beta_{0}, \phi_{0},\chi_e, \chi_m,\ell\}= \{4,400,1,\frac{\pi}{6},0,0,-1\}$. As in the previous case, with increasing $\tau$, $E$ decreases much faster than $B$.
\par
In Fig. \ref{fig3}, we have plotted the spatial components of rotating magnetic and electric fields $B^{\mu}=(0,\mathbf{B})$ and $E^{\mu}=(0,\mathbf{E})$
\begin{eqnarray}\label{D22}
\frac{\mathbf{B}}{B_{0}}&=&\left(\frac{\tau_{0}}{\tau}\right)^{a}\left(\cos\phi(\eta),\sin\phi(\eta),0\right),\nonumber\\
\hspace{-.5cm}\frac{\mathbf{E}}{E_{0}}&=&\left(\frac{\tau_{0}}{\tau}\right)^{a}\left(\cos\zeta(\eta),\sin\zeta(\eta),0\right),
\end{eqnarray}
with $a=1-\ell \omega_{0}\beta_{0}$ and $\phi(\eta)$ as well as $\zeta(\eta)$ from (\ref{A35}). Here, $\omega_{0}$ from (\ref{D20}) are to be inserted into $a$,   $\phi(\eta)$ and $\zeta(\eta)$. Let us remind that for $n=0$, $\frac{E}{B}=\frac{E_{0}}{B_{0}}=\mbox{const}$. Hence $B$ and $E$ decrease with the same slope as $\tau$ increases. In Fig. \ref{fig3}(a), where the set of free parameters are chosen to be $\{\sigma, \beta_{0}, \phi_{0},\chi_e, \chi_m,\ell\}= \{400,1,\frac{\pi}{6},0,0,+1\}$, $\mathbf{B}$ and $\mathbf{E}$ are parallel, and a clockwise rotation occurs with increasing $\eta$.
\par
In Fig. \ref{fig3}(b), $\mathbf{B}$ (blue arrows) and $\mathbf{E}$ (red arrows) are anti-parallel. As expected, a counterclockwise rotation occurs with increasing $\eta$, and $B$ as well as $E$ decrease with increasing $\tau$ with the same slope. Here, we have worked with $\{\sigma, \beta_{0}, \phi_{0},\chi_e, \chi_m,\ell\}= \{400,1,\frac{\pi}{6},0,0,-1\}$.
\par
As concerns $e^{\frac{\cal{L}}{\kappa}}$ from (\ref{A34}), in the limit $n\to 0$, it is given by
\begin{eqnarray}\label{D23}
\lefteqn{\hspace{-1cm}
e^{\frac{\cal{L}}{\kappa}}=1+
\frac{\sigma\tau_0 E_0^{2}}{(c_s^2-2(a-1))\epsilon_0}\left(\left(\frac{\tau_0}{\tau}\right)^{-c_s^2+2(a-1)}-1\right)
}\nonumber\\
&&\hspace{-1cm}+\frac{a\left(\chi_e E_{0}^{2}+\chi_m B_0^{2}\right)}{(c_s^2+1-2a)\epsilon_0}\left(\left(\frac{\tau_0}{\tau}\right)^{-c_s^2-1+2a}-1\right),
\end{eqnarray}
where $a=1-\ell\omega_{0}\beta_{0}$ with $\omega_{0}$ given in (\ref{D20}). Plugging this expression into (\ref{A29}), (\ref{A30}) and (\ref{A31}), we arrive at the evolution of thermodynamic fields $T,p$ and $\epsilon$ in the case, where $\mathbf{B}$ and $\mathbf{E}$ evolve as presented in (\ref{D22}).
%%%%%%%%%%%%%%%%%%%%%%%%%%
\subsubsection{Case 2: Slowly rotating $E$ and $B$ fields}\label{sec4B2}
%%%%%%%%%%%%%%%%%%%%%%%%%%
In this case, the angular velocity $\omega_{0}$ is assumed to be small ($\omega_{0}\ll 1$). Consequently, ${\cal{M}}$ may be approximated by
\begin{eqnarray}\label{D24}
{\cal{M}}(u)\sim\omega_{0}f(u),
\end{eqnarray}
with $f(u)$ satisfying the differential equation
\begin{eqnarray}\label{D25}
f''(u)+Ae^{u}f'(u)=0.
\end{eqnarray}
Here, $A\equiv \frac{\sigma\tau_{0}}{1+\chi_{e}}$. This differential equation arises by inserting the Ansatz (\ref{D24}) into the master equation (\ref{A37}), and neglecting terms proportional to $\omega_{0}^{2}$. To solve (\ref{D25}), we use the first relation in (\ref{D17}), and arrive for $f'(u)$ at
\begin{eqnarray}\label{D26}
f'(u)=\ell\beta_{0}e^{-A(e^{u}-1)}.
\end{eqnarray}
The final result for ${\cal{M}}(\tau)$ then reads
\begin{eqnarray}\label{D27}
\lefteqn{\hspace{-1cm}
{\cal{M}}(\tau)\sim\omega_{0}f(\tau)=\ell\omega_{0}\beta_{0}e^{\frac{\sigma\tau_{0}}{1+\chi_{e}}}
}
\nonumber\\
&&\hspace{-0.5cm}\times
\bigg\{\Gamma\left(0,\frac{\sigma\tau_{0}}{1+\chi_{e}}\right)-\Gamma\left(0,\frac{\sigma\tau}{1+\chi_{e}}\right)\bigg\}.
\end{eqnarray}
To perform the integration over $\tau$, (\ref{D9}) is used. Using (\ref{A53}) together with (\ref{D26}), ${\cal{N}}$ is given by
\begin{eqnarray}\label{D28}
{\cal{N}}(\tau)={\cal{M}}(\tau)-\frac{\sigma(\tau-\tau_{0})}{1+\chi_{e}},
\end{eqnarray}
with ${\cal{M}}$ from (\ref{D27}). The proper time evolution of the magnetic and electric fields thus reads
\begin{eqnarray}\label{D29}
 B&=&B_{0}\left(\frac{\tau_{0}}{\tau}\right)e^{\cal{M}(\tau)},\nonumber\\
E&=&E_{0}\left(\frac{\tau_{0}}{\tau}\right)e^{\cal{N}(\tau)},
\end{eqnarray}
with ${\cal{M}}$ and ${\cal{N}}$ from (\ref{D27}) and (\ref{D28}), respectively.  The ratio $E/B$ is then given by
\begin{eqnarray}\label{D30}
\frac{E}{B}=\beta_{0}e^{-\frac{\sigma(\tau-\tau_{0})}{1+\chi_{e}}},
\end{eqnarray}
with $\beta_{0}=E_{0}/B_{0}$. The above results can be studied in two different limits $\sigma(\tau-\tau_{0})\ll (1+\chi_{e})$ and $\sigma(\tau-\tau_{0})\gg (1+\chi_{e})$, by using
\begin{eqnarray}\label{D31}
\Gamma(0,z)&\stackrel{z\to 0}{\approx}& -\gamma_{E}-\ln z+z,\nonumber\\
\Gamma(0,z)&\stackrel{z\to \infty}{\approx}& \frac{e^{-z}}{z}.
\end{eqnarray}
For small conductivity, $\sigma\ll \frac{(1+\chi_{e})}{\tau-\tau_{0}}$, we thus arrive at
\begin{eqnarray}\label{D32}
\lefteqn{\hspace{-1cm}
B\approx B_{0}\left(\frac{\tau_0}{\tau}\right)^{a}
}\nonumber\\
&&\hspace{-1cm}
\times\bigg\{1-\frac{\sigma(1-a)}{1+\chi_{e}}\bigg[\tau_{0}\ln\left(\frac{\tau_0}{\tau}\right)+(\tau-\tau_{0})\bigg]\bigg\},
\end{eqnarray}
and
\begin{eqnarray}\label{D33}
\hspace{-1cm}
E&\approx& E_{0}\left(\frac{\tau_0}{\tau}\right)^{a}\bigg\{1-\frac{\sigma}{1+\chi_{e}}\bigg[(1-a)\tau_{0}\ln\left(\frac{\tau_{0}}{\tau}\right)
\nonumber\\
&&+(2-a)(\tau-\tau_{0})\bigg]\bigg\},
\end{eqnarray}
with $a=1-\ell\omega_{0}\beta_{0}$, as in the previous case. Hence, as it turns out, (\ref{D32}) and (\ref{D33}) represent a deviation from the power-law solution (\ref{D19}).
\par
In the case of large conductivity, $\sigma\gg \frac{(1+\chi_{e})}{\tau-\tau_{0}}$,
the magnetic field behaves as
\begin{eqnarray}\label{D34}
B\approx B_{0}\left(\frac{\tau_{0}}{\tau}\right)\left(1+\frac{\ell\omega_{0}\beta_{0}(1+\chi_{e})}{\sigma\tau_{0}}\right),
\end{eqnarray}
while, as expected, the electric field vanishes
\begin{eqnarray}\label{D35}
E=B\beta_{0}e^{-\frac{\sigma(\tau-\tau_{0})}{1+\chi_{e}}}\to 0.
\end{eqnarray}
In Fig. \ref{fig4}, we have plotted the spatial components of rotating magnetic and electric fields $B^{\mu}=(0,\mathbf{B})$ and $E^{\mu}=(0,\mathbf{E})$
\begin{eqnarray}\label{D36}
\frac{\mathbf{B}}{B_{0}}&=&\left(\frac{\tau_{0}}{\tau}\right)e^{\cal{M}(\tau)}\left(\cos\phi(\eta),\sin\phi(\eta),0\right),\nonumber\\
\frac{\mathbf{E}}{E_{0}}&=&\left(\frac{\tau_{0}}{\tau}\right)e^{\cal{N}(\tau)}\left(\cos\zeta(\eta),\sin\zeta(\eta),0\right),
\end{eqnarray}
with ${\cal{M}}$ and ${\cal{N}}$ from (\ref{D27}) and (\ref{D28}), respectively. The angles $\phi(\eta)$ and $\zeta(\eta)$ are given in (\ref{D35}). In Fig. \ref{fig4}(a), the vectors of $\mathbf{B}/B_{0}$ (blue arrows) are plotted with the set of free parameters $\{\sigma,\omega_{0},\beta_{0},\phi_{0},\chi_{e},\chi_{m},\ell\}=\{40,0.1,0.1,\frac{\pi}{6},0,0,+1\}$. The electric vectors $\mathbf{E}/E_{0}$ are parallel to $\mathbf{B}/B_{0}$, and are not demonstrated in this plot. The vectors are slowly rotating with increasing $\eta$. In Fig. \ref{fig4}(b), anti-parallel vectors $\mathbf{B}/B_{0}$ (blue arrows) and $\mathbf{E}/E_{0}$ (red arrows) are plotted with the set of free parameters $\{\sigma,\omega_{0},\beta_{0},\phi_{0},\chi_{e},\chi_{m},\ell\}=\{40,0.1,0.1,\frac{\pi}{6},0,0,-1\}$. A slow rotation is set up with increasing $\eta$.
%%%%%%%%%%%%%%%%%%%%%%%%%%
\section{Numerical results}\label{sec5}
\setcounter{equation}{0}
%%%%%%%%%%%%%%%%%%%%%%%%%%
As it is demonstrated in previous sections, the combination of five partial differential equations (\ref{A15}), (\ref{A18}) and (\ref{A19}), arising from the energy conservation law and Maxwell equations of motion, leads to two different series of solutions for the evolution of electromagnetic and thermodynamic fields. They are essentially characterized by $\frac{d{\cal{M}}}{du}=0$ and $\frac{d{\cal{M}}}{du}\neq 0$. Here, non-vanishing ${\cal{M}}$ describes the deviation from frozen flux theorem $\partial_{\mu}(Bu^{\mu})=0$ of ideal transverse MHD [see (\ref{A25}) and the most general solution of the magnetic field $B$ in non-ideal transverse MHD from (\ref{A32})]. Whereas the solution corresponding to $\frac{d{\cal{M}}}{du}=0$ leads to non-rotating parallel or anti-parallel electric and magnetic fields, the solutions corresponding to $\frac{d{\cal{M}}}{du}\neq 0$ describe rotating $\mathbf{B}$ and $\mathbf{E}$ fields. The proper time evolution of the magnitudes of these fields are shown to be determined by exact and approximate analytical solutions to (\ref{A36}) and (\ref{A37}). We have, in particular, shown that, apart from $B$ and $E$, the proper time evolution of thermodynamic fields $T, p$ and $\epsilon$ are also affected by vanishing or non-vanishing ${\cal{M}}$.
\begin{figure}[hbt]
\includegraphics[width=6.5cm,height=6cm]{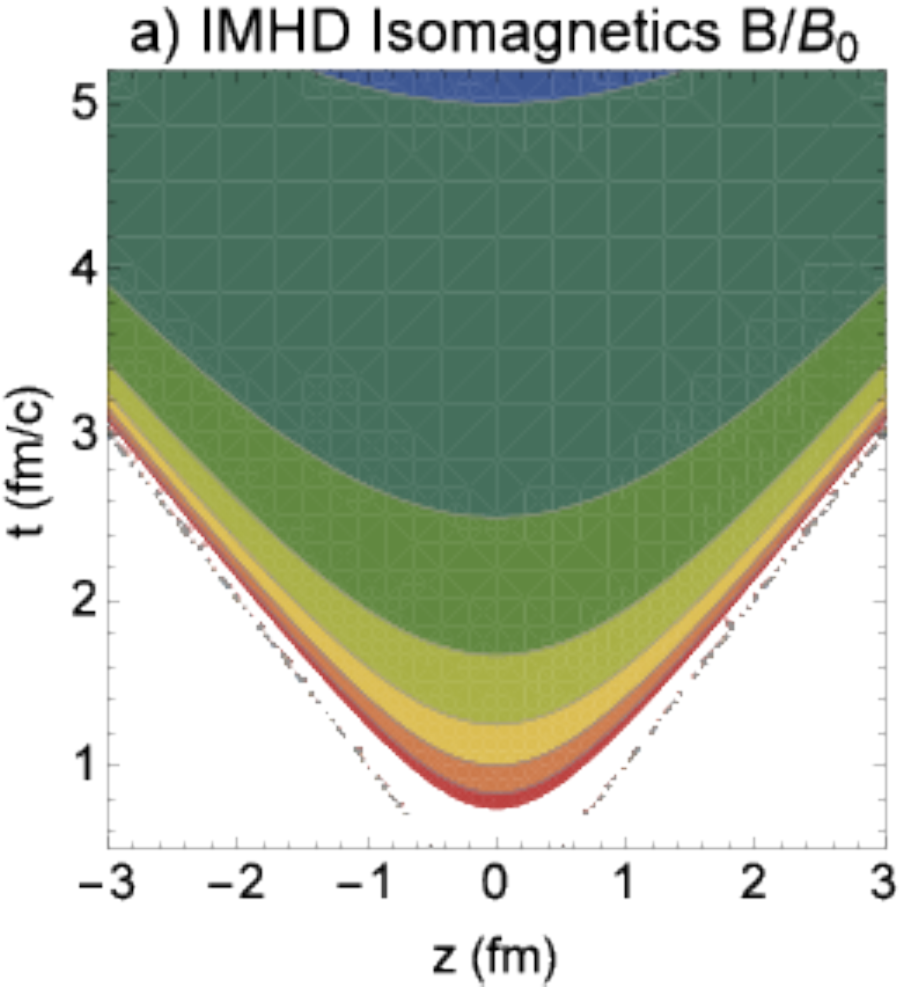}
\includegraphics[width=1cm,height=6cm]{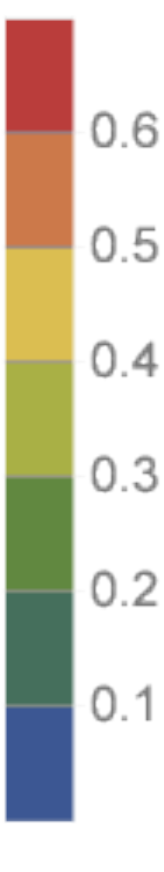}
\includegraphics[width=6.5cm,height=6cm]{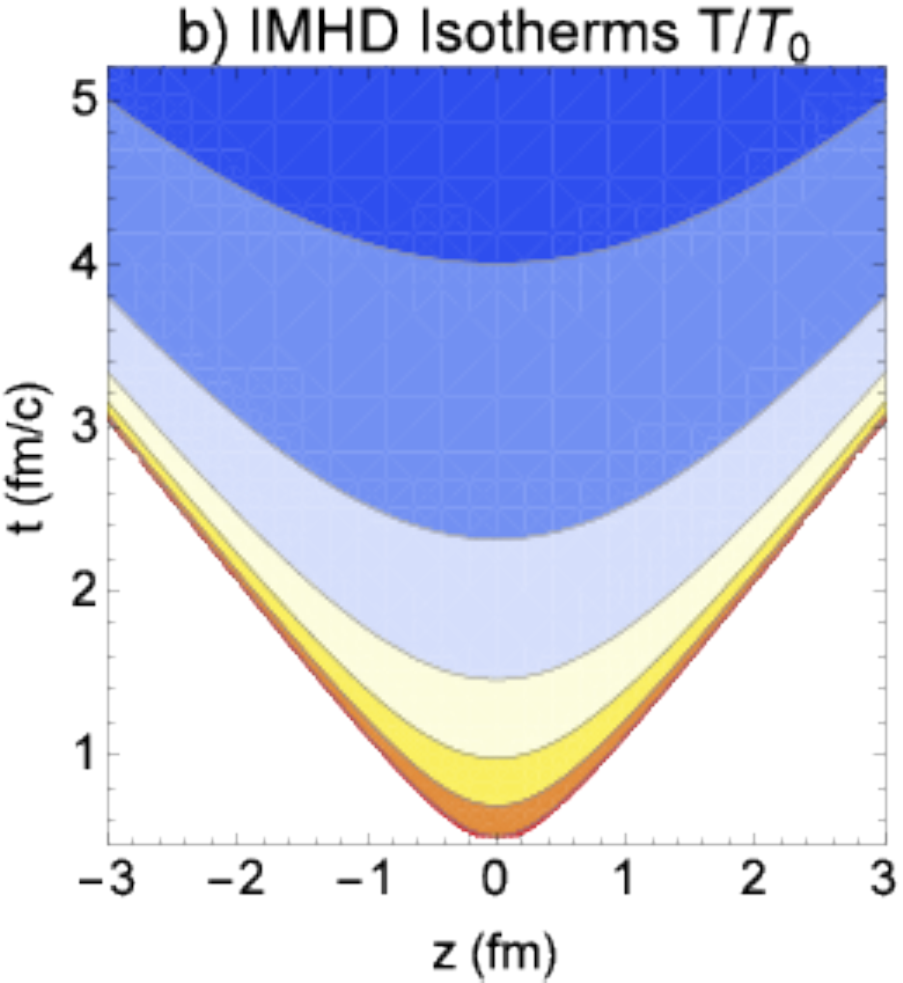}
\includegraphics[width=1cm,height=6cm]{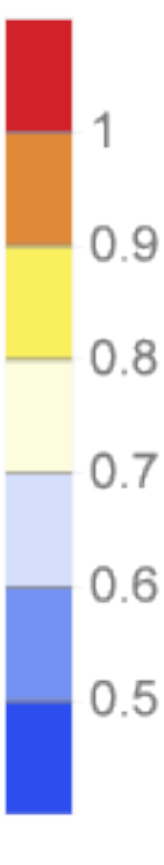}
\caption{(color online). Space-time evolution of $B$ and $T$ in ideal transverse MHD. (a) Isomagnetic fluxes of $B/B_{0}$, (b) isothermal fluxes of $T/T_0$. }\label{fig5}
\end{figure}
\begin{figure*}
\includegraphics[width=5.3cm,height=5cm]{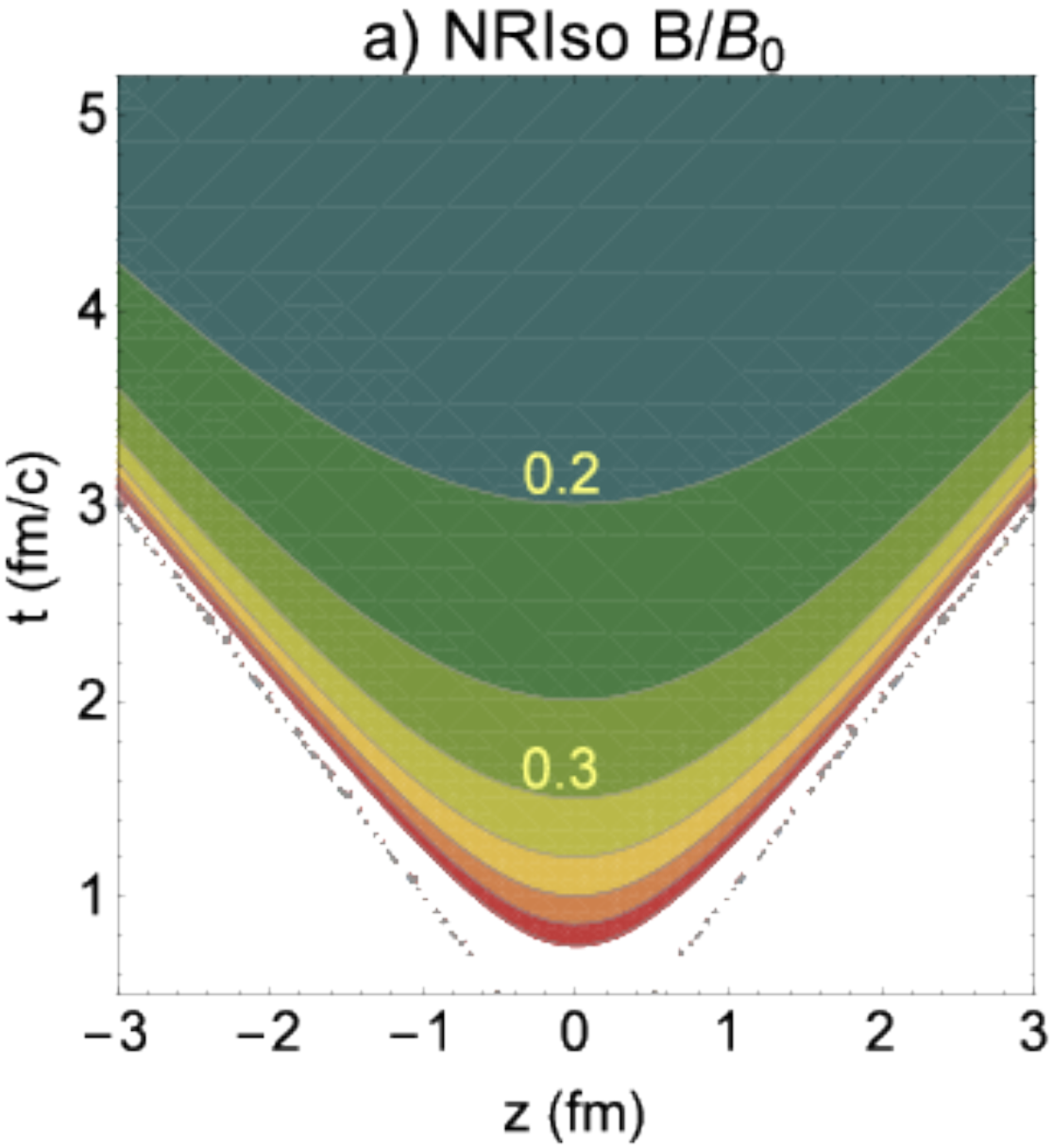}
\includegraphics[width=5.3cm,height=5cm]{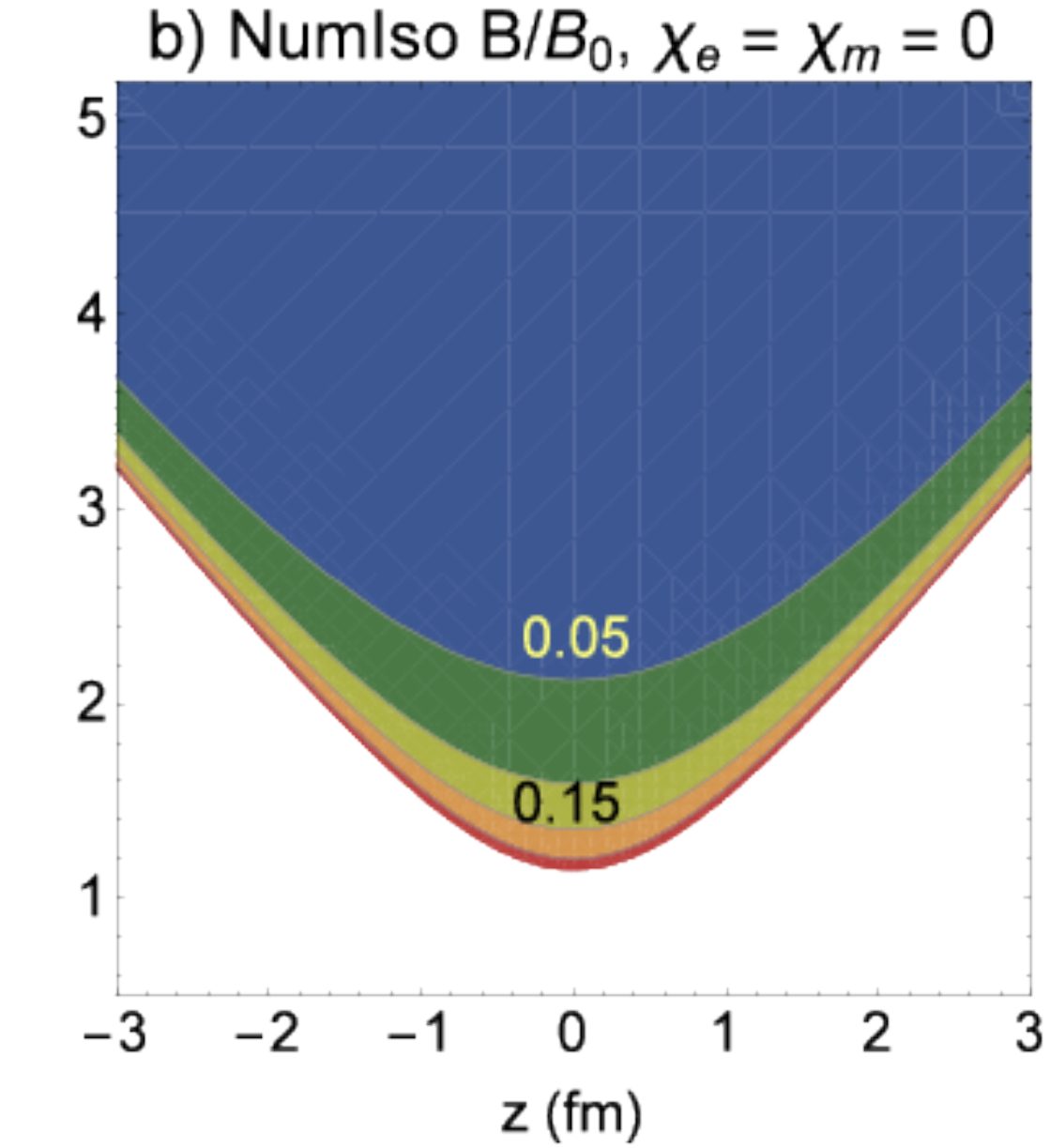}
\includegraphics[width=5.3cm,height=5cm]{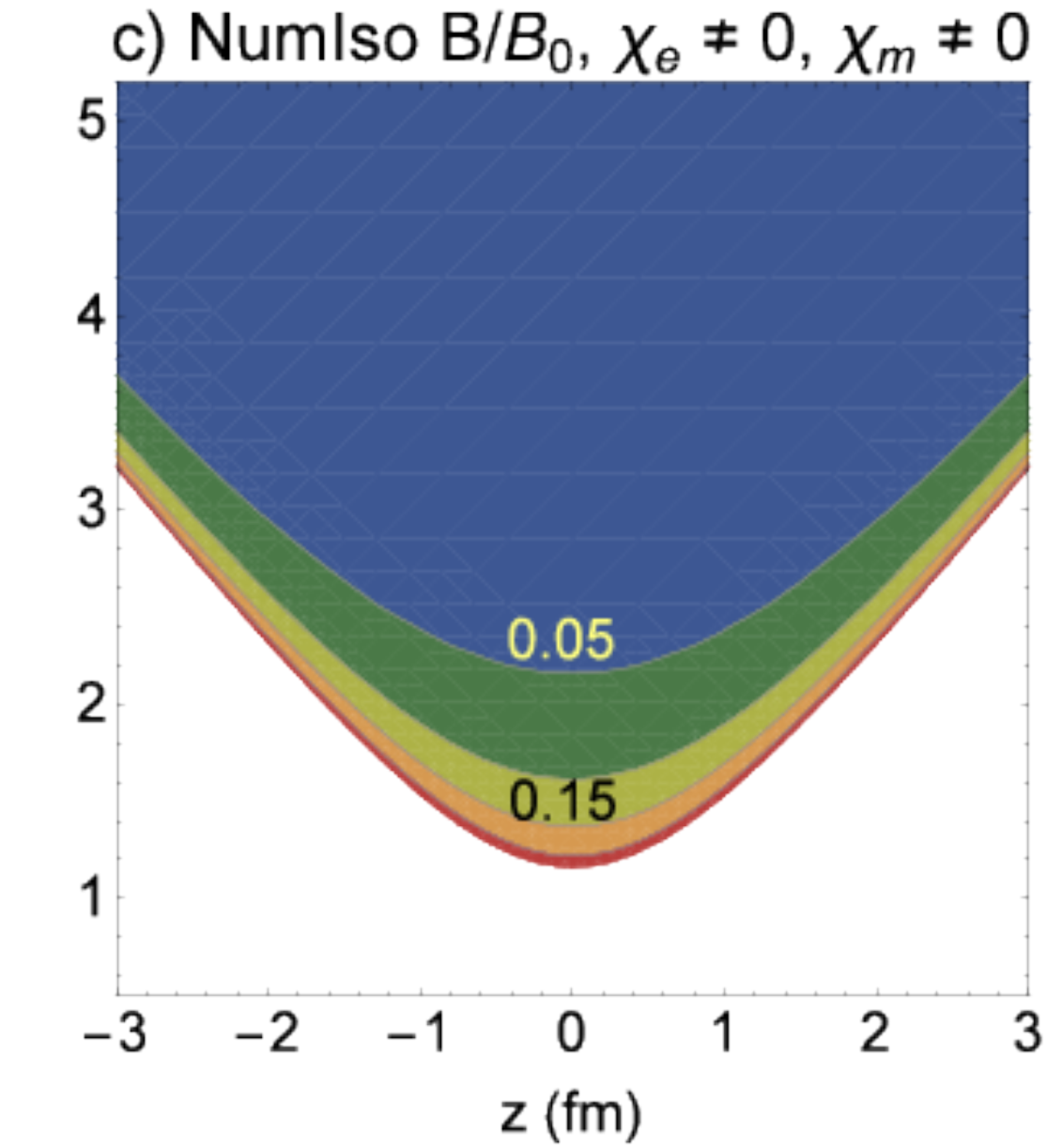}
%%%
\includegraphics[width=5.3cm,height=5cm]{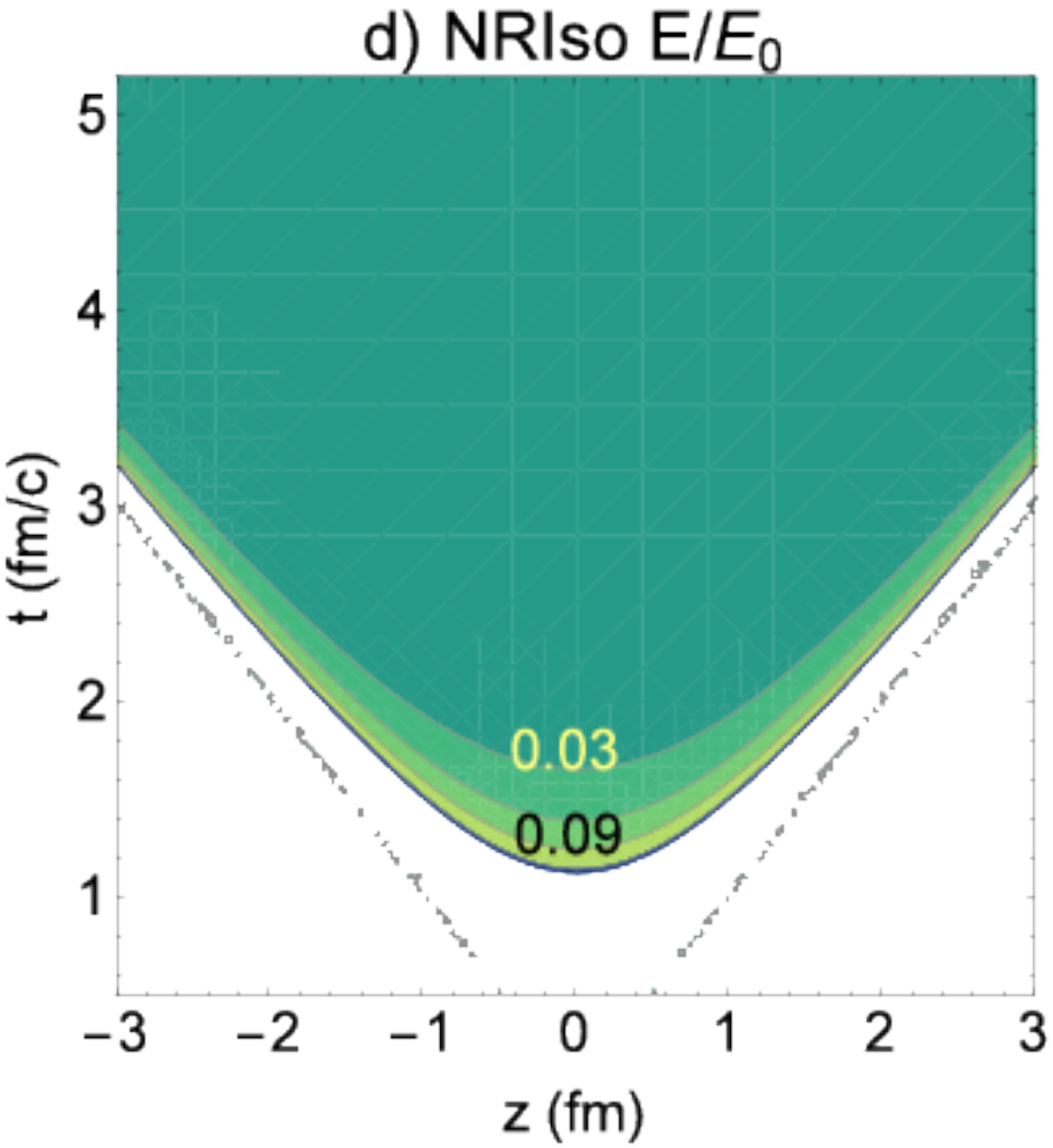}\hspace{-0cm}
\includegraphics[width=5.3cm,height=5cm]{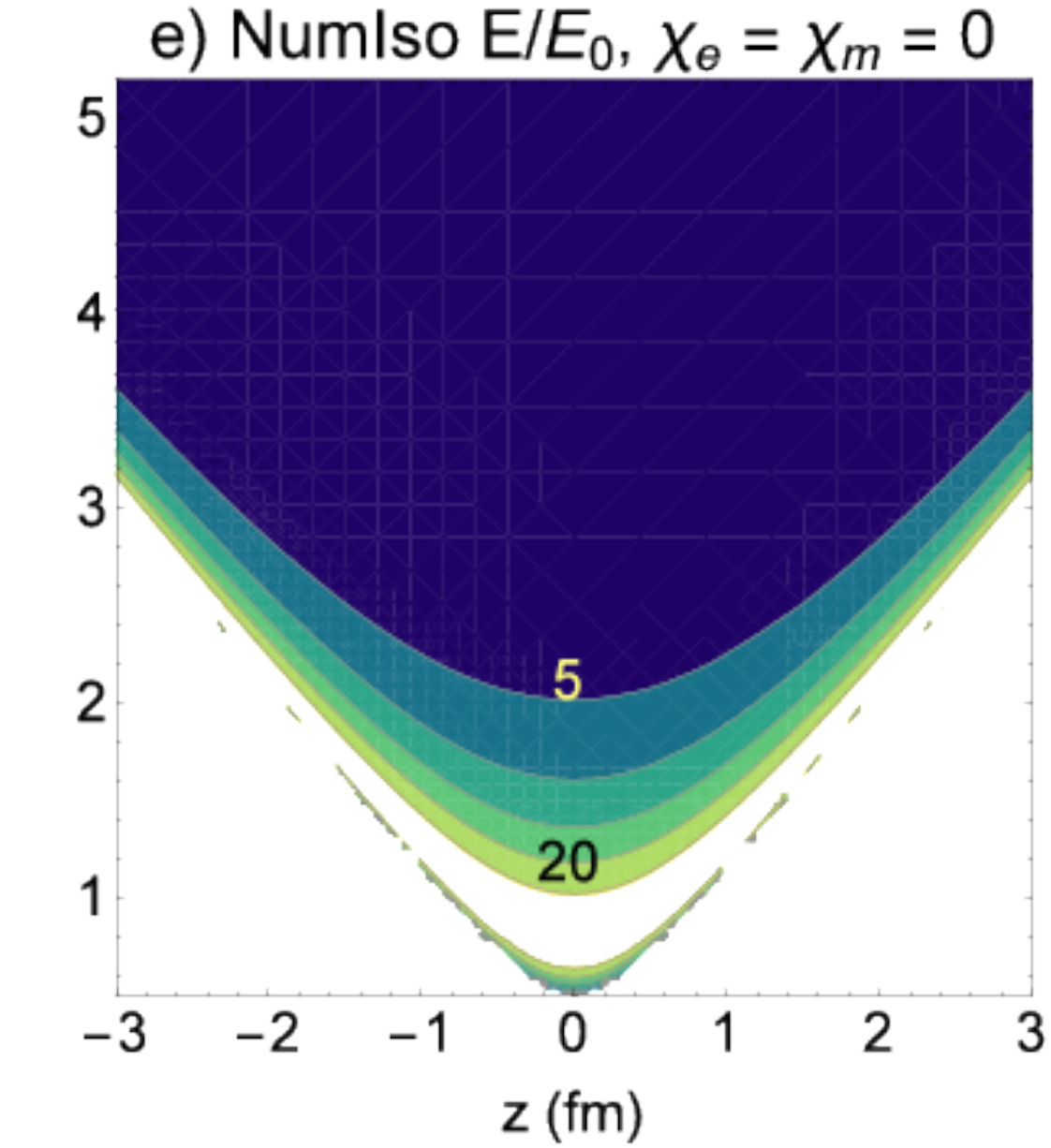}\hspace{-0cm}
\includegraphics[width=5.3cm,height=5cm]{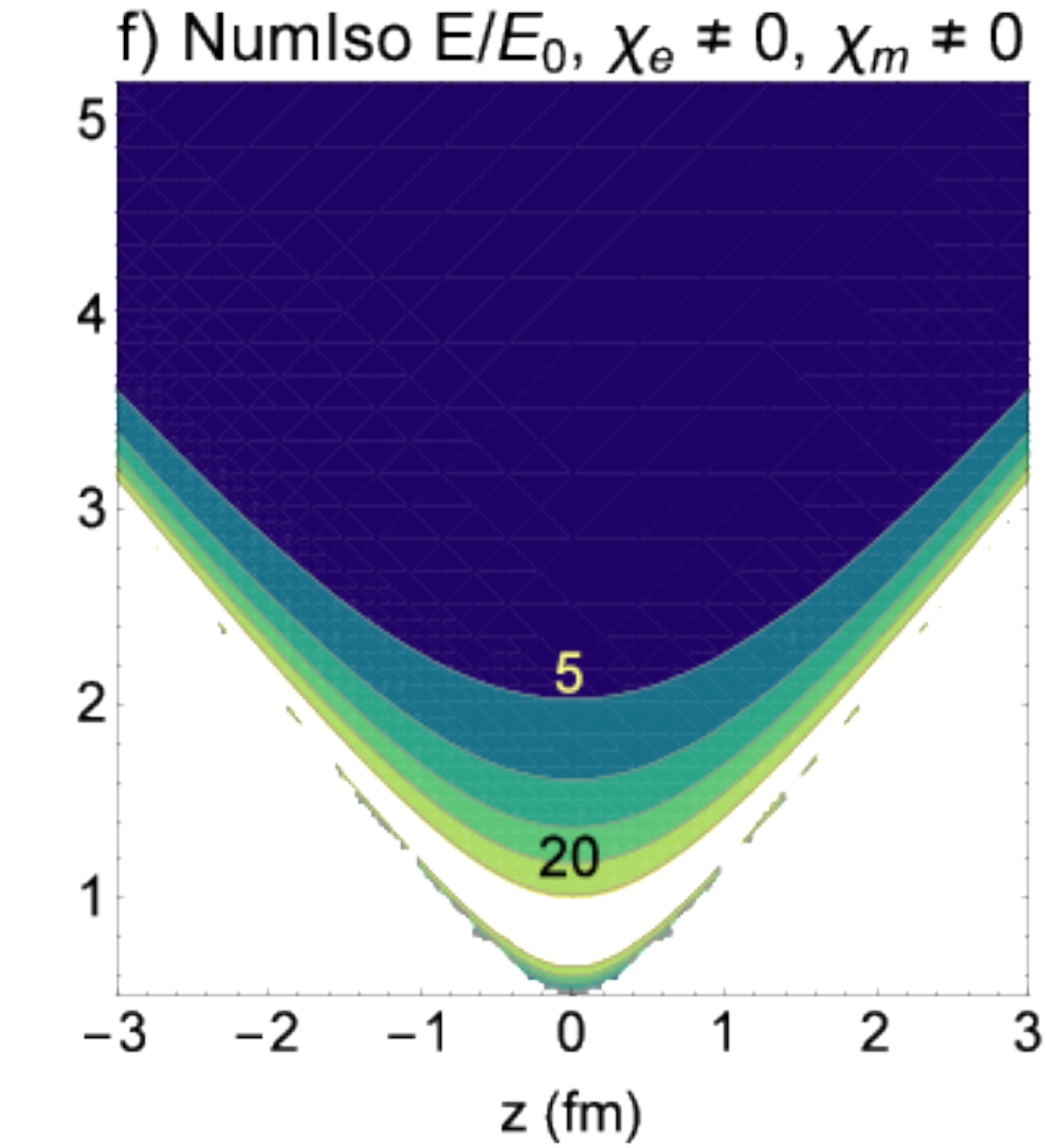}
%%%
\includegraphics[width=5.3cm,height=5cm]{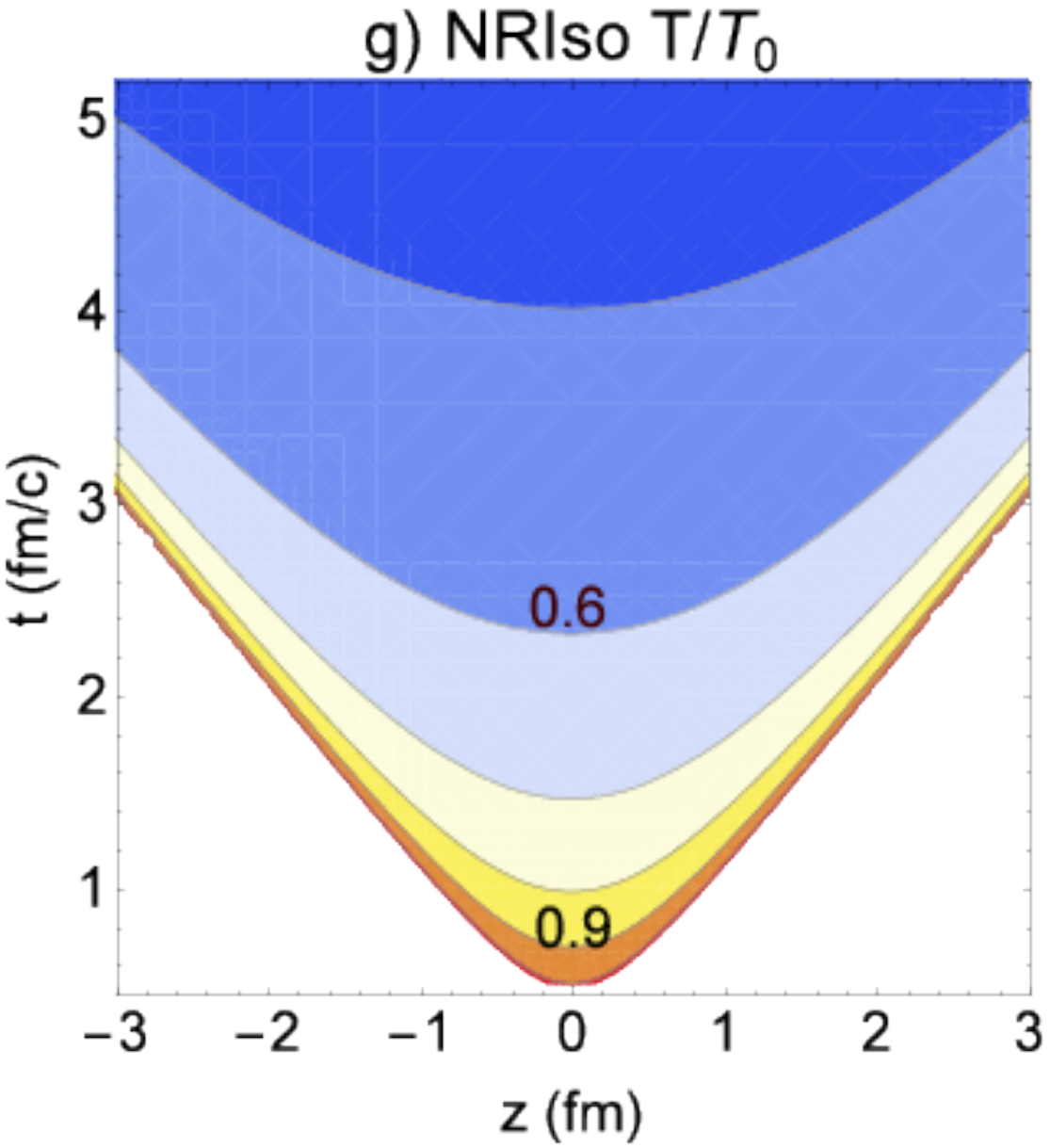}
\includegraphics[width=5.3cm,height=5cm]{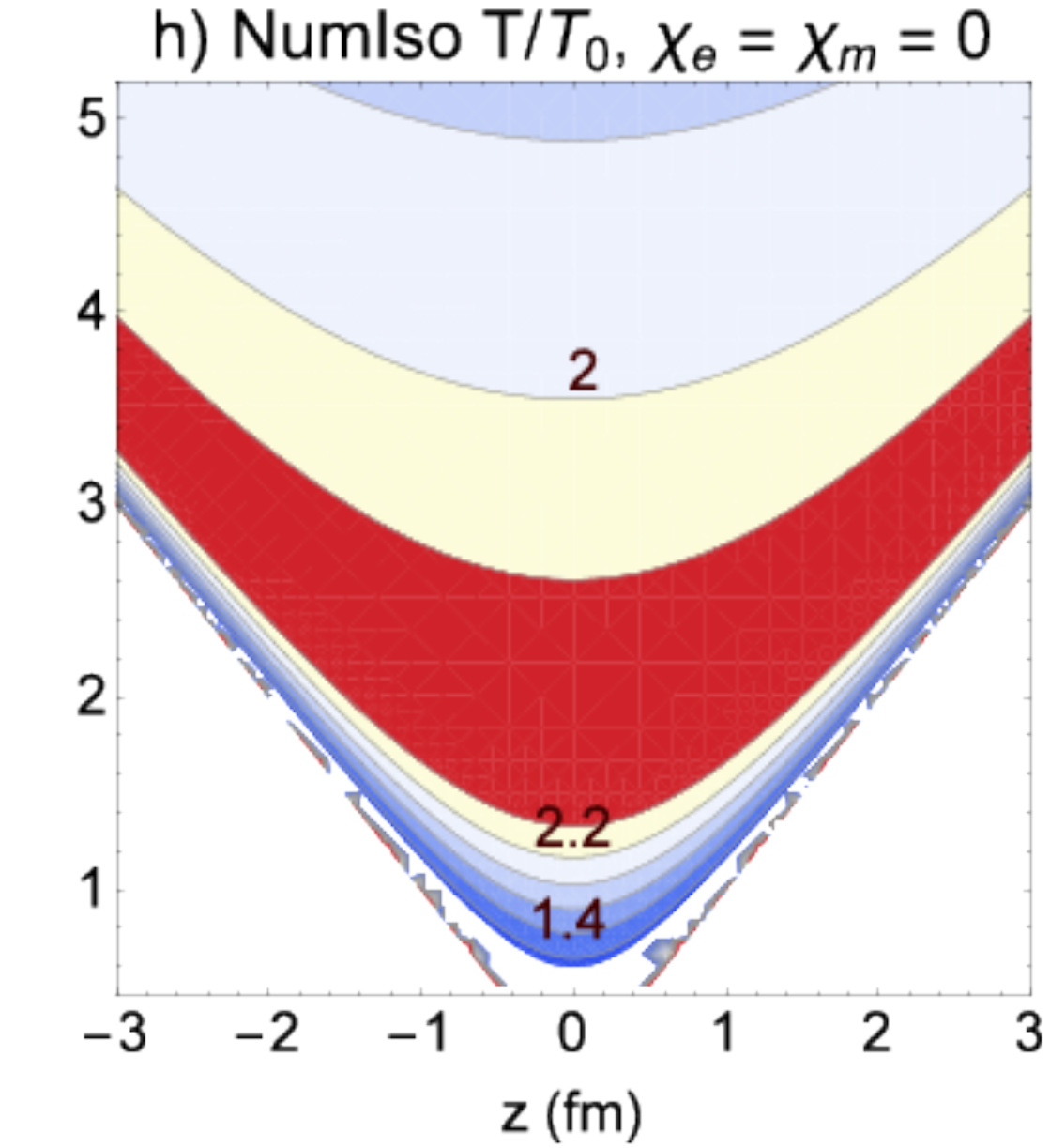}
\includegraphics[width=5.3cm,height=5cm]{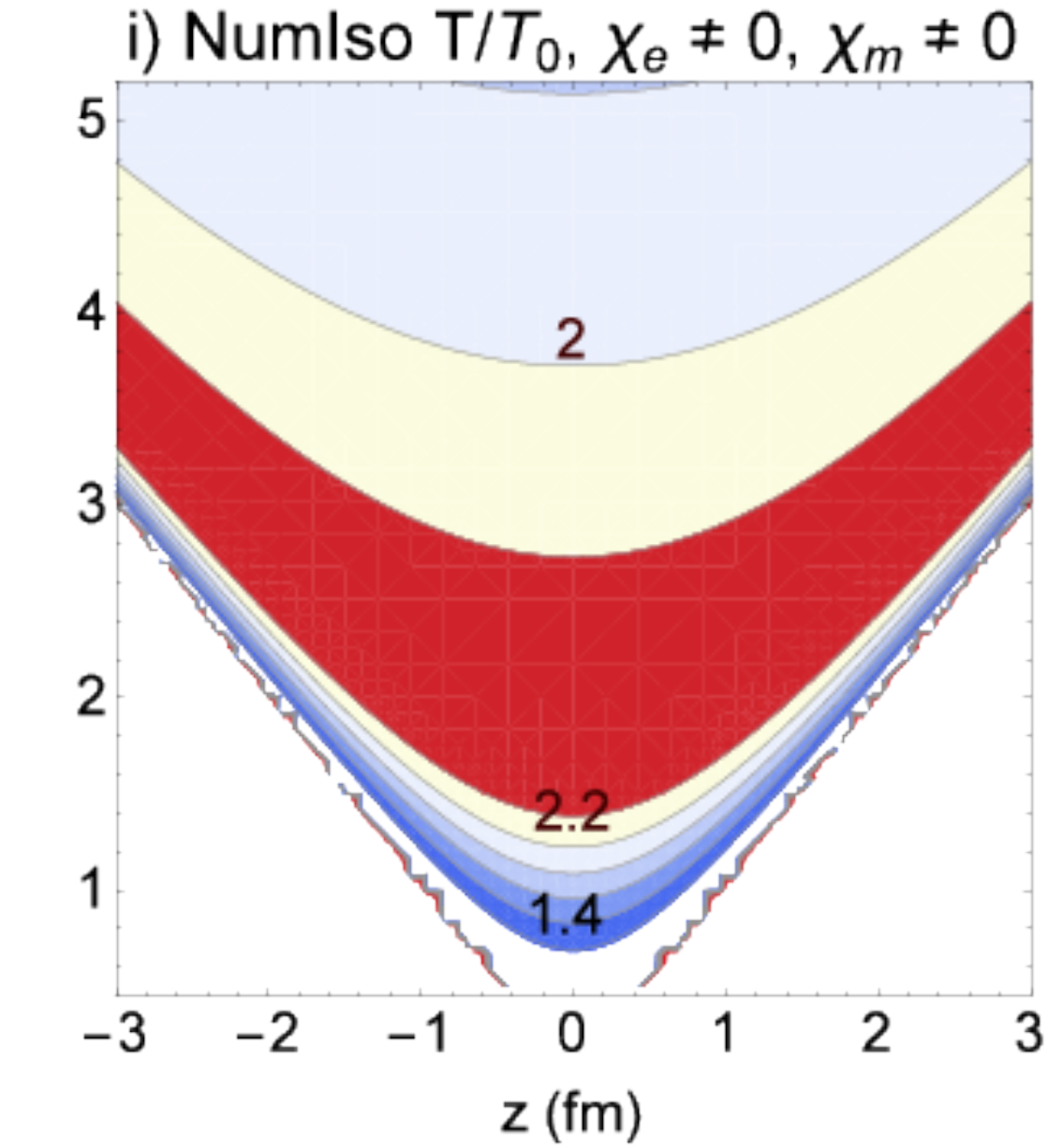}
\caption{(color online). Contour plots for $B/B_{0}$, $E/E_{0}$ and $T/T_{0}$ arising from non-rotating (NR) (panels a, d, and g) and numerical solutions of rotating electromagnetic fields (panels b-c, e-f and h-i).
The set of free parameters  (\ref{OO1}), (\ref{OO2}) and (\ref{OO3}) are chosen to determine the NR (panels a, d, and g) and rotating solutions corresponding to vanishing (panels b, e and h)  and non-vanishing  (panels c, f and i)  susceptibilities. Qualitatively speaking, non-vanishing $\omega_{0}$ affects the space-time evolutions of $B,E$ and $T$. Moreover, non-vanishing electric and magnetic susceptibilities, $\chi_{e}$ and $\chi_{m}$, slightly affect the numerical rotating solutions for $B,E$ and $T$ fluxes. }\label{fig6}
\end{figure*}
\par
In this section, we will use the numerical solution to the master equation (\ref{A37}), and will numerically determine the time evolution of $E=|\mathbf{E}|,B=|\mathbf{B}|$ and $T$.\footnote{The time evolution of $p$ and $\epsilon$ are similar to that of $T$, and will not be presented here.} To demonstrate the effect of rotation, we will qualitatively compare the space-time evolution of rotating and non-rotating (NR) solutions of non-ideal transverse MHD for $E,B$ and $T$ [Sec. \ref{sec5A}]. The cases of vanishing and non-vanishing susceptibilities will be discussed separately. In Sec. \ref{sec5B}, we will present a quantitative analysis on the reliability of approximate solutions corresponding to (\ref{A37}) presented in Sec. \ref{sec4B}. This will be done by comparing the power-law (PL) and slowly rotating (SR) solutions, from (\ref{D22}) and (\ref{D29}), with the numerical solutions for $B,E$, and $T$ arising from (\ref{A37}) and (\ref{A53}), from which we particularly determine ${\cal{M}}$ and ${\cal{N}}$, in combination with (\ref{A32}), (\ref{A33}) and (\ref{A34}).
\par
In Sec. \ref{sec5C}, we will study the effect of various free parameters $\{\Omega_{0},\sigma,\chi_{m}\}$ on the proper time evolution of electromagnetic and thermodynamic fields  $B,E$ and $T$. We will focus on potentially different effects of $\Omega_{0}>0$ and $\Omega_{0}<0$, as well as
$\chi_{m}<0$ and $\chi_{m}>0$, corresponding to dia- and paramagnetic fluids. We will show that with our specific choices of free parameters,\footnote{Free parameters are chosen particularly with regard to the realistic example of QGP.} $\Omega_{0}>0$ leads to negative $E/E_{0}$, and is therefore unphysical. We will therefore consider only the case of  $\Omega_{0}<0$ along with other free parameters.
The effect of $\sigma_{0}\equiv \frac{B_{0}^{2}}{\epsilon_{0}}$, with $B_{0}= B(\tau_{0})$ and $\epsilon_{0}=\epsilon(\tau_{0})$, on the proper time evolution of $T$ will also be discussed in detail. Reformulating $\sigma_{0}$ in terms of $eB_{0}/m_{\pi}^{2}$, with the pion mass $m_{\pi}\sim 0.140$ GeV, it will be then possible to plot $T/T_{0}$ in terms of $eB_{0}/m_{\pi}^{2}$, and compare, in this way, the effect of $B_{0}$ on $T/T_{0}$ at RHIC and LHC. We will choose different sets of $\{\chi_{e}, \chi_{m}\}$ as well as $\Omega_{0}$, and will scrutinize the interplay between these parameters on the gradient of temperature once $eB_{0}$ increases from its value at RHIC ($eB_{0}\sim 1.5 m_{\pi}^{2}$) to its value at LHC ($eB_{0}\sim 15 m_{\pi}^{2}$).
%%%
\subsection{Space-time evolution of $E, B$ and $T$}\label{sec5A}
%%%
In Sec. \ref{sec2}, we studied the proper time evolution of ideal transverse MHD. We argued that in this case, $B$ evolves as (\ref{S23}) with ${\cal{B}}=1$, while $E=0$. We have also shown that the evolution of $T$ is given by (\ref{S19}). Using these relations, we have presented, in Fig. \ref{fig5}, the contour plots of $B/B_{0}$ [Fig. \ref{fig5}(a)] and $T/T_{0}$ [Fig. \ref{fig5}(b)] with ${\cal{U}}=1$. A qualitative comparison shows that the magnetic field decays much faster than the temperature.\footnote{This is related with the fact that $c_{s}<1$. } It declines within $t=5$ fm/c down to 10 percent of its original value at $\tau_{0}=0.5$ fm/c.
\par
In Fig. \ref{fig6}(a)-(i), the contour plots of $B/B_{0}$ [Figs. \ref{fig6}(a)-\ref{fig6}(c)], $E/E_{0}$ [Figs. \ref{fig6}(d)-\ref{fig6}(f)] and $T/T_{0}$ [Figs. \ref{fig6}(g)-\ref{fig6}(i)] are demonstrated for non-rotating and rotating electromagnetic fields.
The results for NR solutions, characterized by vanishing $d{\cal{M}}/du$, are demonstrated in Figs. \ref{fig6}(a), \ref{fig6}(d) and \ref{fig6}(g). They correspond to (\ref{D8}) and (\ref{A29}) with $\exp\left(\frac{\cal{L}}{\kappa}\right)$ from (\ref{D10}) and  $\mathbb{V}=1$. Here, the set of free parameters is chosen to be
\begin{eqnarray}\label{OO1}
\hspace{-0.9cm}
\{\sigma_{0},\sigma,\beta_{0},\chi_{e},\chi_{m}\}=\{10, 400~\mbox{MeVc}, 0.01,0,0\}.
\end{eqnarray}
To plot the space-time evolution of the numerical solutions for $B/B_{0}$, $E/E_{0}$ and $T/T_{0}$, denoted by ``NumIso'' $B/B_{0}$, $E/E_{0}$ and $T/T_{0}$, the set of free parameters
\begin{eqnarray}\label{OO2}
\lefteqn{\hspace{-0.9cm}
\{\sigma_{0},\sigma, \Omega_{0},\beta_{0},\chi_{e},\chi_{m}\}
}\nonumber\\
&&=\{10, 400~\mbox{MeVc},-1.5, 0.01,0,0\},
\end{eqnarray}
[Figs. \ref{fig6}(b), \ref{fig6}(e) and \ref{fig6}(h)] as well as
\begin{eqnarray}\label{OO3}
\lefteqn{\hspace{-0.9cm}
\{\sigma_{0},\sigma, \Omega_{0},\beta_{0},\chi_{e},\chi_{m}\}
}\nonumber\\
&&
=\{10, 400~\mbox{MeVc},-1.5, 0.01,0.01,0.01\},
\end{eqnarray}
[Figs. \ref{fig6}(c), \ref{fig6}(f) and \ref{fig6}(i)] are used. First, using these parameters, the master equation (\ref{A37}) is numerically solved. Plugging then ${\cal{M}}$ arising from this equation into (\ref{A53}), ${\cal{N}}$ is determined. The space-time evolution of $B/B_{0}$, $E/E_{0}$ and $T/T_{0}$ are then determined by plugging ${\cal{M}}$ and ${\cal{N}}$
into (\ref{A32}), (\ref{A33}) and (\ref{A34}). The latter leads, in combination with (\ref{A29}) with $\mathbb{V}=1$, to the numerical solution of $T/T_{0}$.
\par
A comparison between NR result for the space-time evolution $B/B_{0}$ from Fig. \ref{fig6}(a) shows that the magnetic field decays faster in the case of non-vanishing $\frac{d{\cal{M}}}{du}$ (rotating electromagnetic fields) with vanishing [Fig. \ref{fig6}(b)] and non-vanishing susceptibilities [Fig. \ref{fig6}(c)]. In all these cases $B/B_{0}$ decays monotonically with $t$ to values of $B\ll B_{0}$. In contrast, the numerical results for $E/E_{0}$ and $T/T_{0}$ exhibit a completely different behavior. Qualitatively, $E/E_{0}$ and $T/T_{0}$ increase rapidly with increasing $t\leq 2$ fm/c to values of $E$ and $T$ larger than their original values $E_{0}$ and $T_{0}$. Then, in the interval $2\lesssim t\lesssim 5$ fm/c, they decay slowly to values which are still larger than $E_{0}$ and $T_{0}$ [see Sec. \ref{sec5C} and Appendix]. Non-vanishing susceptibilities do not affect this qualitative picture too much.
\par
In Sec. \ref{sec5C}, we will present, among others, a careful quantitative analysis of the effect of $\Omega_{0}$ and $\chi_{m}$ on the proper time evolution of $B/B_{0}$, $E/E_{0}$ and $T/T_{0}$.
%%%%%%%%%%%%%%%%%%%%%%%%
\subsection{Reliability of analytical solutions of the master equation (\ref{A37}): A qualitative analysis}\label{sec5B}
%%%%%%%%%%%%%%%%%%%%%%%%
In Sec. \ref{sec4B}, two approximate analytical solutions for the master equation (\ref{A37}) have been presented. The first case, for which $\frac{E}{B}\sim \frac{E_{0}}{B_{0}}$ was assumed, leads to PL, and the second one, for which $\omega_{0}$ was assumed to be small, leads to SR solutions for the proper time evolution of $B,E$ and $T$.
%%%%%%%%%%%%%%%% fig 7
\begin{figure}[hbt]
\includegraphics[width=8cm,height=6cm]{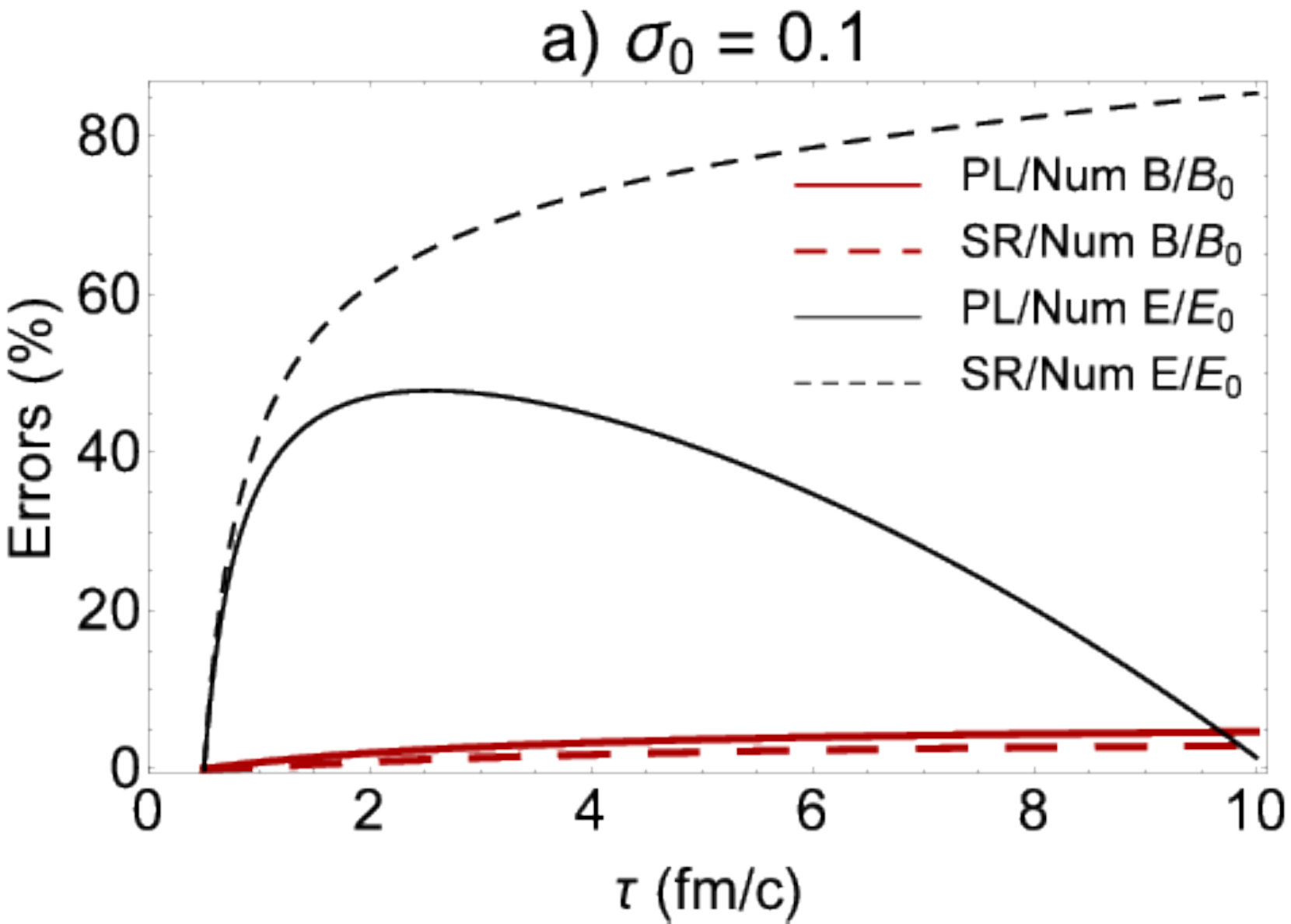}
\includegraphics[width=8cm,height=6cm]{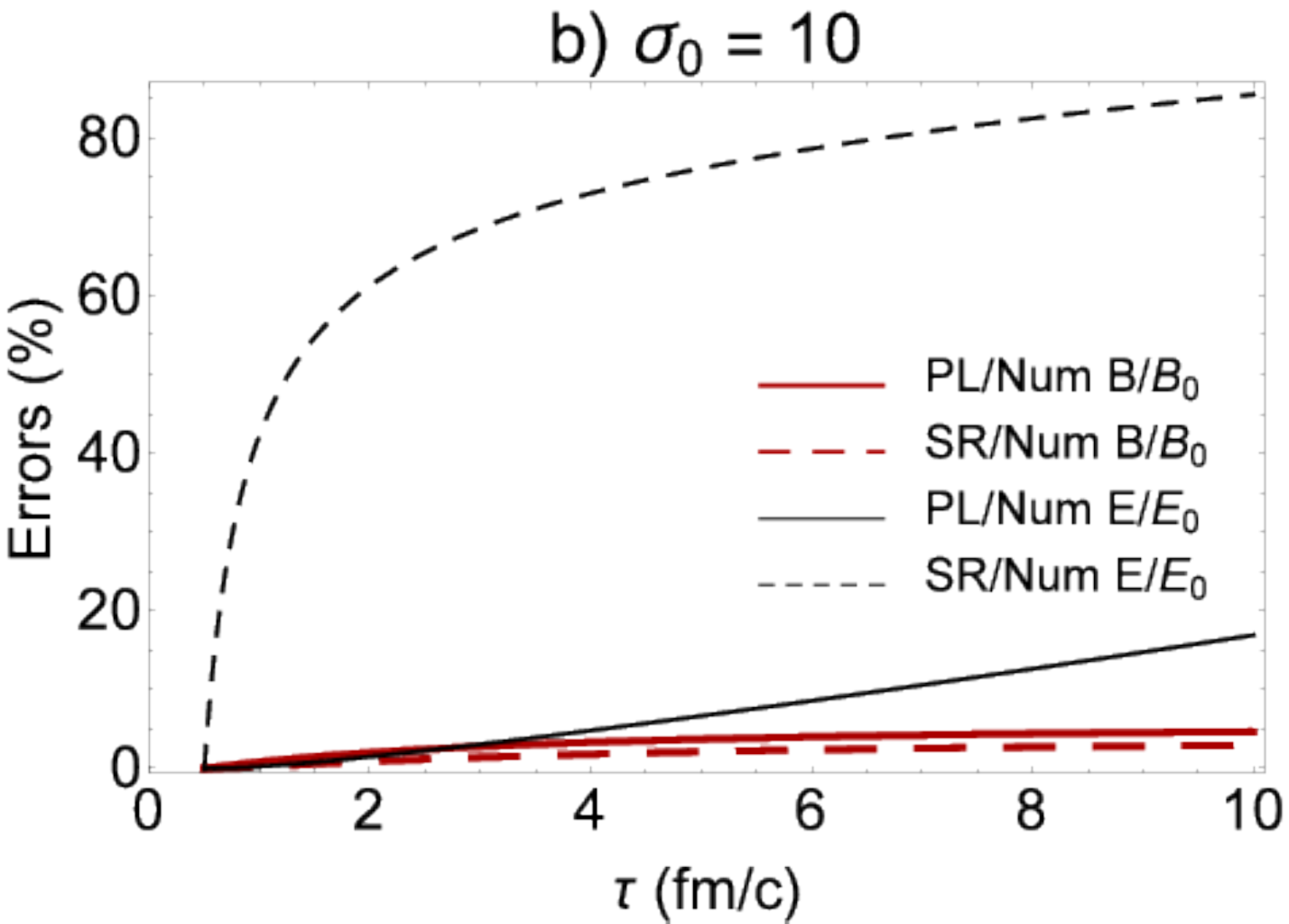}
\caption{(color online). The $\tau$ dependence of the deviation of power-law (PL) and slowly rotating (SR) approximate analytical solutions of $B/B_{0}$ (red solid and dashed curves) and $E/E_{0}$ (thin black solid and dashed curves) from the corresponding numerical solutions are plotted for two sets of free parameters (\ref{OO5}) (panel a) and (\ref{OO6}) (panel b).  The results for these deviations are strongly affected by the choice of free parameters, and increase, in general, with increasing $\tau$. For $B$ ($E$), the SR (PL) solution is more reliable than the PL (SR) solution.}\label{fig7}
\end{figure}
%%%%%%%%%%%%%%%% fig 8
\begin{figure}[hbt]
\includegraphics[width=8cm,height=6cm]{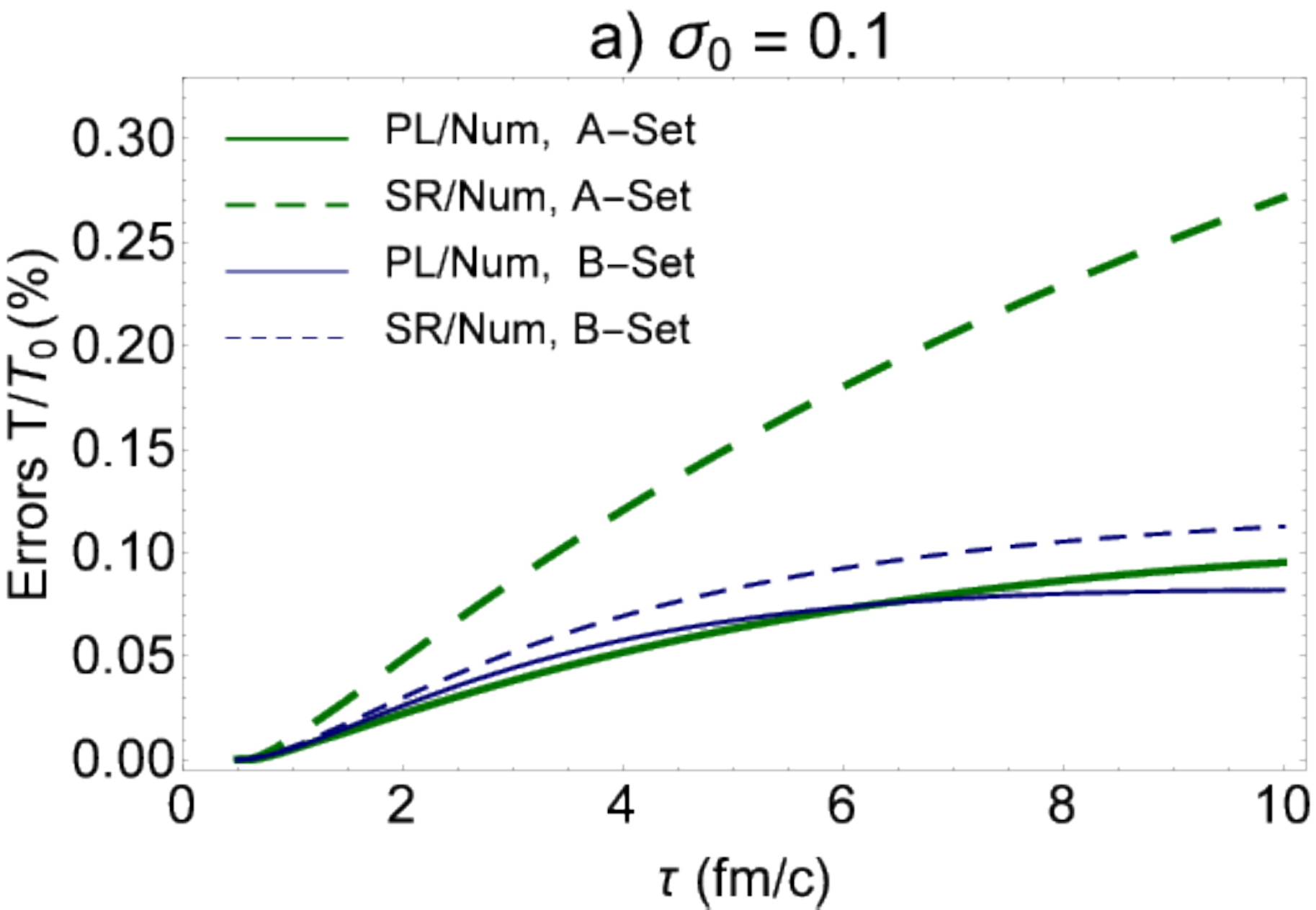}
\includegraphics[width=8cm,height=6cm]{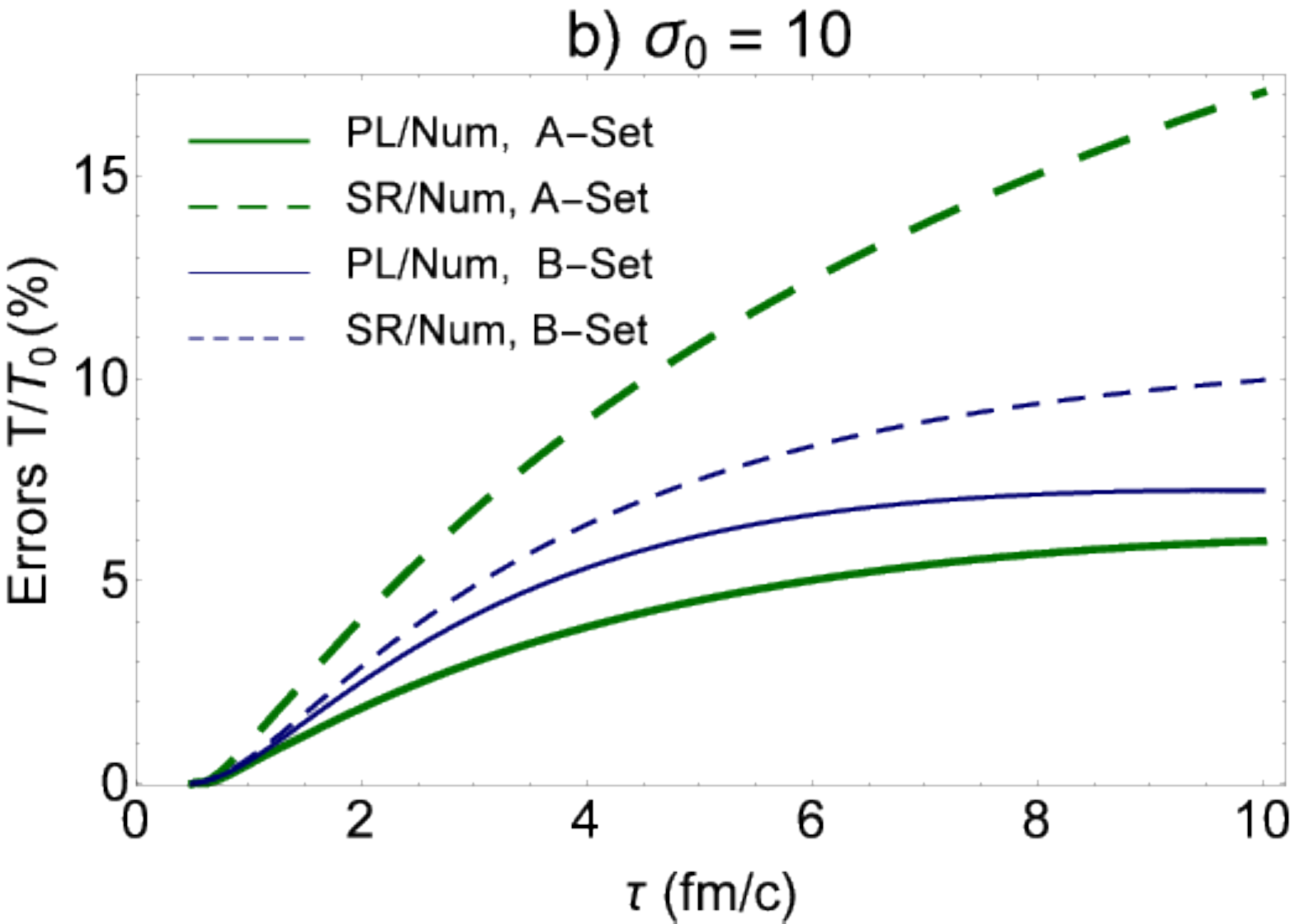}
\caption{(color online). The $\tau$ dependence of the deviation of power-law (PL) and slowly rotating (SR) approximate analytical solutions of $T/T_{0}$ from the corresponding numerical solutions to $T/T_{0}$ is plotted for two sets of free parameters (\ref{OO7}) (panel a) and  (\ref{OO8}) (panel b).  The green (blue) solid and dashed curves correspond to A- (B-) sets defined in (\ref{OO7}) and (\ref{OO8}).  As expected, the deviations of PL and SR solutions from numerical solution for $T/T_{0}$ are strongly affected by the choice of parameters.  }\label{fig8}
\end{figure}
\par
In this section, we will quantitatively determine the deviation of these approximate analytical solutions from the numerical solutions of these fields.
This deviation is determined from
\begin{eqnarray}\label{OO4}
\lefteqn{\mbox{Error in \%}}\nonumber\\
&\equiv& \bigg|\frac{\mbox{approx. analytical}-\mbox{numerical solution}}{\mbox{numerical solution}}\bigg|\times 100.\nonumber\\
\end{eqnarray}
The $\tau$ dependence of these errors for $B/B_{0}$ and $E/E_{0}$ are presented in Fig. \ref{fig7}. The sets of free  parameters
\begin{eqnarray}\label{OO5}
\lefteqn{\hspace{-0.5cm}
\{\tau_{0},\sigma_{0}, \sigma, \Omega_{0}, \beta_{0},\chi_{e},\chi_{m}\}}\nonumber\\
&&\hspace{-0.5cm}=\{0.5~\mbox{fm/c}, 0.1, 40~\mbox{MeVc}, - 0.1,0.1, 0,0\},
\end{eqnarray}
and
\begin{eqnarray}\label{OO6}
\lefteqn{\hspace{-0.5cm}
\{\tau_{0},\sigma_{0}, \sigma, \Omega_{0}, \beta_{0},\chi_{e},\chi_{m}\}}\nonumber\\
&&\hspace{-0.5cm}=\{0.5~\mbox{fm/c}, 10, 40~\mbox{MeVc}, - 0.1,0.1, 0,0\},
\end{eqnarray}
are used in Figs. \ref{fig7}(a) and \ref{fig7}(b), respectively.
As it turns out from the results of Fig. \ref{fig7}, for $B/B_{0}$, the SR approximate solution (red dashed curves) is more reliable than the PL solution (red solid curves).  In contrary, for $E/E_{0}$, the errors for PL solutions (black solid curves) in the whole range of $\tau\in[0,10]$ fm/c is smaller than those for SR solution (black dashed curves) in the same proper time interval. Except the deviation of PL from numerical solutions for $E/E_{0}$ in Fig. \ref{fig7}(a) for $\sigma_{0}=0.1$ (black solid curve), the errors increase with increasing $\tau$. Moreover, in contrast to the deviations of $B/B_{0}$ from the numerical solutions, which is in general lower than $20\%$, the deviations of $E/E_{0}$ are larger, and, depending on the choice of free parameters raise up to $80\%$. Increasing $\sigma_{0}$ from $\sigma_{0}=0.1$ to $\sigma_{0}=10$ does not change this picture too much.
\par
In Fig. \ref{fig8}, the $\tau$ dependence of the deviation of PL (solid curves) and SR (dashed curves) solutions of $T/T_{0}$ from its numerical solution are plotted for two set of free parameters
\begin{eqnarray}\label{OO7}
\lefteqn{\hspace{-0.9cm}\mbox{A-Set:}~~
\{\tau_{0},\sigma_{0}, \sigma, \Omega_{0}, \beta_{0},\chi_{e},\chi_{m}\}
}\nonumber\\
&&\hspace{-0.5cm}=\{0.5~\mbox{fm/c}, 0.1, 4~\mbox{MeVc}, - 0.1,1, 0,0\}.\nonumber\\
\lefteqn{\hspace{-0.9cm}\mbox{B-Set:}~~
\{\tau_{0},\sigma_{0}, \sigma, \Omega_{0}, \beta_{0},\chi_{e},\chi_{m}\}
}\nonumber\\
&&\hspace{-0.5cm}=\{0.5~\mbox{fm/c}, 0.1, 40~\mbox{MeVc}, - 0.1,0.1, 0,0\},
\end{eqnarray}
in Fig. \ref{fig8}(a), and
\begin{eqnarray}\label{OO8}
\lefteqn{
\mbox{\hspace{-0.9cm}A-Set:}~~
\{\tau_{0},\sigma_{0}, \sigma, \Omega_{0}, \beta_{0},\chi_{e},\chi_{m}\}
}\nonumber\\
&&\hspace{-0.5cm}=\{0.5~\mbox{fm/c}, 10, 4~\mbox{MeVc}, - 0.1,1, 0,0\}.\nonumber\\
\lefteqn{\hspace{-0.9cm}\mbox{B-Set:}~~
\{\tau_{0},\sigma_{0}, \sigma, \Omega_{0}, \beta_{0},\chi_{e},\chi_{m}\}
}\nonumber\\
&&\hspace{-0.5cm}=\{0.5~\mbox{fm/c}, 10, 40~\mbox{MeVc}, - 0.1,0.1, 0,0\},
\end{eqnarray}
in Fig. \ref{fig8}(b). The green (blue) solid and dashed curves correspond to A- (B-) sets defined in (\ref{OO7}) and (\ref{OO8}).
Comparing with the errors of $B/B_{0}$ and $E/E_{0}$, presented in Fig. \ref{fig7}, the errors for $T/T_{0}$ are much smaller. They increase with increasing $\tau$, and strongly depend on the choice of free parameters, especially $\sigma_{0}$ [compare Figs. \ref{fig8}(a) with \ref{fig8}(b)] and $\{\sigma,\beta_{0}\}$ pairs (compare the results for A- and B-sets). In general, similar to the case of $E/E_{0}$, the PL solution for $T/T_{0}$ is more reliable than the SR solution.
\par
In summary, the above analysis shows that the deviation of analytical PL and SR solutions to (\ref{A37}) from the numerical solution to the same equation depends strongly on the choice of the set of free parameters $\{\tau_{0},\sigma_{0}, \sigma, \Omega_{0}, \beta_{0},\chi_{e},\chi_{m}\}$. In what follows, we will focus solely on $B,E$ and $T$, arising from numerical solution of (\ref{A37}).
%%%%%%%%%%%%%%%% fig 9
\begin{figure*}[hbt]
\includegraphics[width=5.6cm,height=4.4cm]{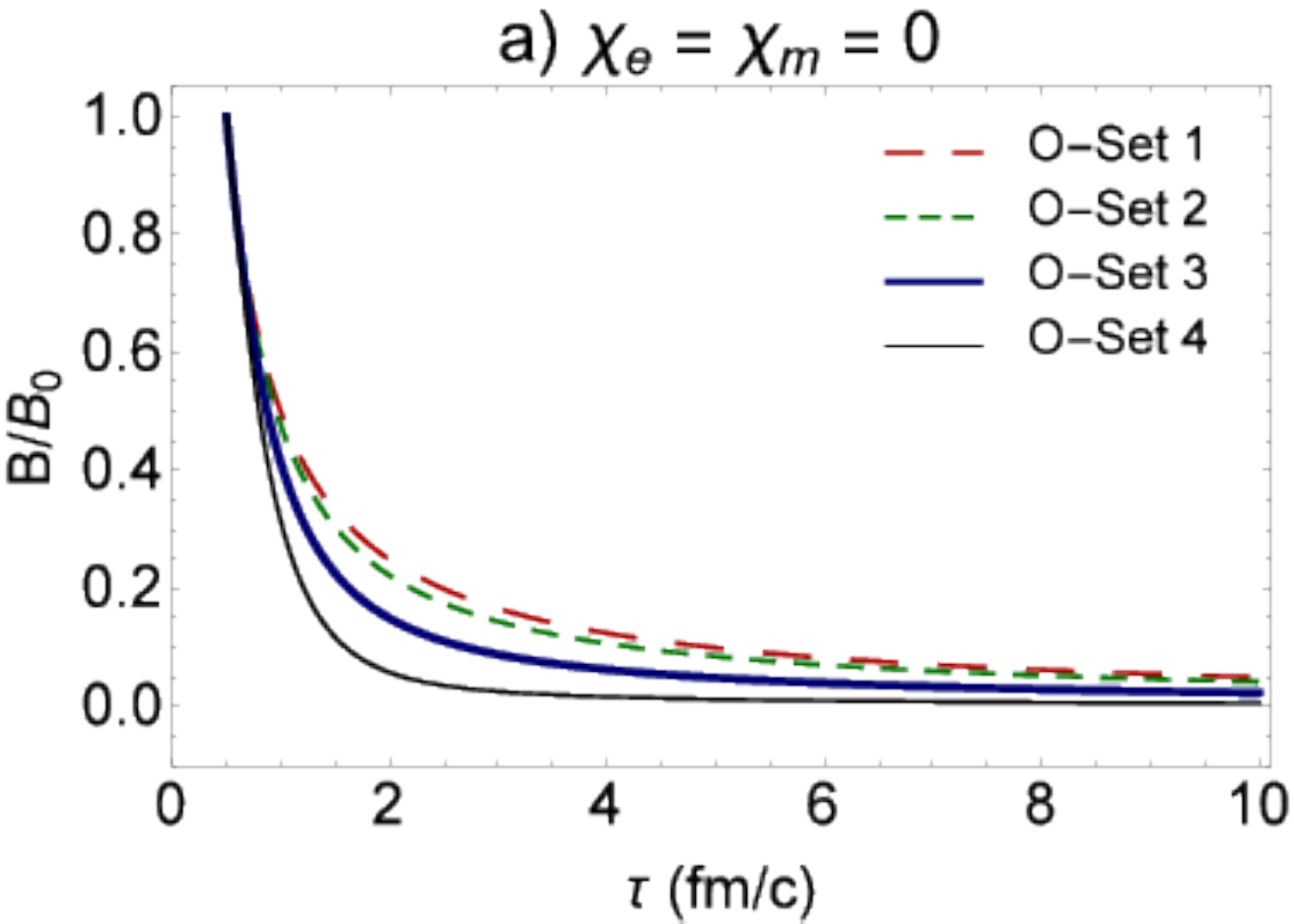}
\includegraphics[width=5.6cm,height=4.4cm]{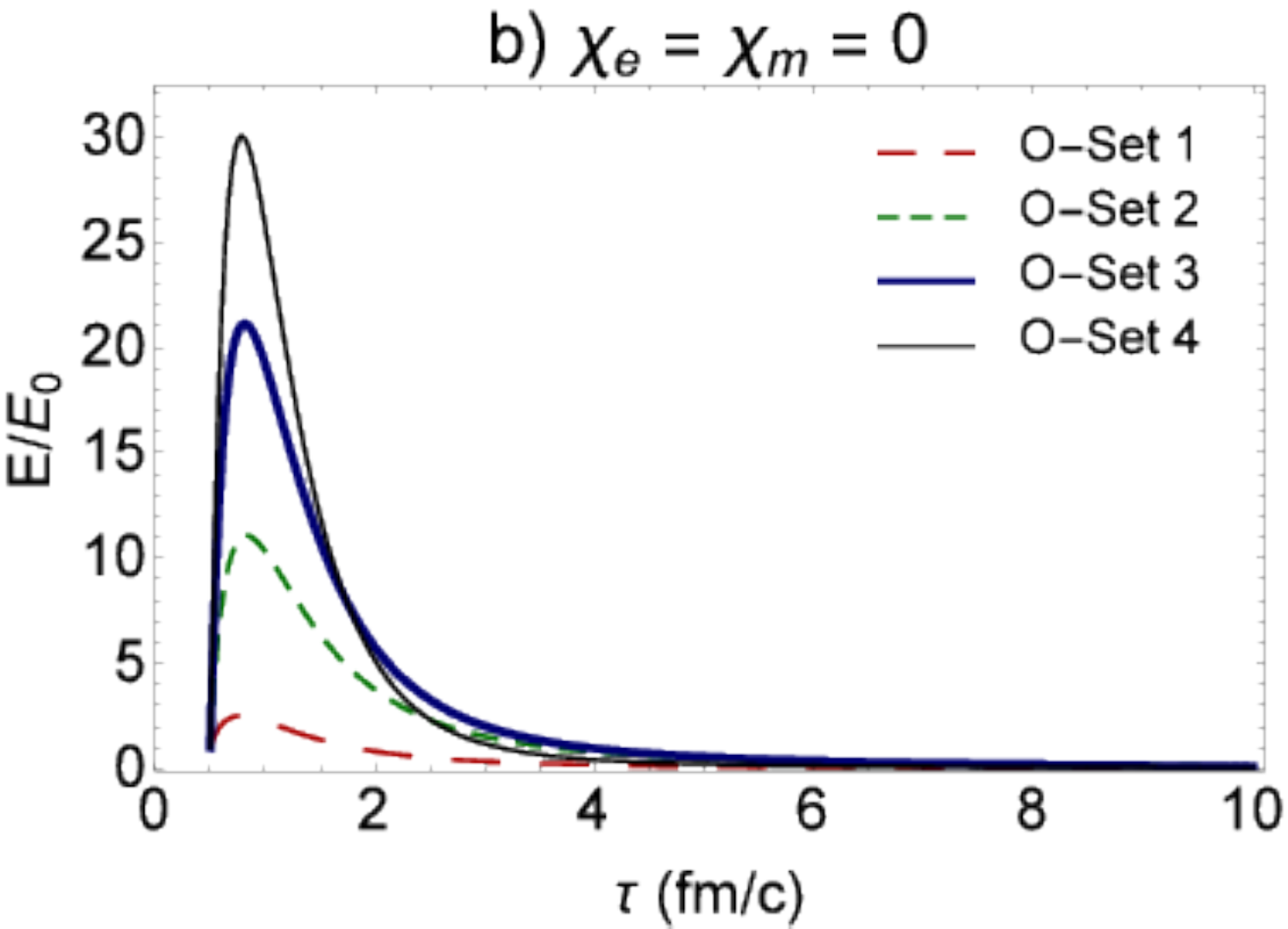}
\includegraphics[width=5.6cm,height=4.4cm]{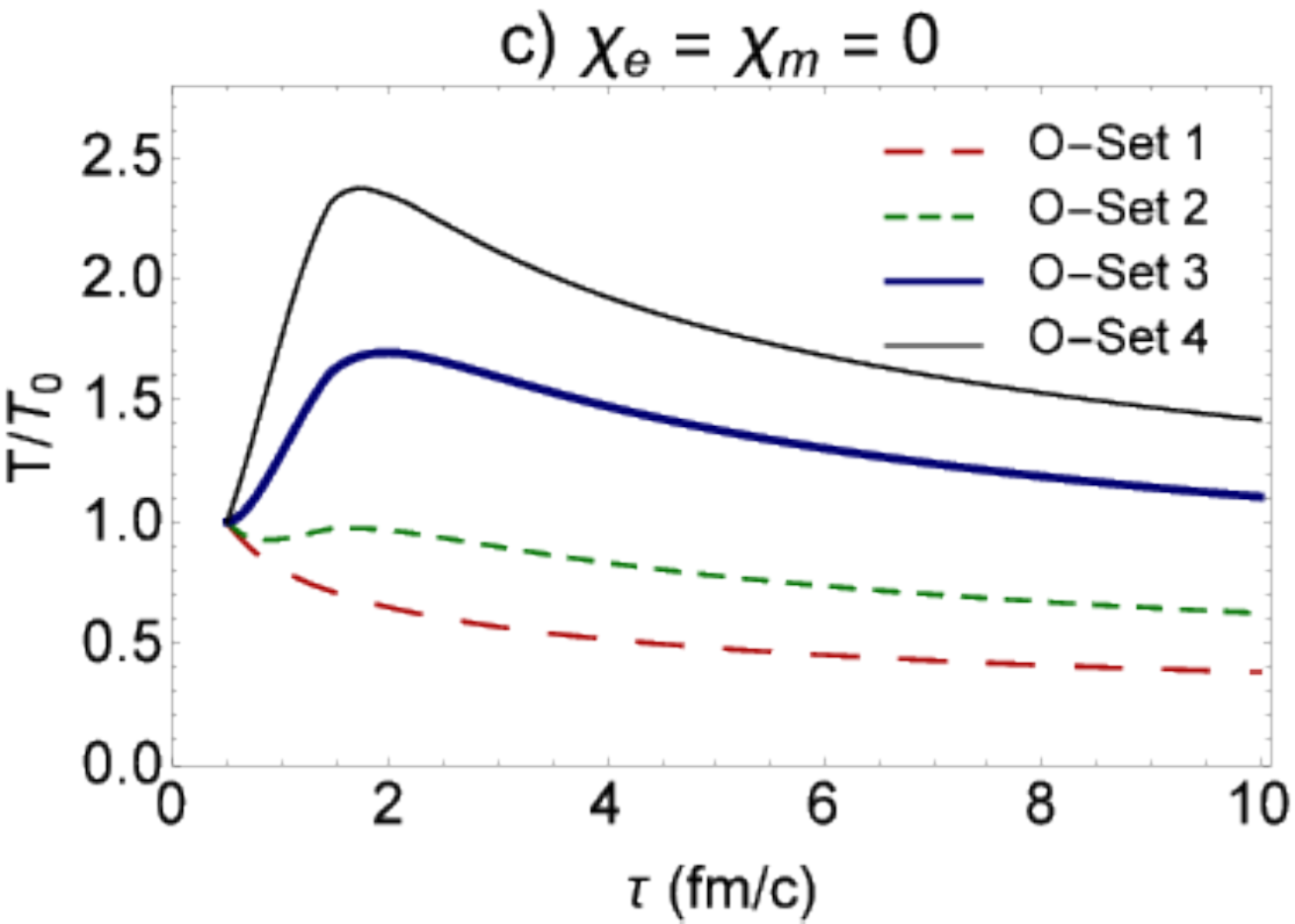}
\caption{(color online). The $\tau$ dependence of $B/B_{0}$ (panel a), $E/E_{0}$ (panel b) and $T/T_{0}$ (panel c), arising from the numerical solution of (\ref{A37}), are plotted. The O-Set $i$'s, $i=1,2,3,4$ correspond to $\Omega_{0}$ sets from (\ref{OO9}) and (\ref{OO10}). Non-vanishing $\Omega_{0}$ strongly affects the $\tau$ dependence and the lifetime of $B,E$ and $T$. }\label{fig9}
\end{figure*}
%%%%%%%%%%%%%%%% fig10 new
\begin{figure}[hbt]
\includegraphics[width=8cm,height=6cm]{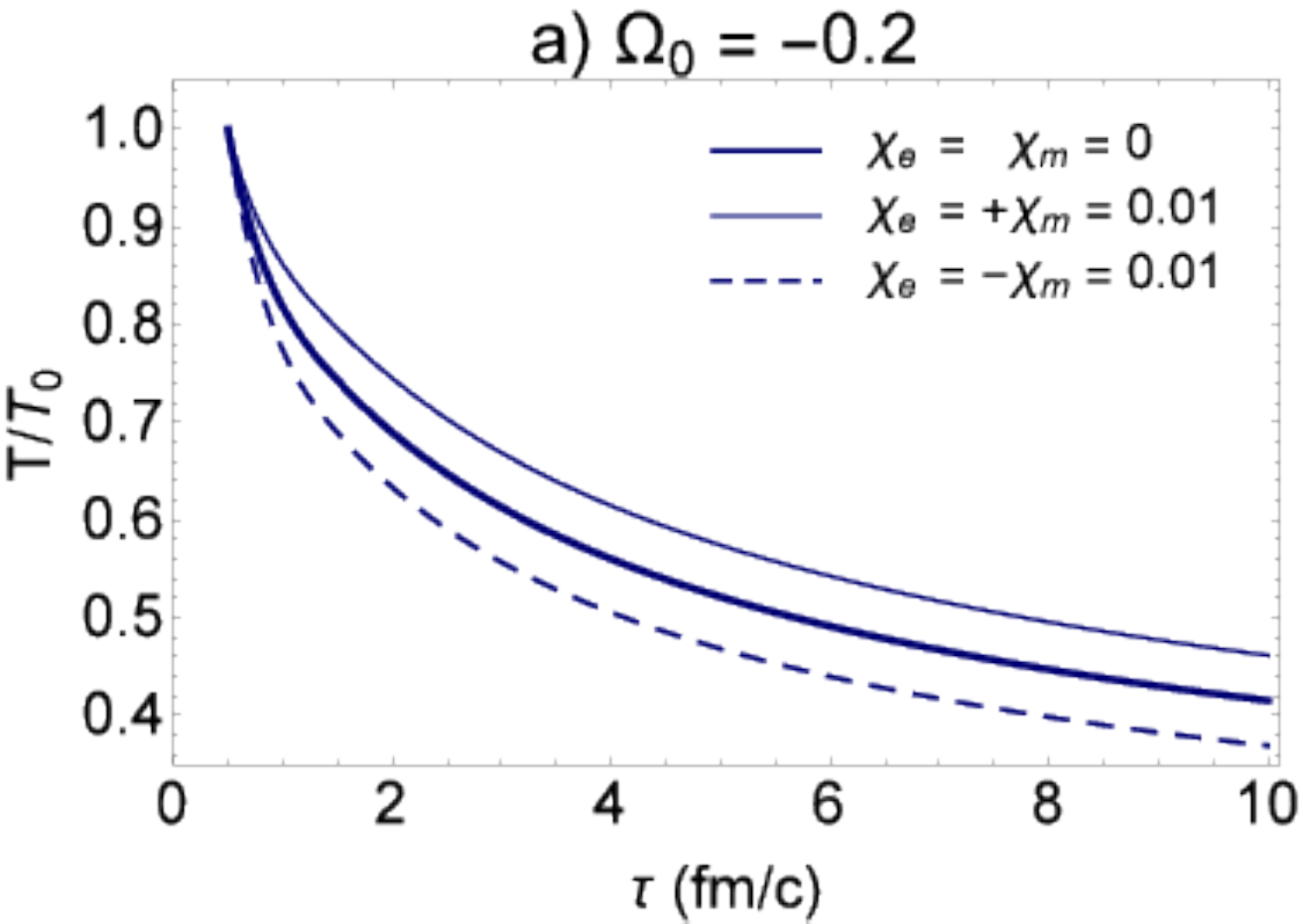}
\includegraphics[width=8cm,height=6cm]{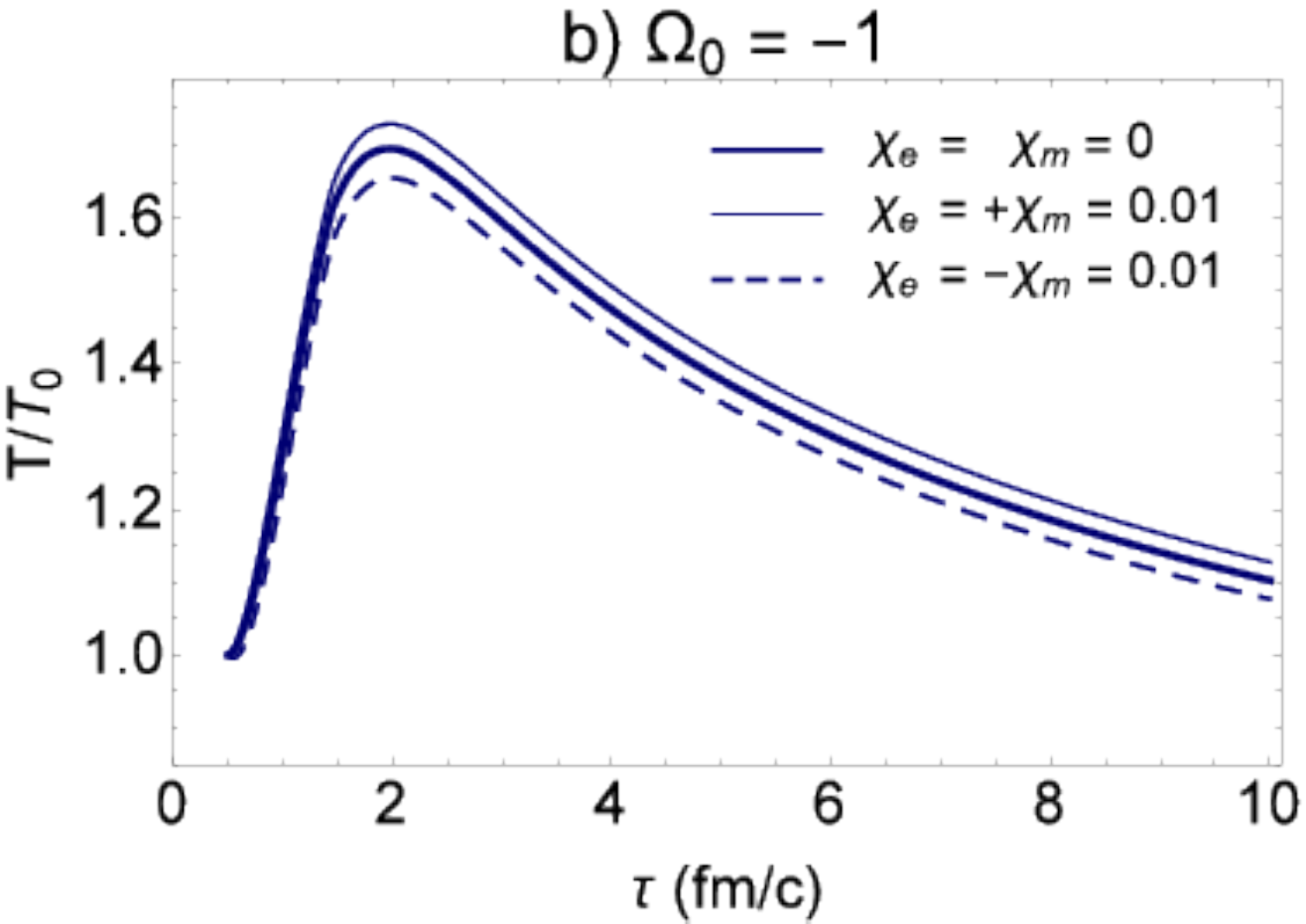}
\caption{(color online). The $\tau$ dependence of $T/T_{0}$ arising from the numerical solution of (\ref{A37}) is plotted for the set of free parameters (\ref{OO11}) with $\chi_{m}=0$ (thick solid curve), $\chi_{m}=+0.01$ (thin solid curve) and $\chi_{m}=-0.01$ (dashed curve) and $\Omega_{0}=-0.2$ (panel a) as well as $\Omega_{0}=-1$ (panel b). As it turns out, a diamagnetic fluid cools faster than a paramagnetic fluid. }\label{fig10}
\end{figure}
%%%%%%%%%%%%%%%% fig 11 new
\begin{figure*}[hbt]
\centering
\includegraphics[width=5.6cm,height=4.4cm]{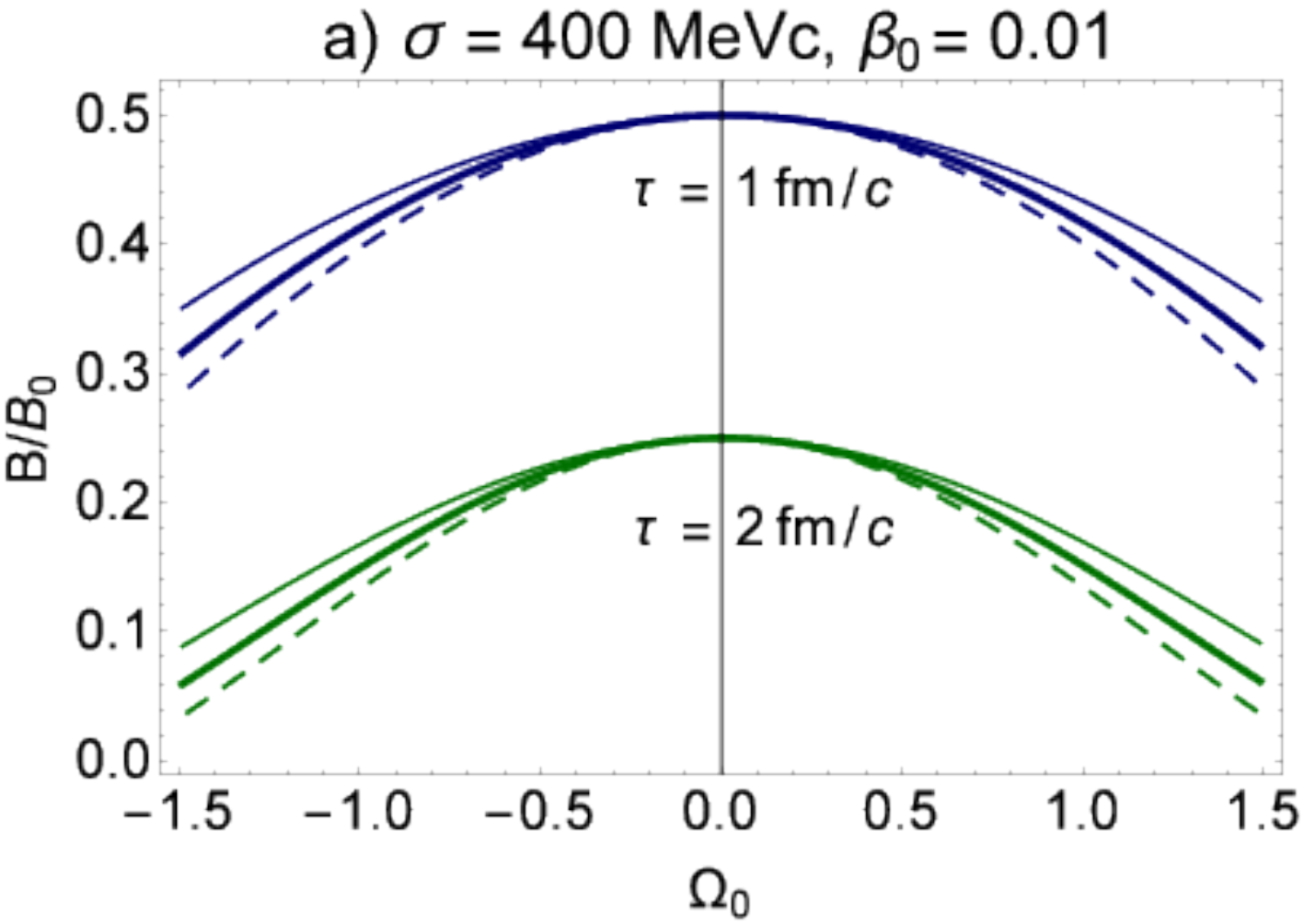}
\includegraphics[width=5.6cm,height=4.4cm]{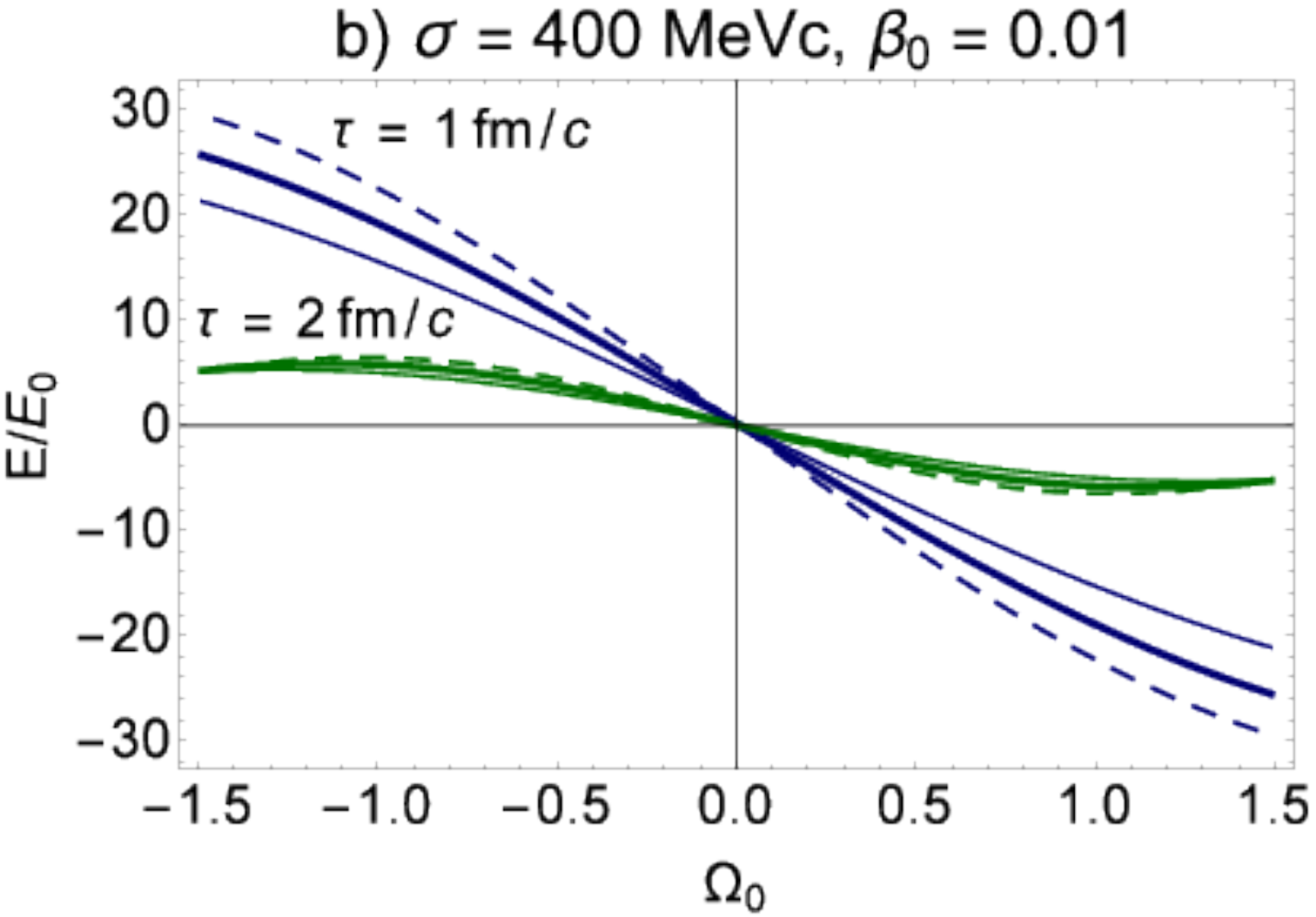}
\includegraphics[width=5.6cm,height=4.4cm]{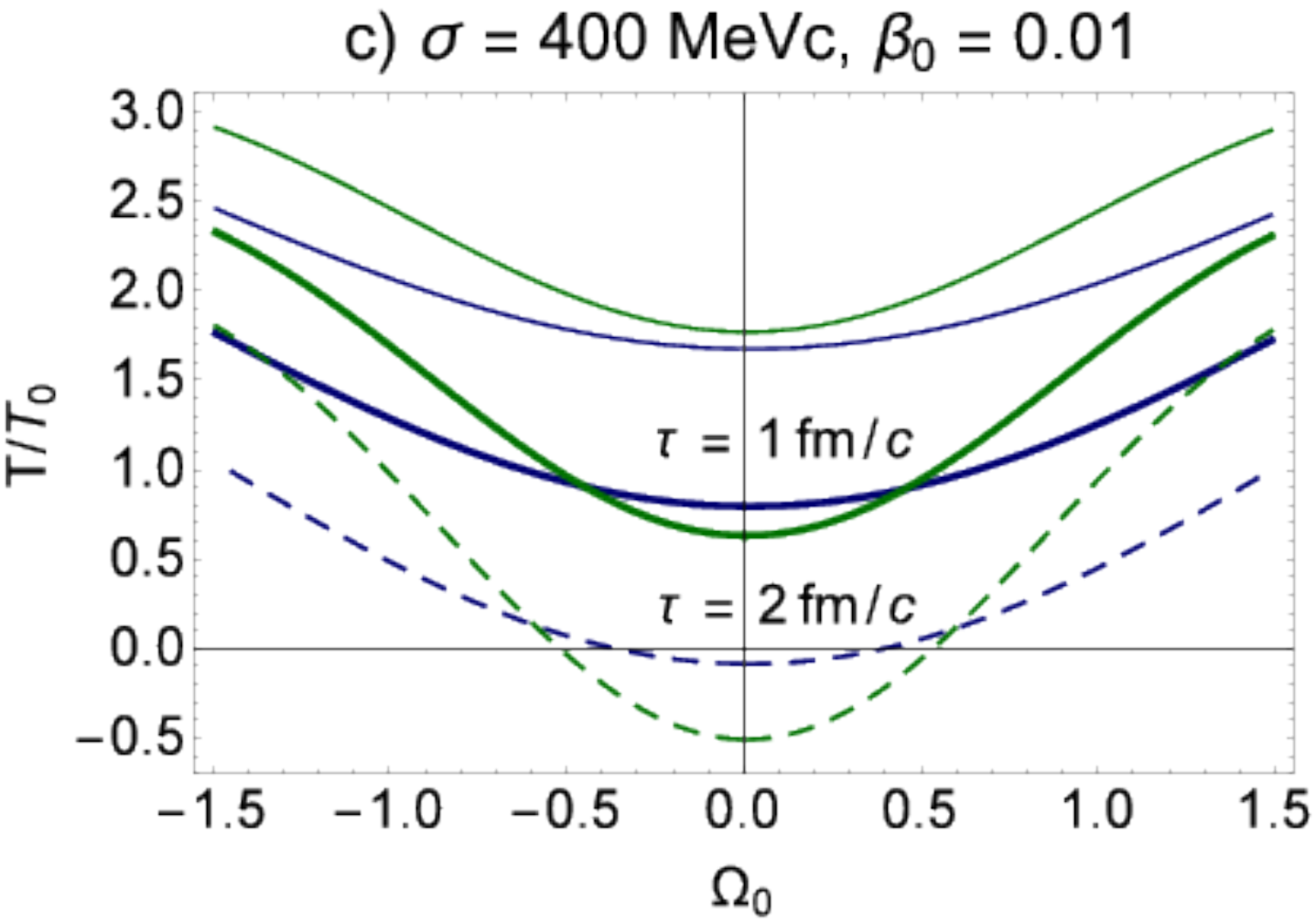}
\caption{(color online). The $\Omega_{0}$ dependence of $B/B_{0}$ (panel a), $E/E_{0}$ (panel b) and $T/T_{0}$ (panel c) are plotted for $\tau=1,2$ fm/c (blue and green solid and dashed curves). The sets of free parameters (\ref{OO13}) with $\chi_{m}=0$ (thick solid curves), $\chi_{m}=+0.2$ (thin solid curves) and $\chi_{m}=-0.2$ (dashed curves) are used to determine $B,E$ and $T$ from numerical solution of (\ref{A37}).}\label{fig11}
\end{figure*}
%%%%%%%%%%%%%%%% fig 12 new
\begin{figure*}[hbt]
\includegraphics[width=5.6cm,height=4.4cm]{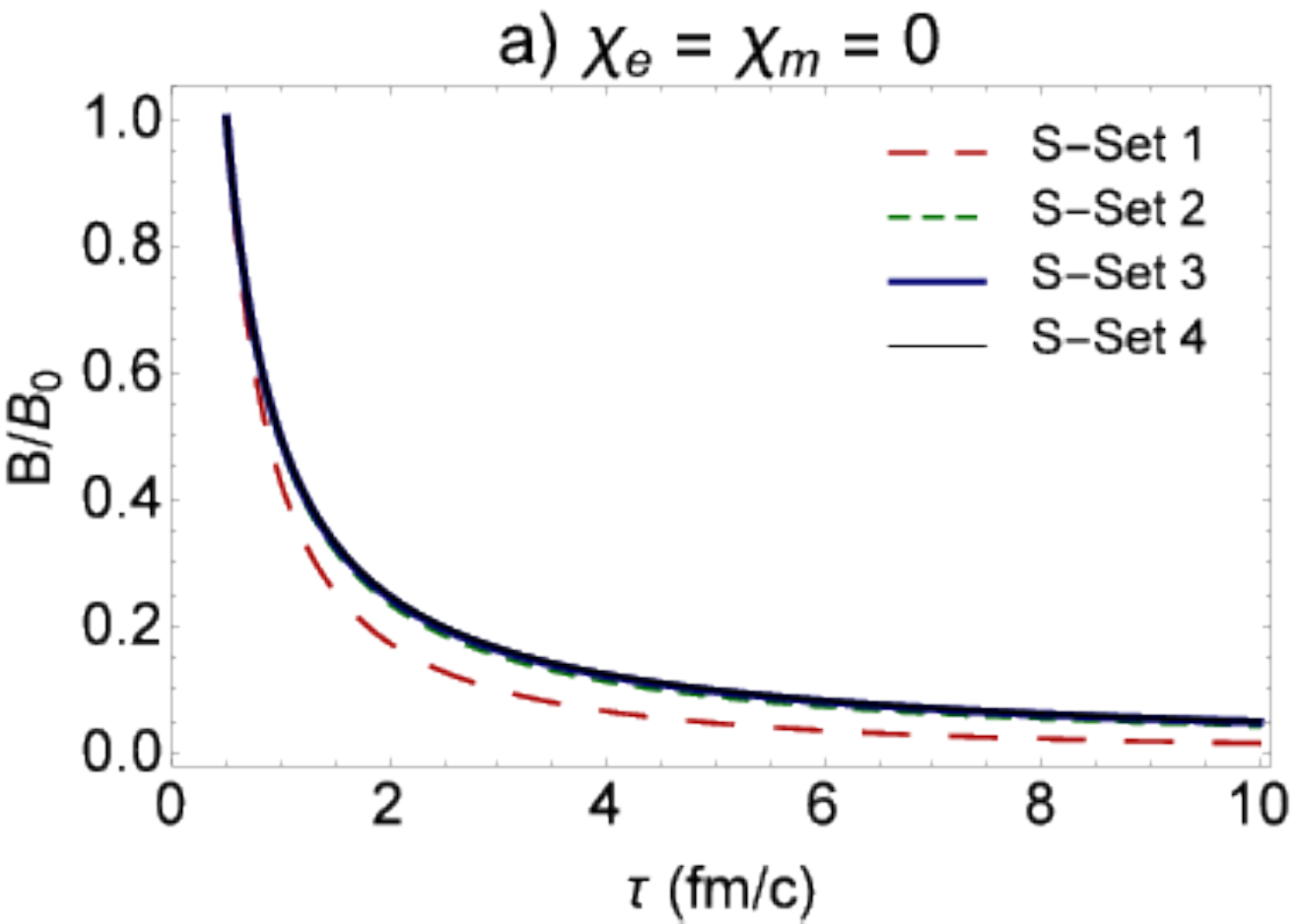}
\includegraphics[width=5.6cm,height=4.4cm]{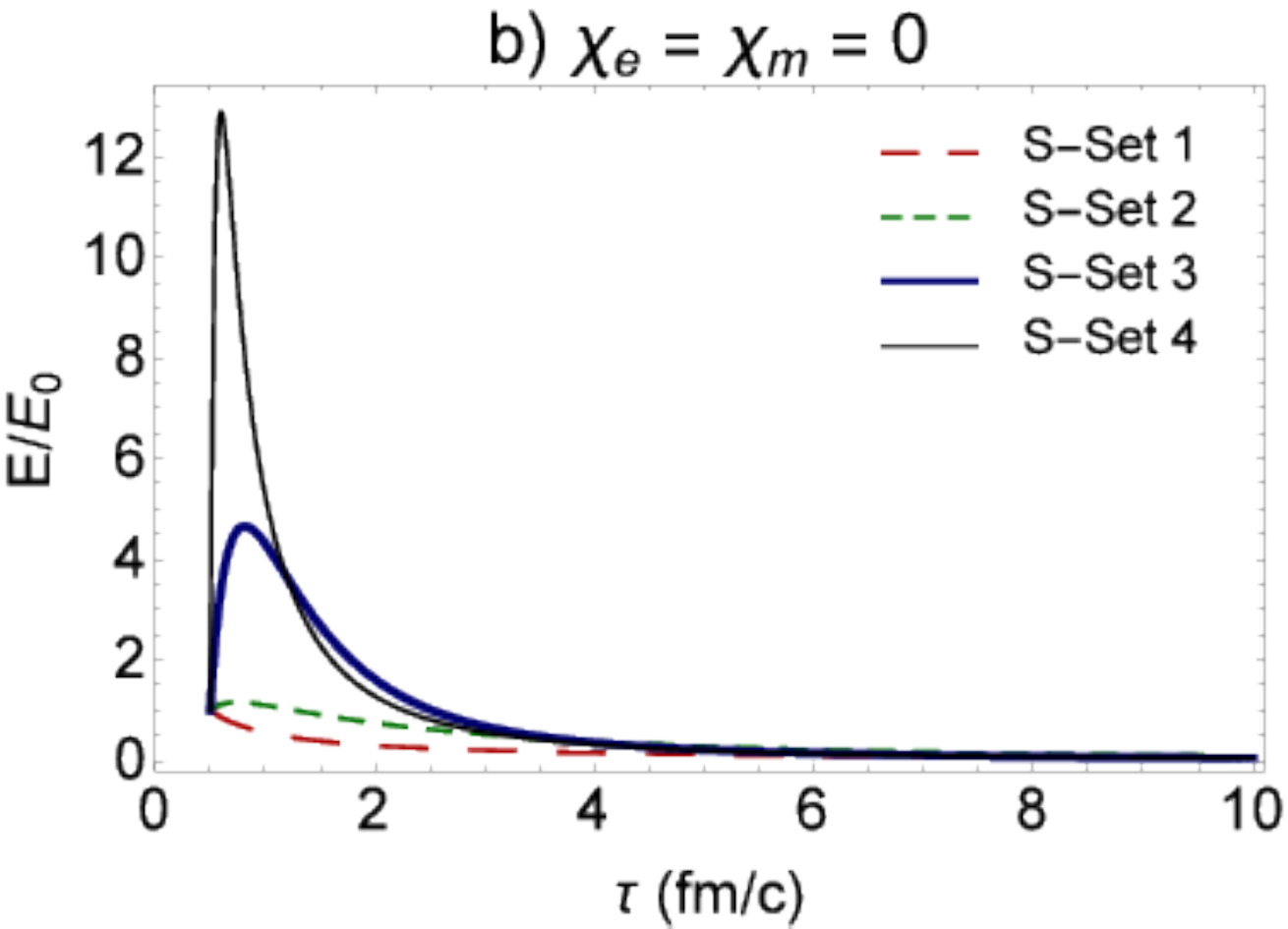}
\includegraphics[width=5.6cm,height=4.4cm]{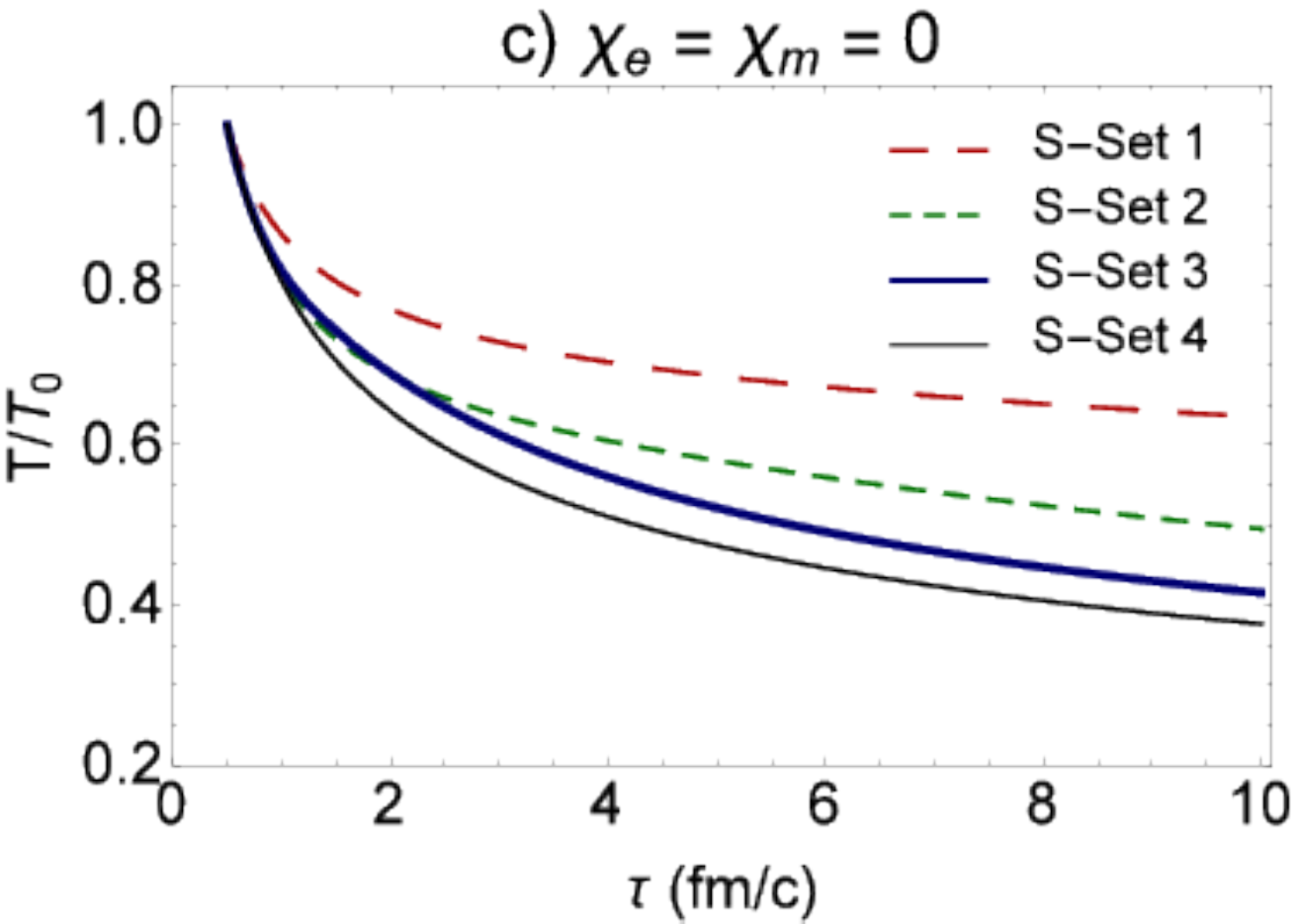}
\caption{(color online). The $\tau$ dependence of $B/B_{0}$ (panel a), $E/E_{0}$ (panel b) and $T/T_{0}$ (panel c) arising from the numerical solution of (\ref{A37}) are plotted. The S-Set $i$'s, with $i=1,2,3,4$ correspond to $\sigma$ sets from (\ref{OO15}) and (\ref{OO16}). Non-vanishing $\sigma$ affects the $\tau$ dependence and the lifetime of $B, E$ and $T$.}\label{fig12}
\end{figure*}
%%%%%%%%%%%%%%%% fig 13
\begin{figure}[hbt]
\includegraphics[width=8cm,height=6cm]{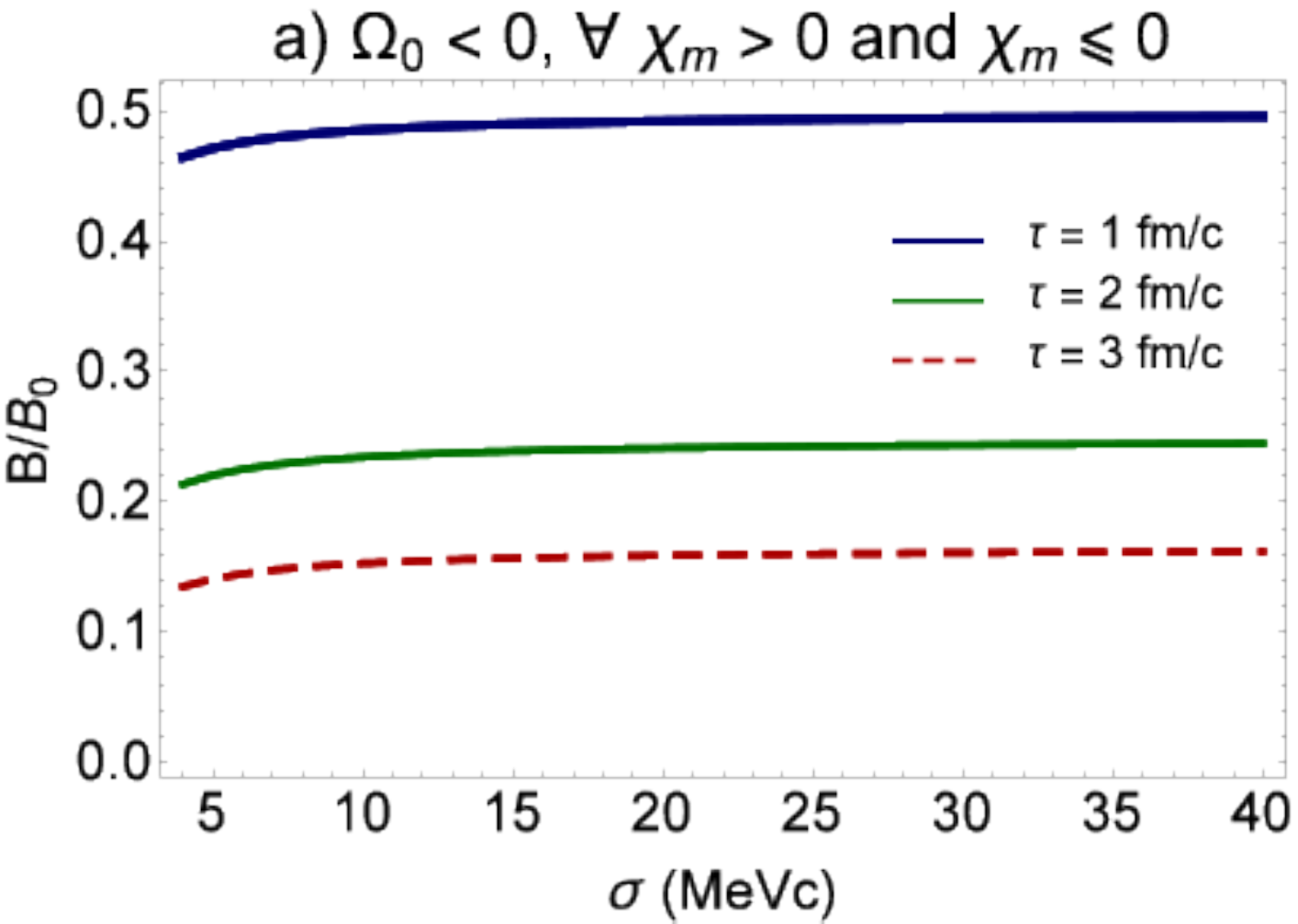}
\includegraphics[width=8cm,height=6cm]{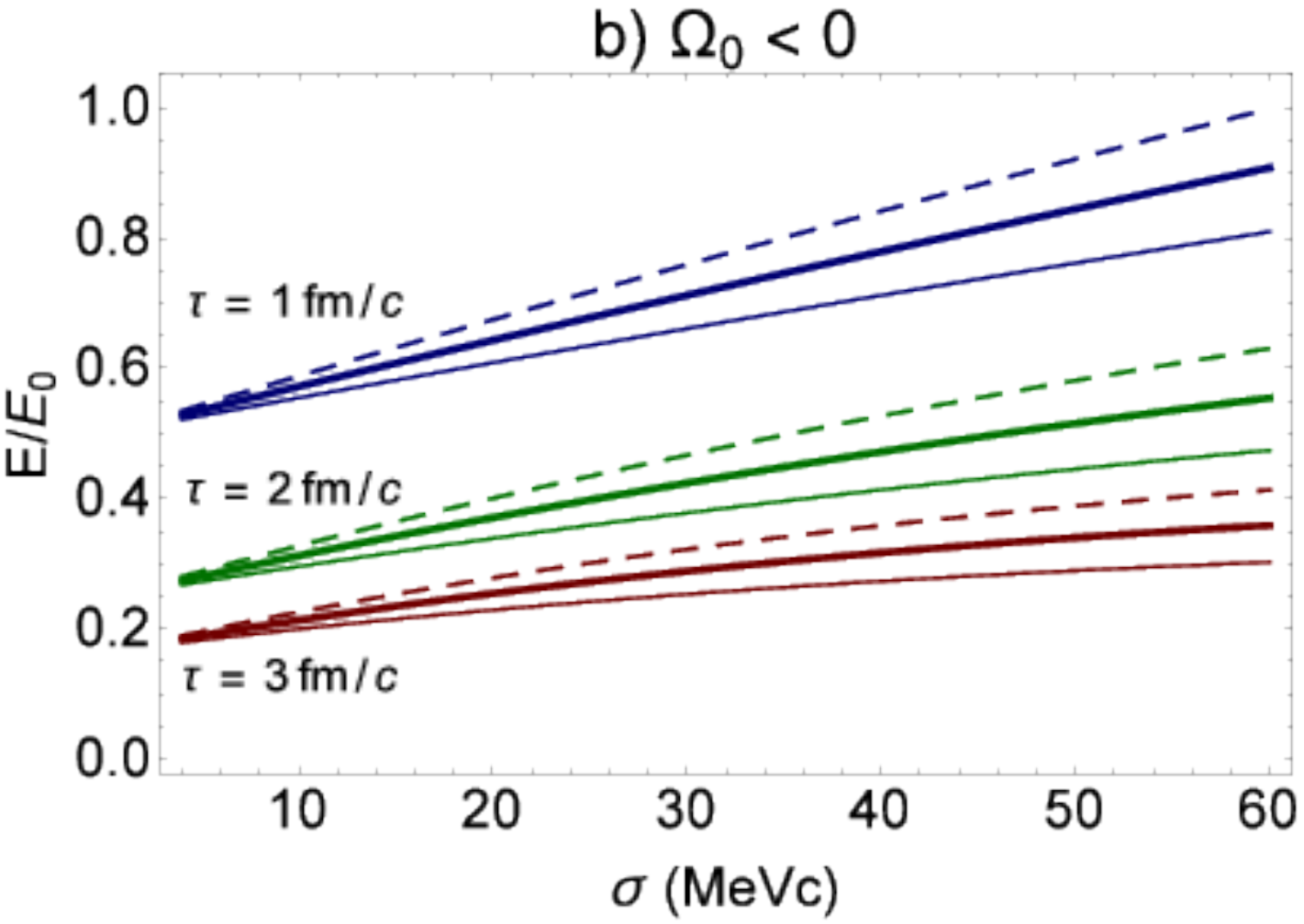}
\caption{(color online).
The $\sigma$ dependence of $B/B_{0}$ (panels a) and $E/E_{0}$ (panels b) are plotted at $\tau=1,2,3$ fm/c (blue, green, and red solid and dashed curves). The sets of free parameters (\ref{OO17}) and (\ref{OO18}) are used to plot these curves. In panel b, thick solid curves correspond to $\{\chi_{e},\chi_{m}\}=\{0,0\}$, thin solid curves to $\{\chi_{e},\chi_{m}\}=\{0.01,+0.2\}$ and dashed curves to $\{\chi_{e},\chi_{m}\}=\{0.01,-0.2\}$.  Whereas different choices of $\chi_{m}$ have no significant effect on the $\sigma$ dependence of $B/B_{0}$ (see also Fig. \ref{fig15}), positive and negative $\chi_{m}$ affect the $\sigma$ dependence of $E/E_{0}$.
 }\label{fig13}
\end{figure}
%%%%%%%%%%%%%%%% fig 14
\begin{figure*}[hbt]
\includegraphics[width=5.6cm,height=4.4cm]{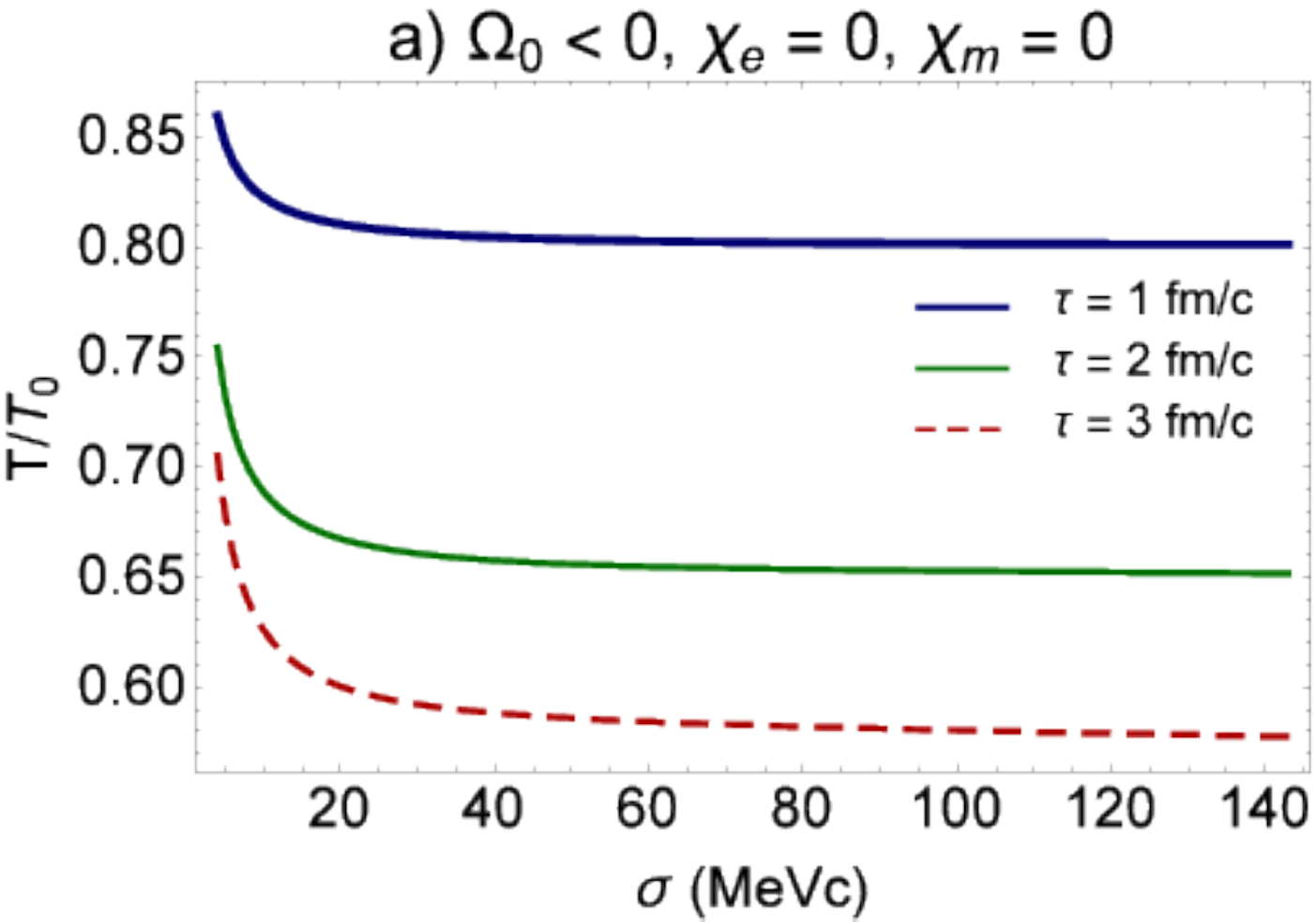}
\includegraphics[width=5.6cm,height=4.4cm]{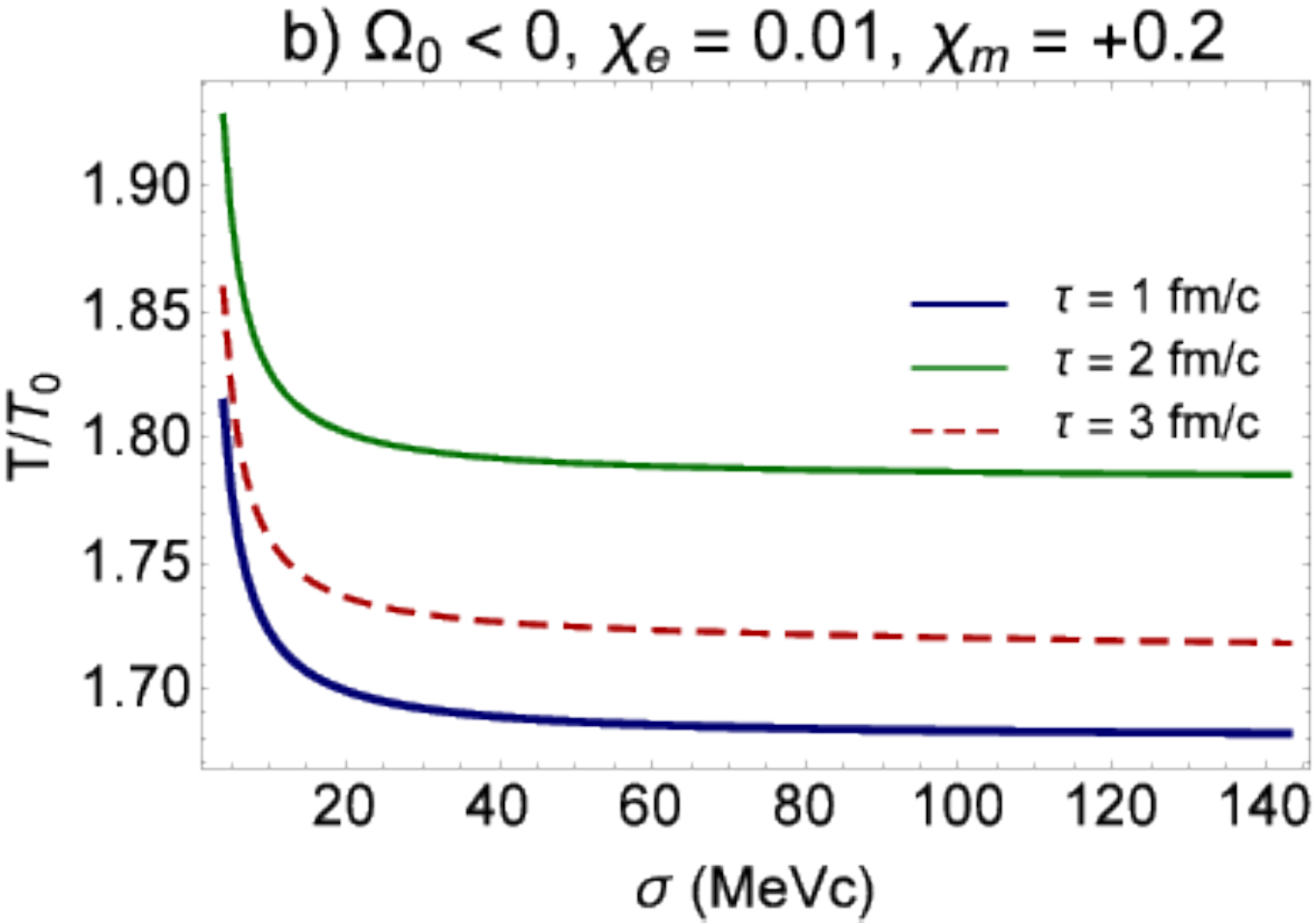}
\includegraphics[width=5.6cm,height=4.4cm]{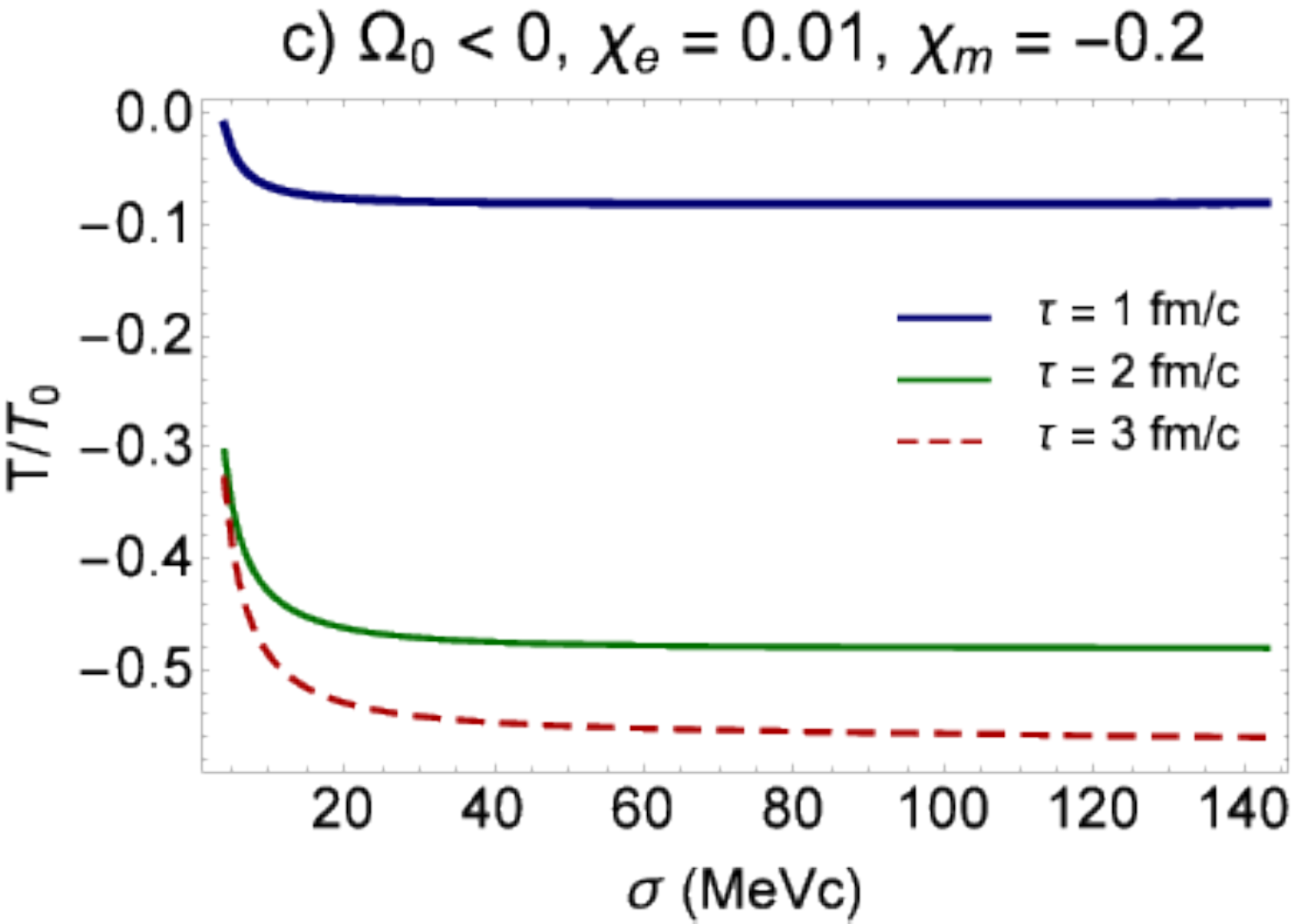}
\caption{(color online). The $\sigma$ dependence of $T/T_{0}$ is plotted at $\tau=1,2,3$ fm/c (blue, green, and red solid and dashed curves). The sets of free parameters (\ref{OO17}) and (\ref{OO18}) with $\chi_{m}=0$ (panel a), $\chi_{m}=+0.2$ (panel b) and $\chi_{m}=-0.2$ (panel c) are used to determine $T/T_{0}$ from the numerical solution of (\ref{A37}).}\label{fig14}
\end{figure*}
%%%%%%%%%%%%%%%% fig 15
\begin{figure*}[hbt]
\includegraphics[width=5.6cm,height=4.4cm]{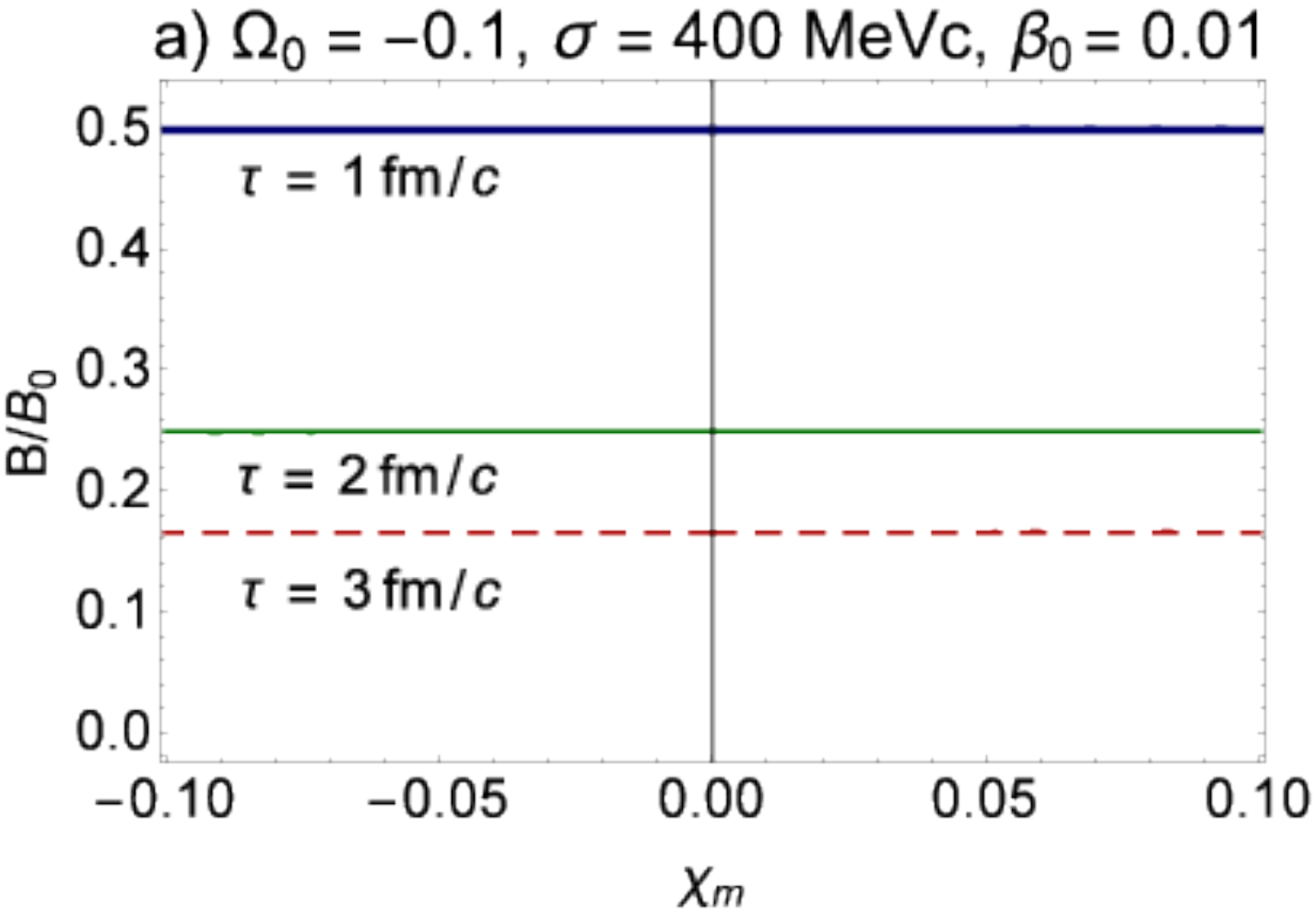}
\includegraphics[width=5.6cm,height=4.4cm]{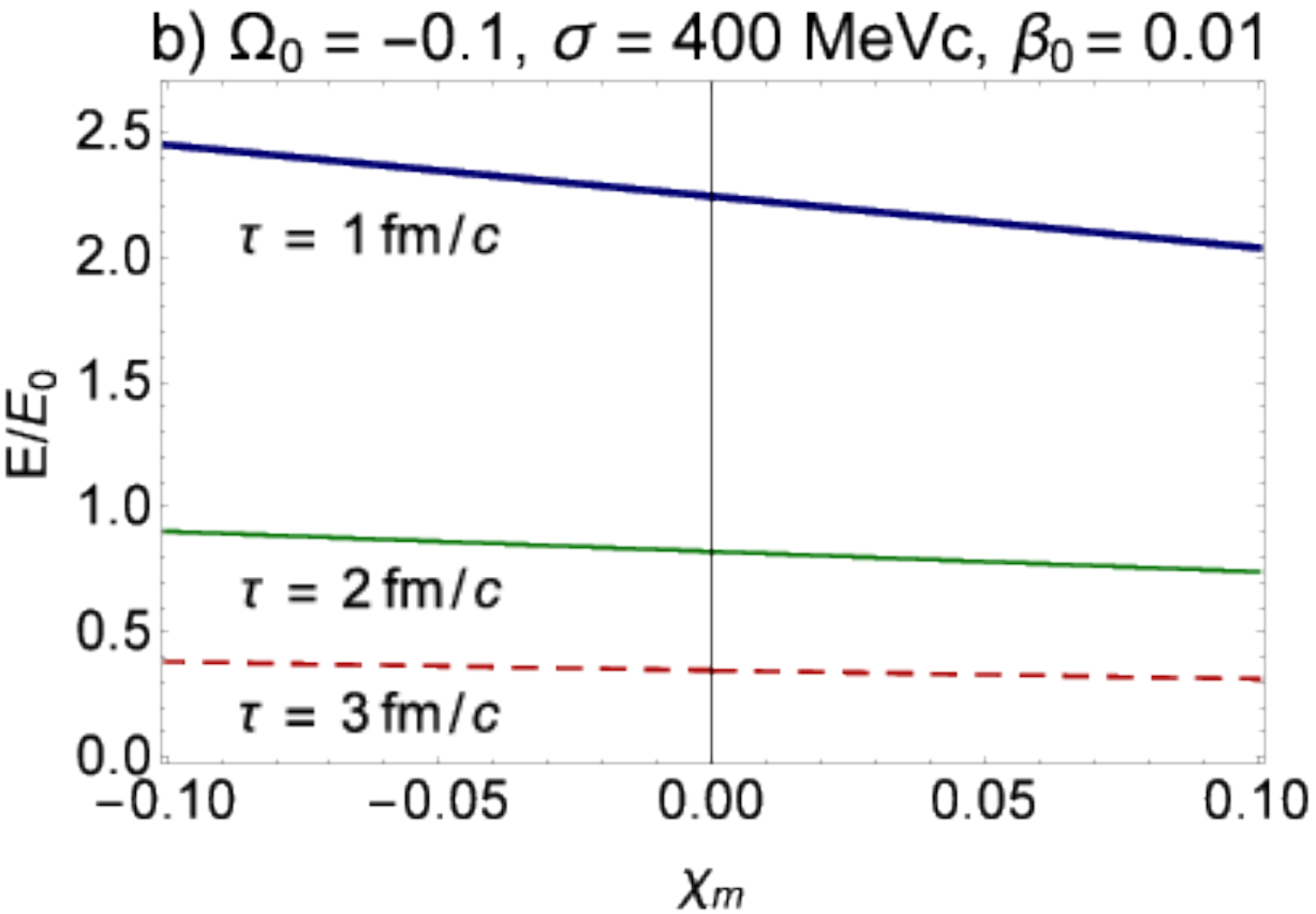}
\includegraphics[width=5.6cm,height=4.4cm]{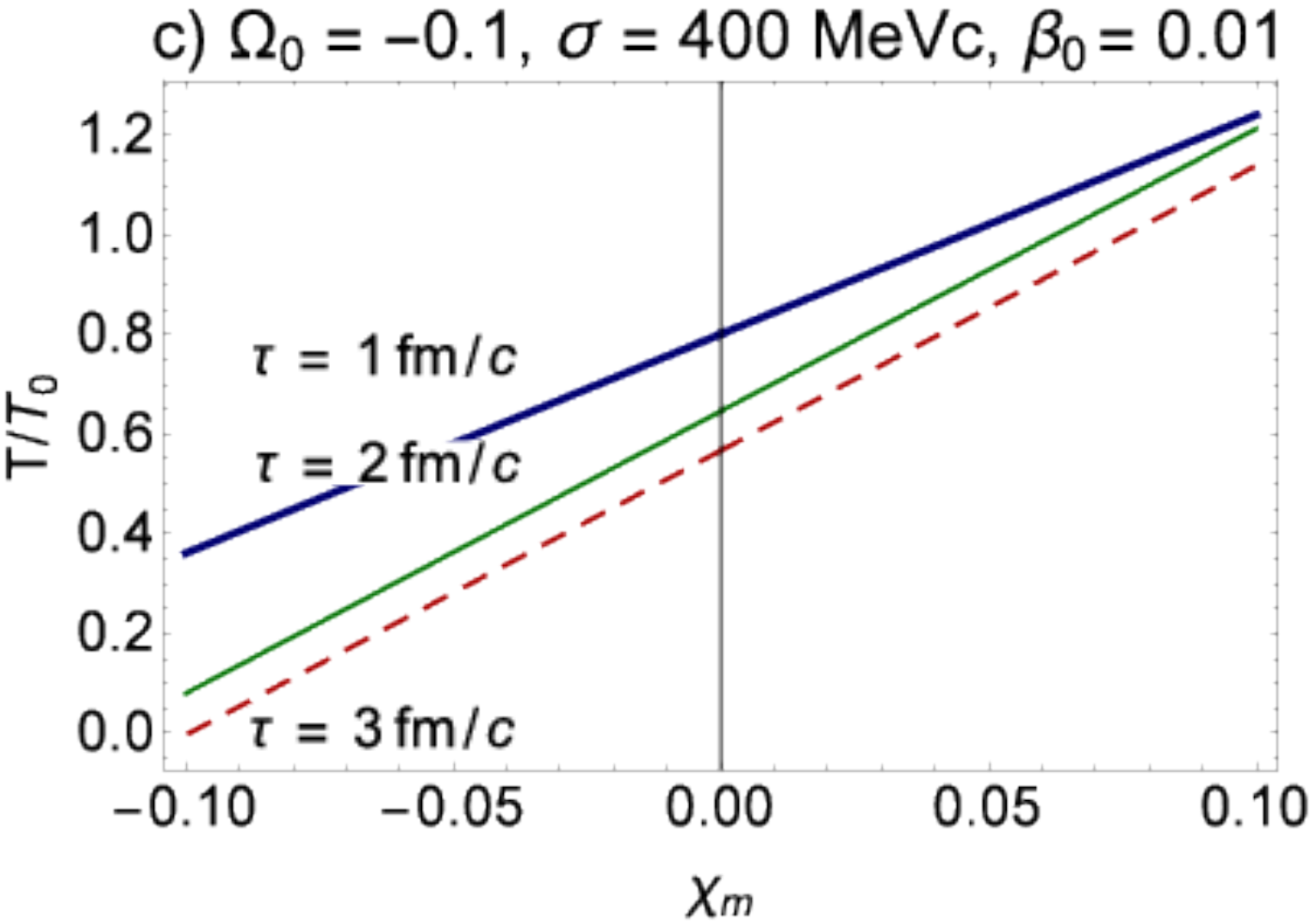}
\includegraphics[width=5.6cm,height=4.4cm]{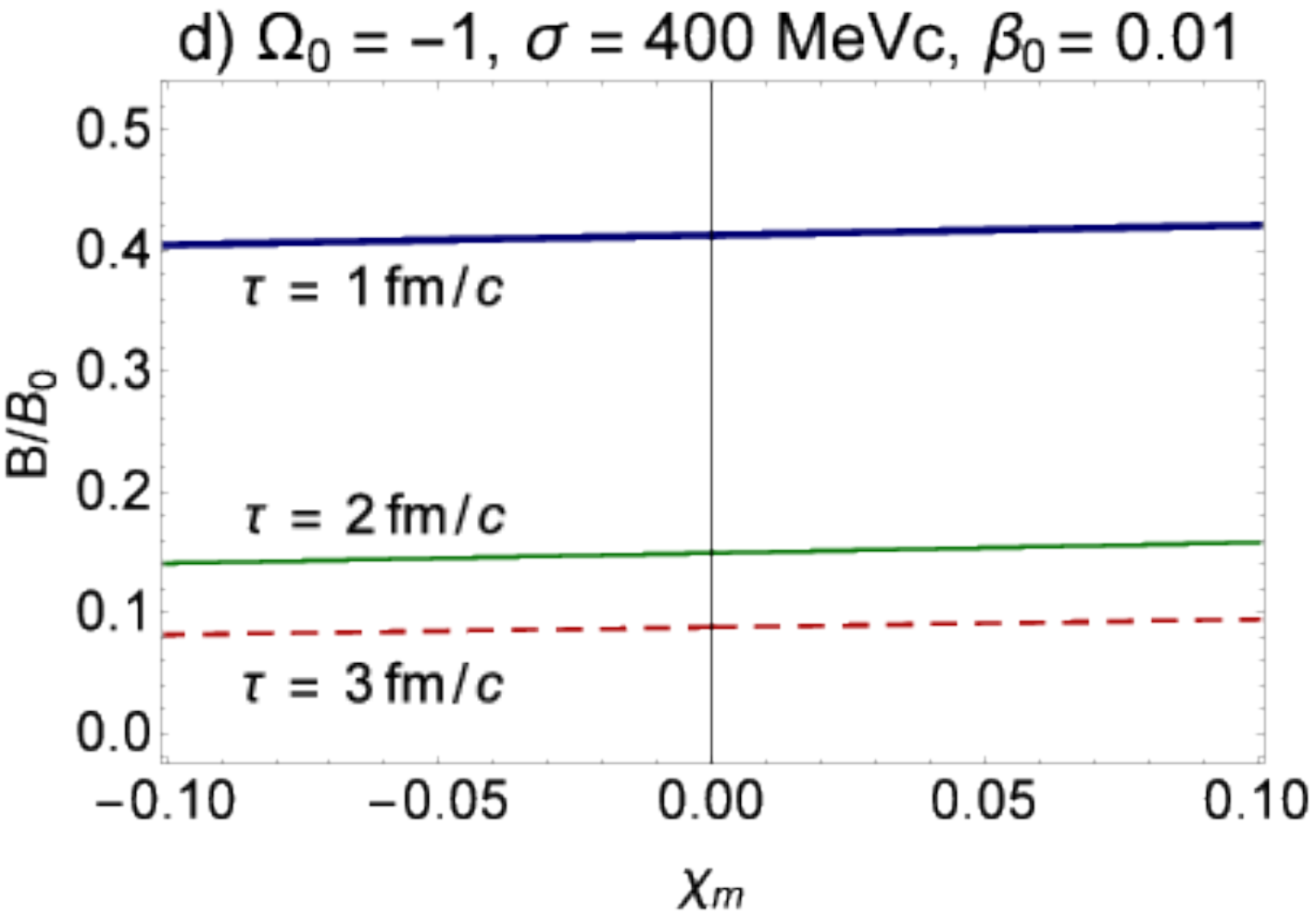}
\includegraphics[width=5.6cm,height=4.4cm]{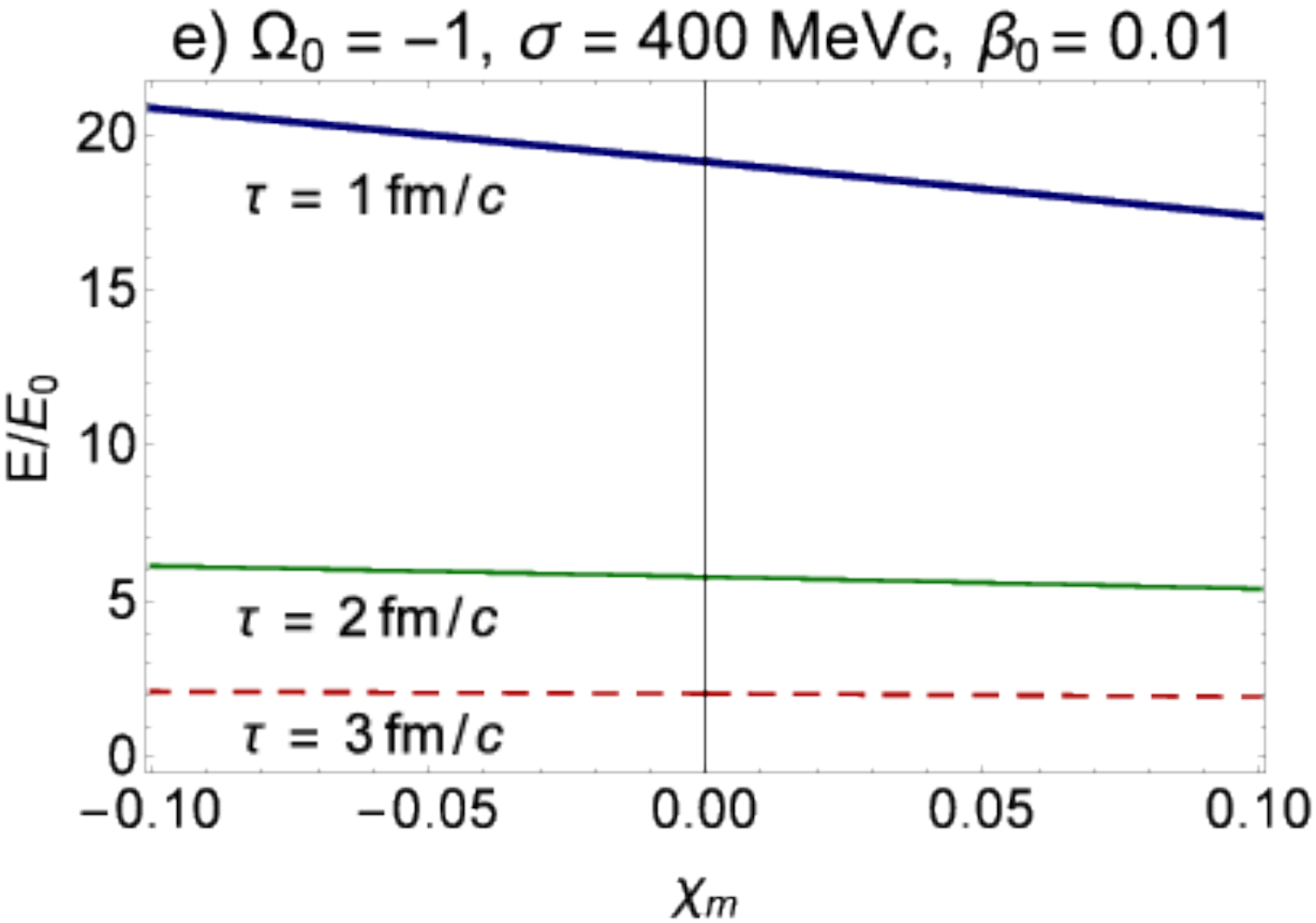}
\includegraphics[width=5.6cm,height=4.4cm]{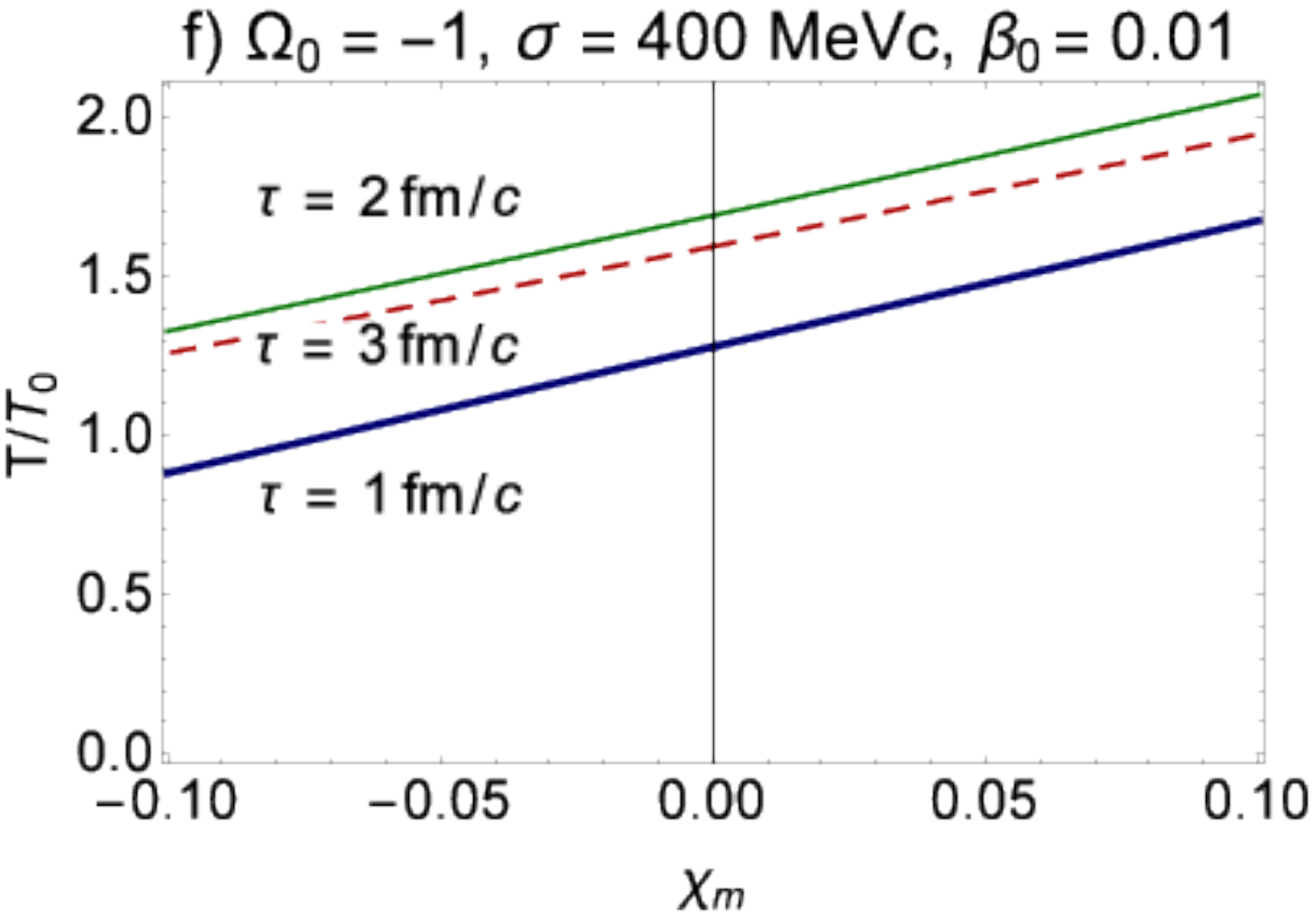}
\caption{(color online). The $\chi_{m}$ dependence of $B/B_{0}$ (panels a and d),  $E/E_{0}$ (panels b and e)  and $T/T_{0}$ (panels c and f) are plotted at $\tau=1,2,3$ fm/c (blue, green solid curves and and red dashed curves). The sets of free parameters  (\ref{OO19}) (panels a-c), and (\ref{OO20}) (panel d-f) are used to determine $B/B_{0}$, $E/E_{0}$ and $T/T_{0}$ from the numerical solution of  (\ref{A37}).}\label{fig15}
\end{figure*}
%%%%%%%%%%%%%%%% fig 16
\begin{figure*}[hbt]
\includegraphics[width=5.6cm,height=4.4cm]{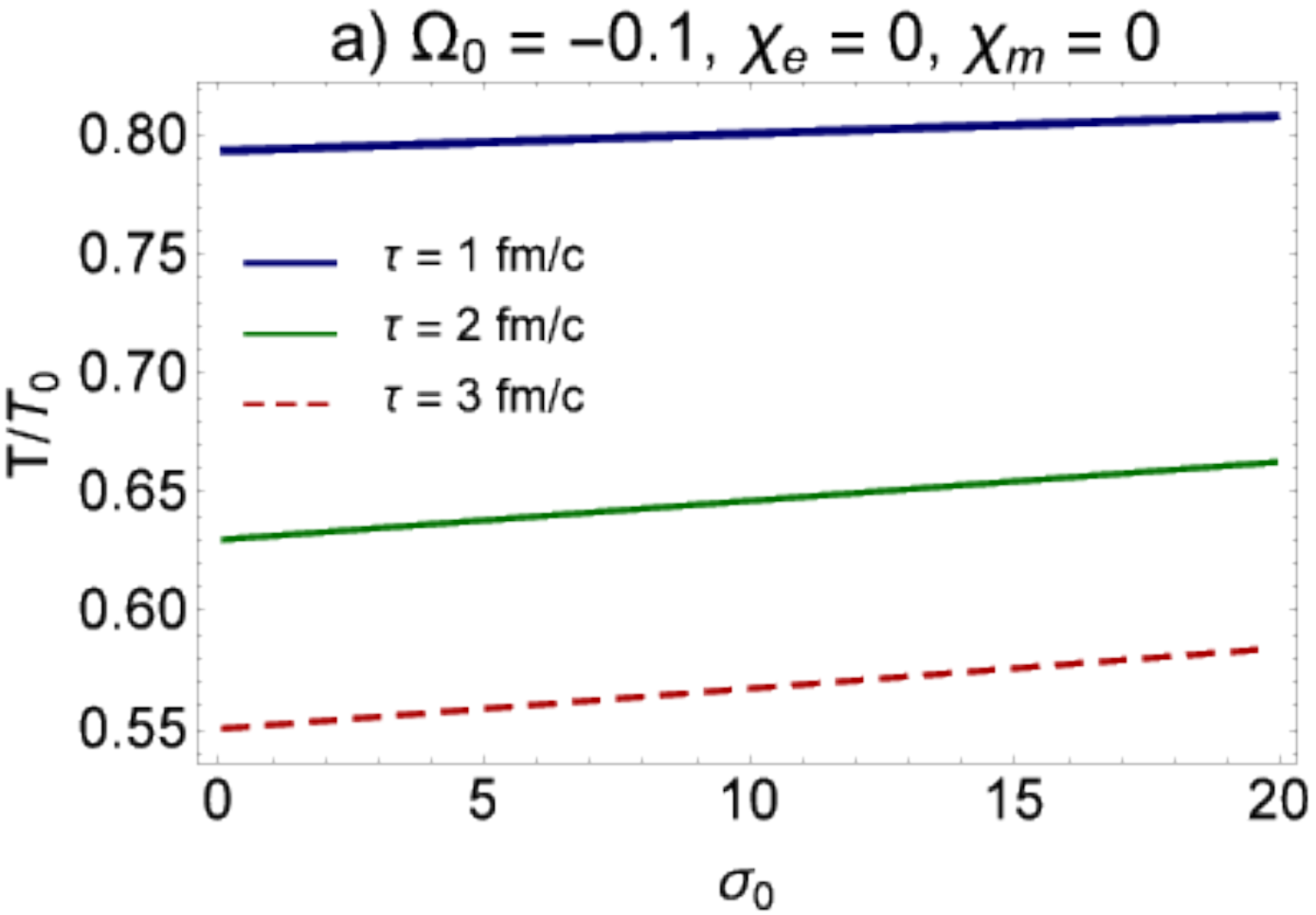}
\includegraphics[width=5.6cm,height=4.4cm]{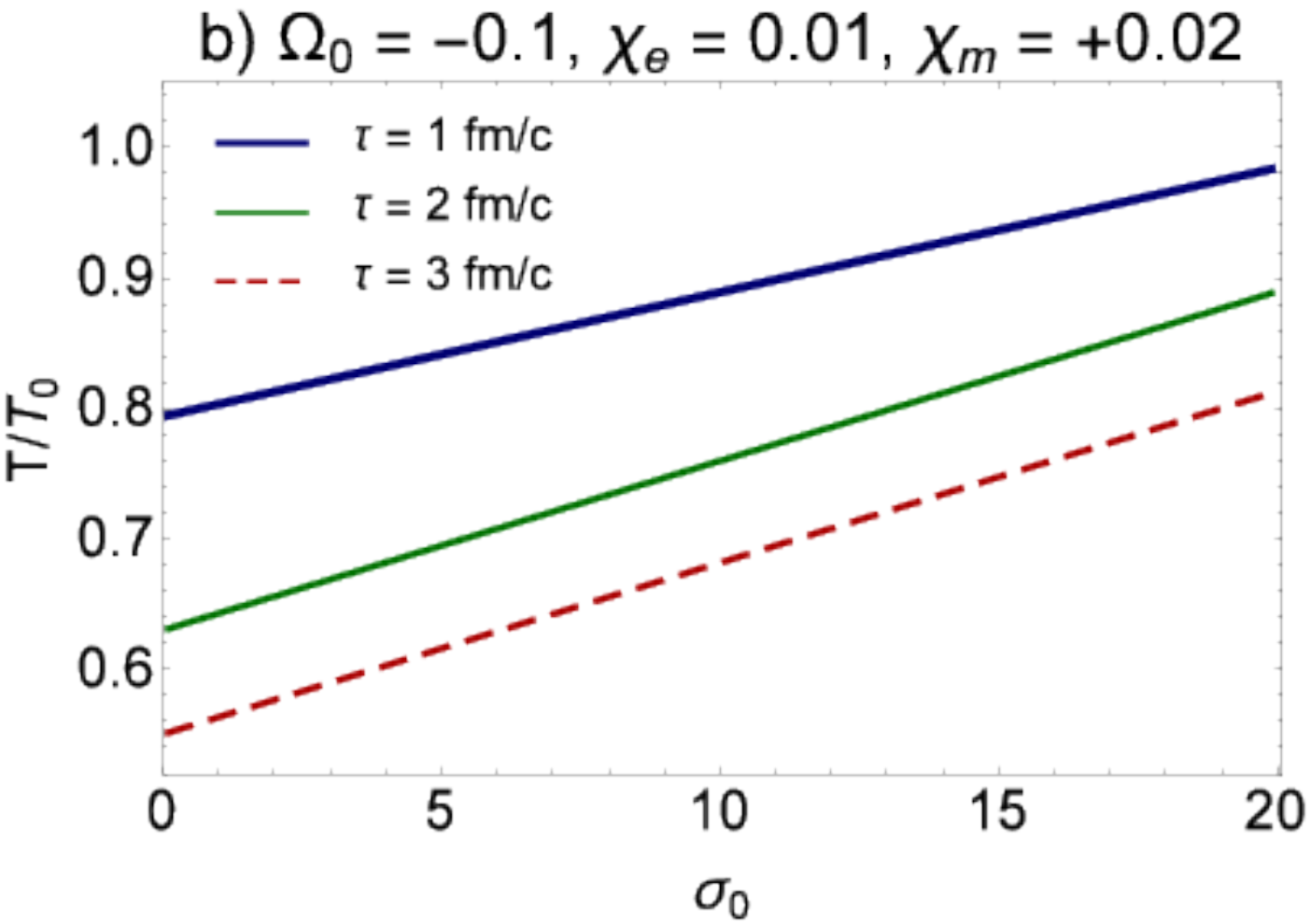}
\includegraphics[width=5.6cm,height=4.4cm]{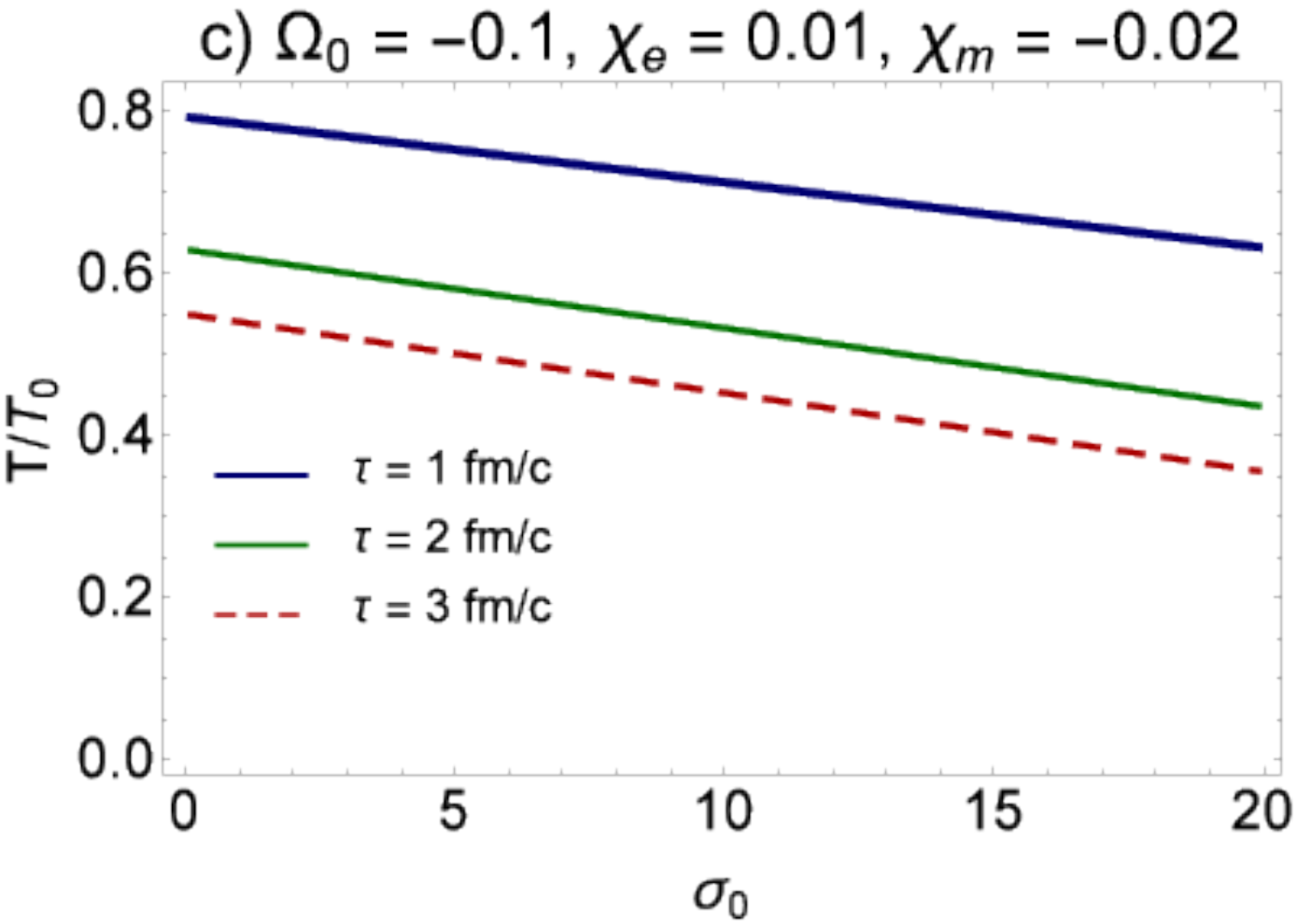}
\includegraphics[width=5.6cm,height=4.4cm]{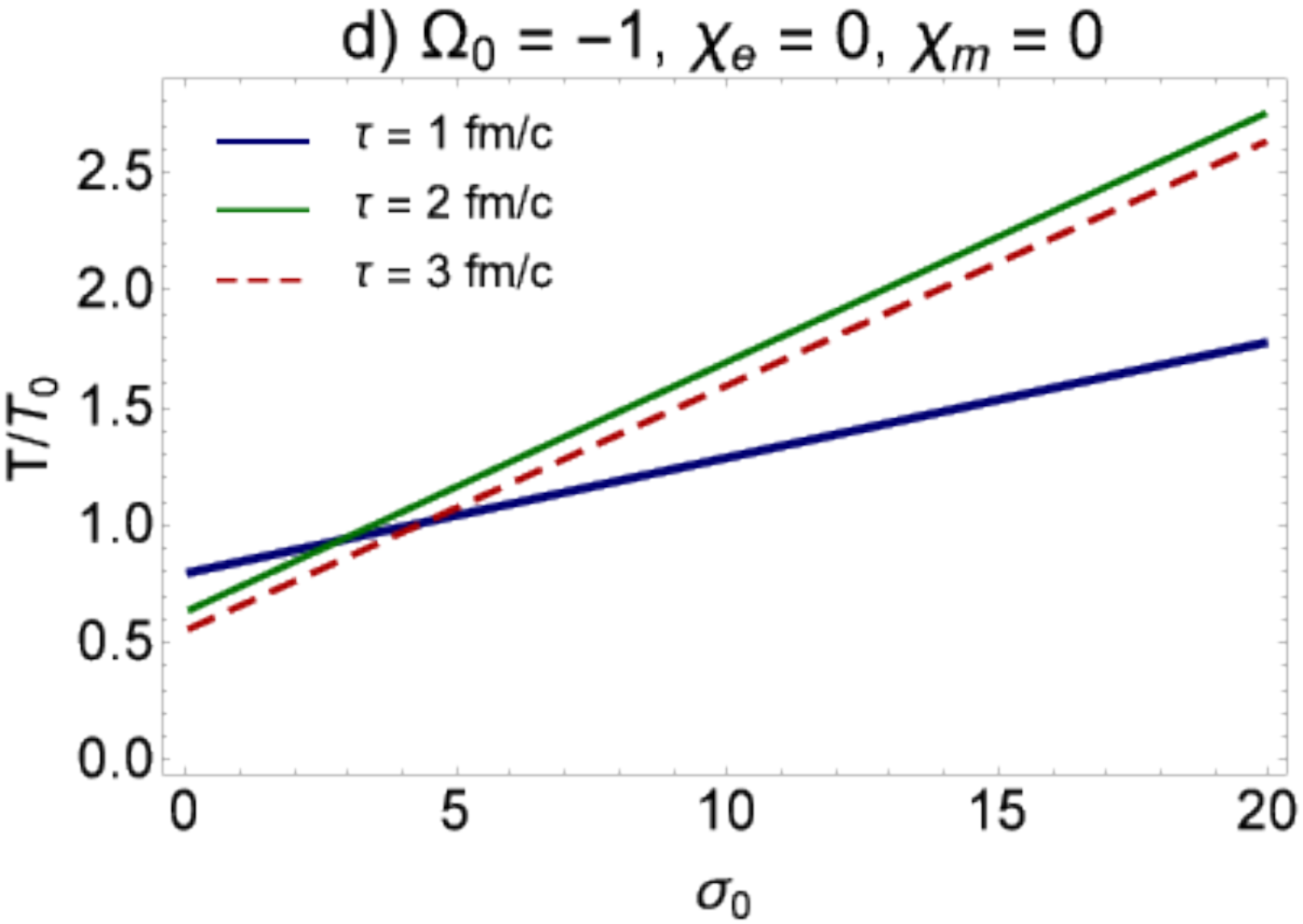}
\includegraphics[width=5.6cm,height=4.4cm]{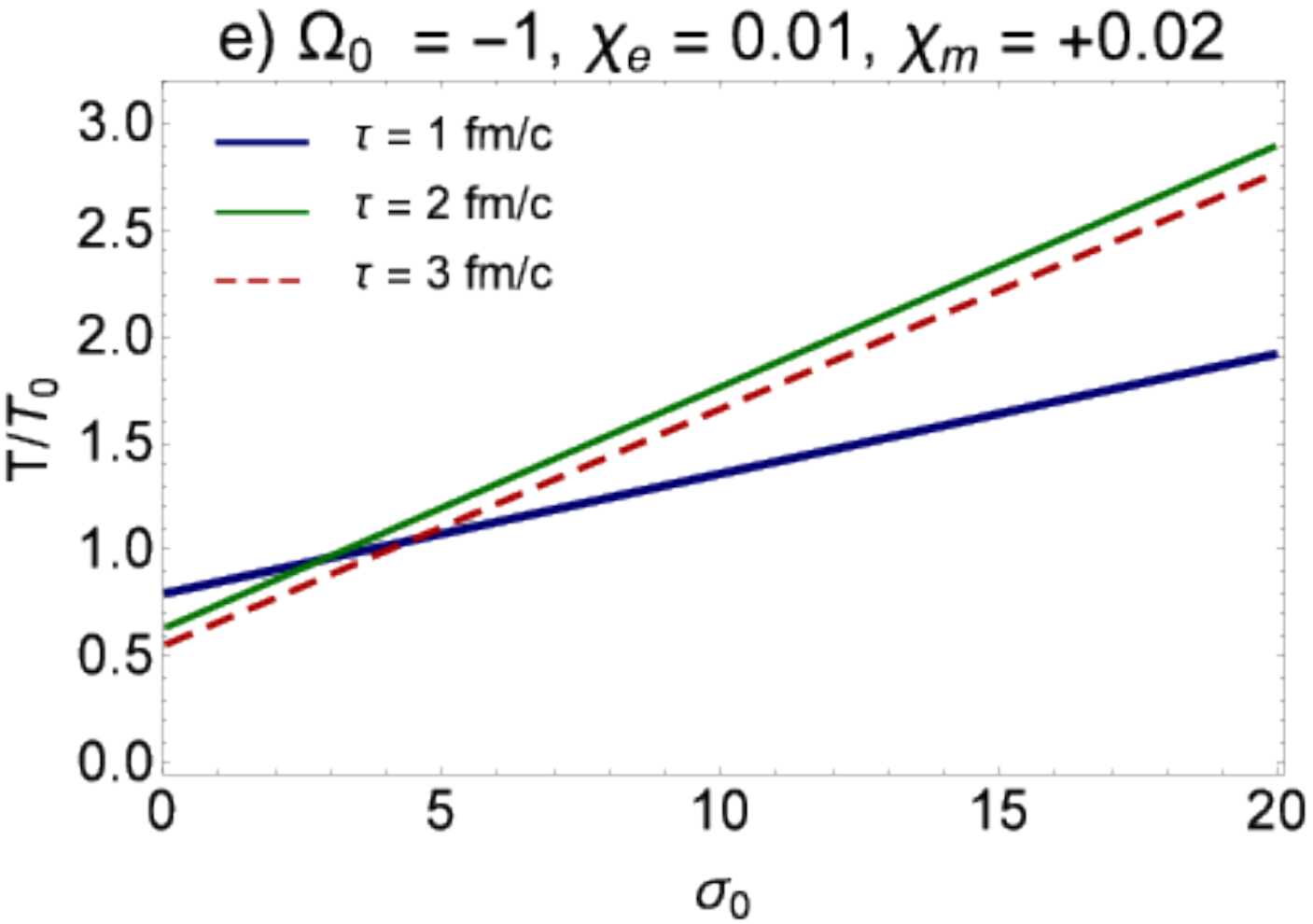}
\includegraphics[width=5.6cm,height=4.4cm]{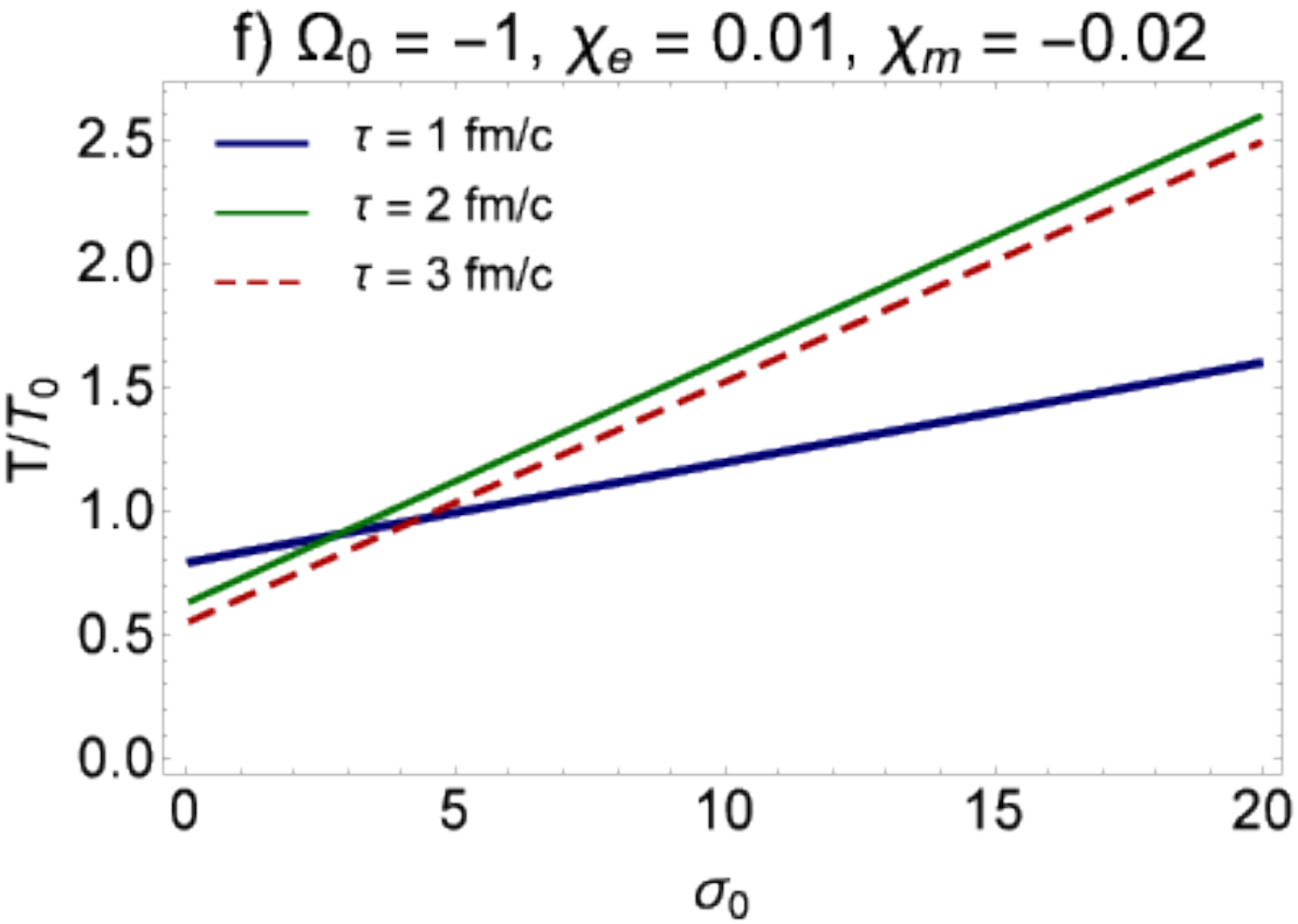}
\caption{(color online). The $\sigma_{0}$ dependence of $T/T_{0}$ is plotted at $\tau=1,2,3$ fm/c (blue, green solid curves and and red dashed curves). The sets of free parameters  (\ref{OO21}) and (\ref{OO22}) are used to determine $T/T_{0}$ from the numerical solution of (\ref{A37}). The amplitude $T/T_{0}$ is strongly affected by angular velocity $\Omega$, electric and magnetic susceptibilities, $\chi_{e}$ and $\chi_{m}$.}\label{fig16}
\end{figure*}
%%%%%%%%%%%%%%%% fig 17
\begin{figure*}[hbt]
\includegraphics[width=5.6cm,height=4.4cm]{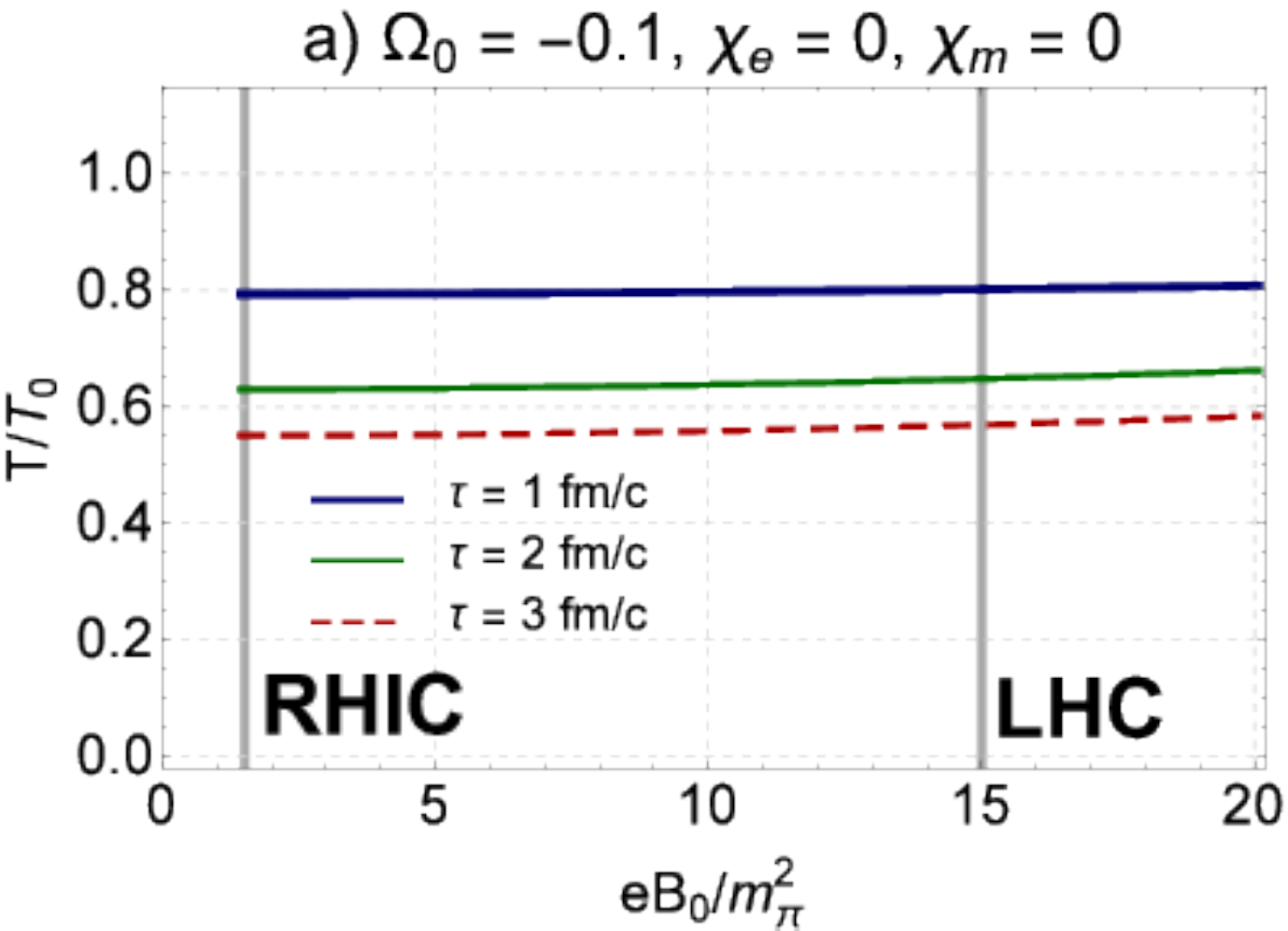}
\includegraphics[width=5.6cm,height=4.4cm]{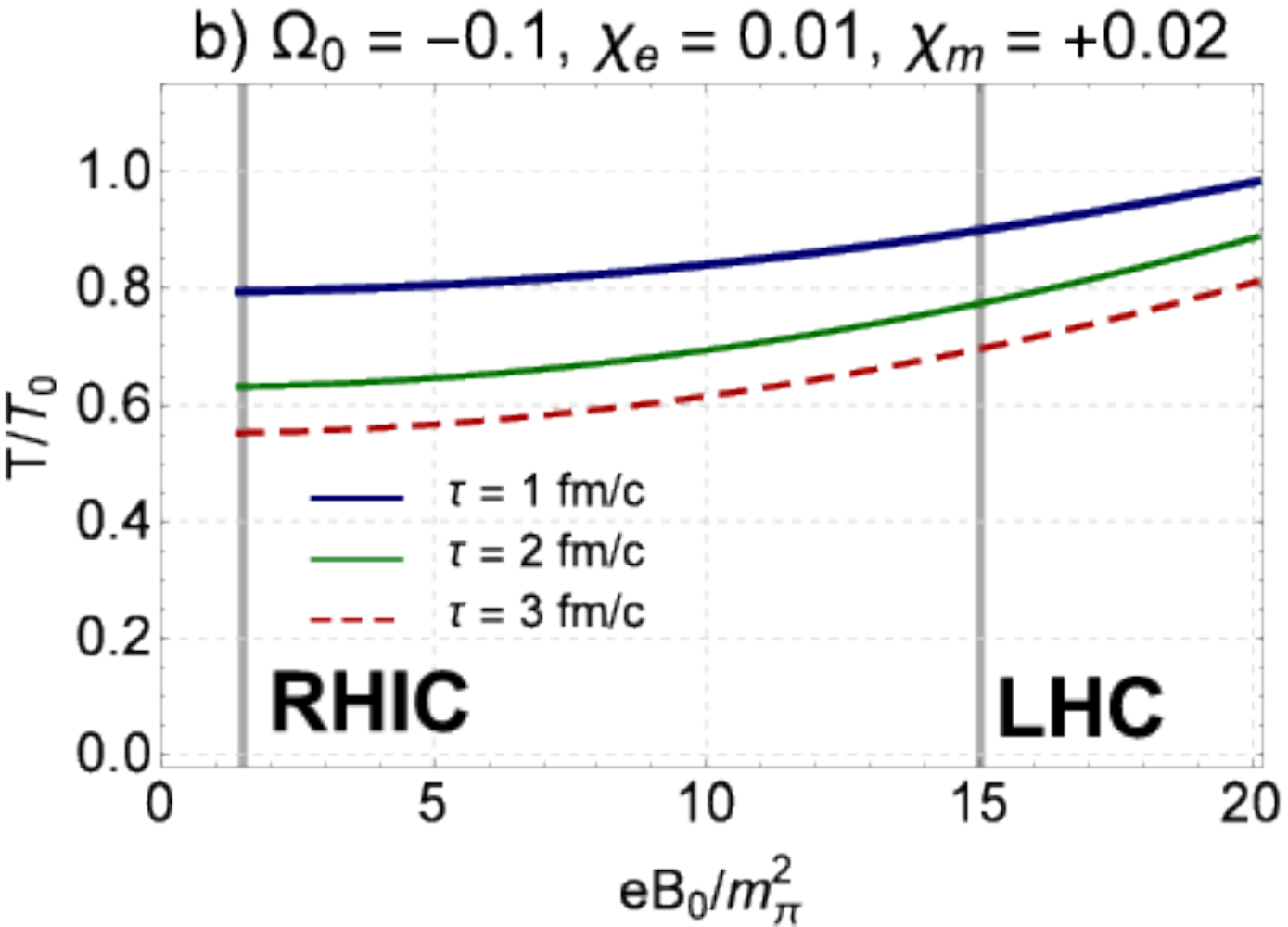}
\includegraphics[width=5.6cm,height=4.4cm]{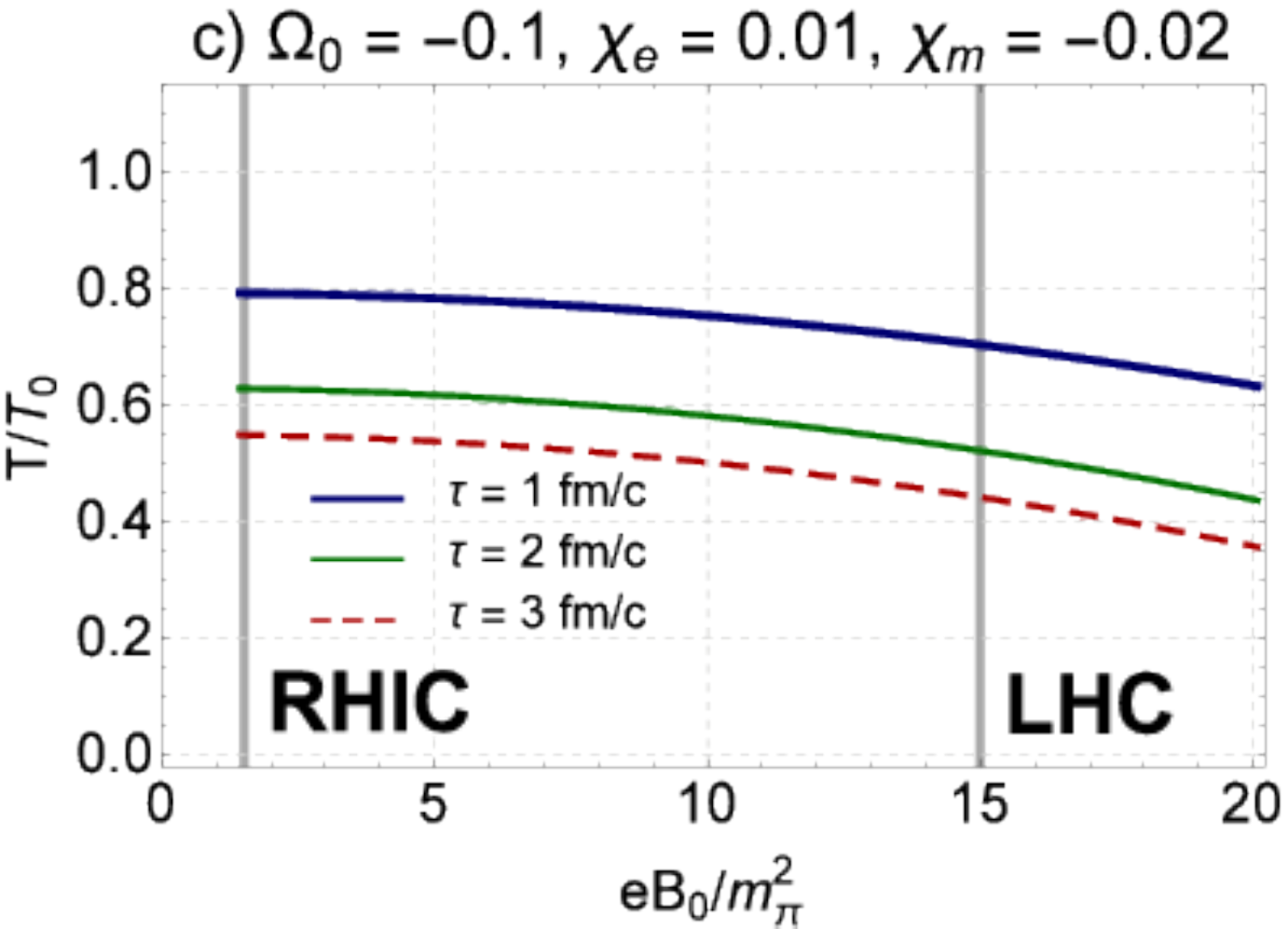}
\includegraphics[width=5.6cm,height=4.4cm]{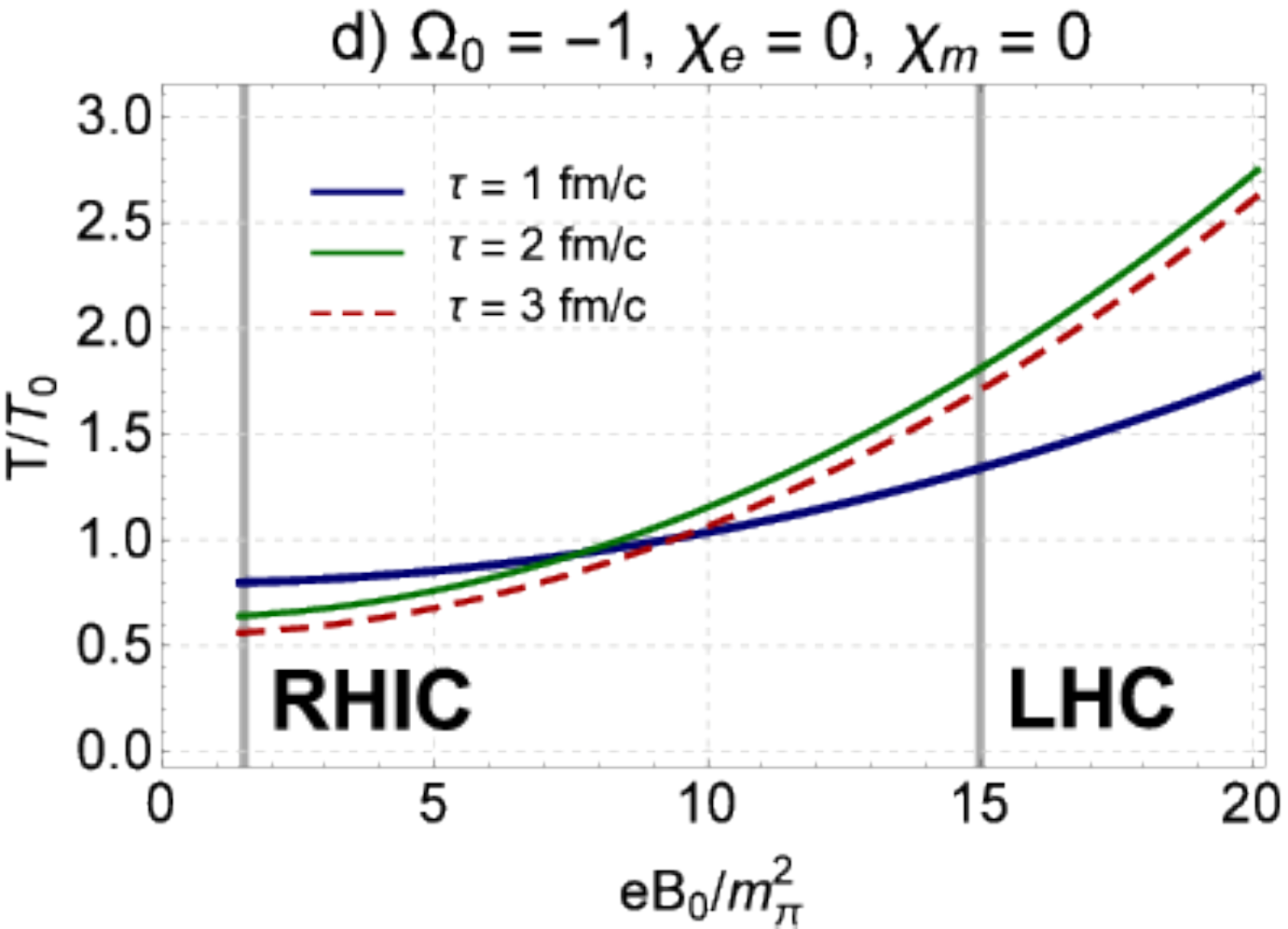}
\includegraphics[width=5.6cm,height=4.4cm]{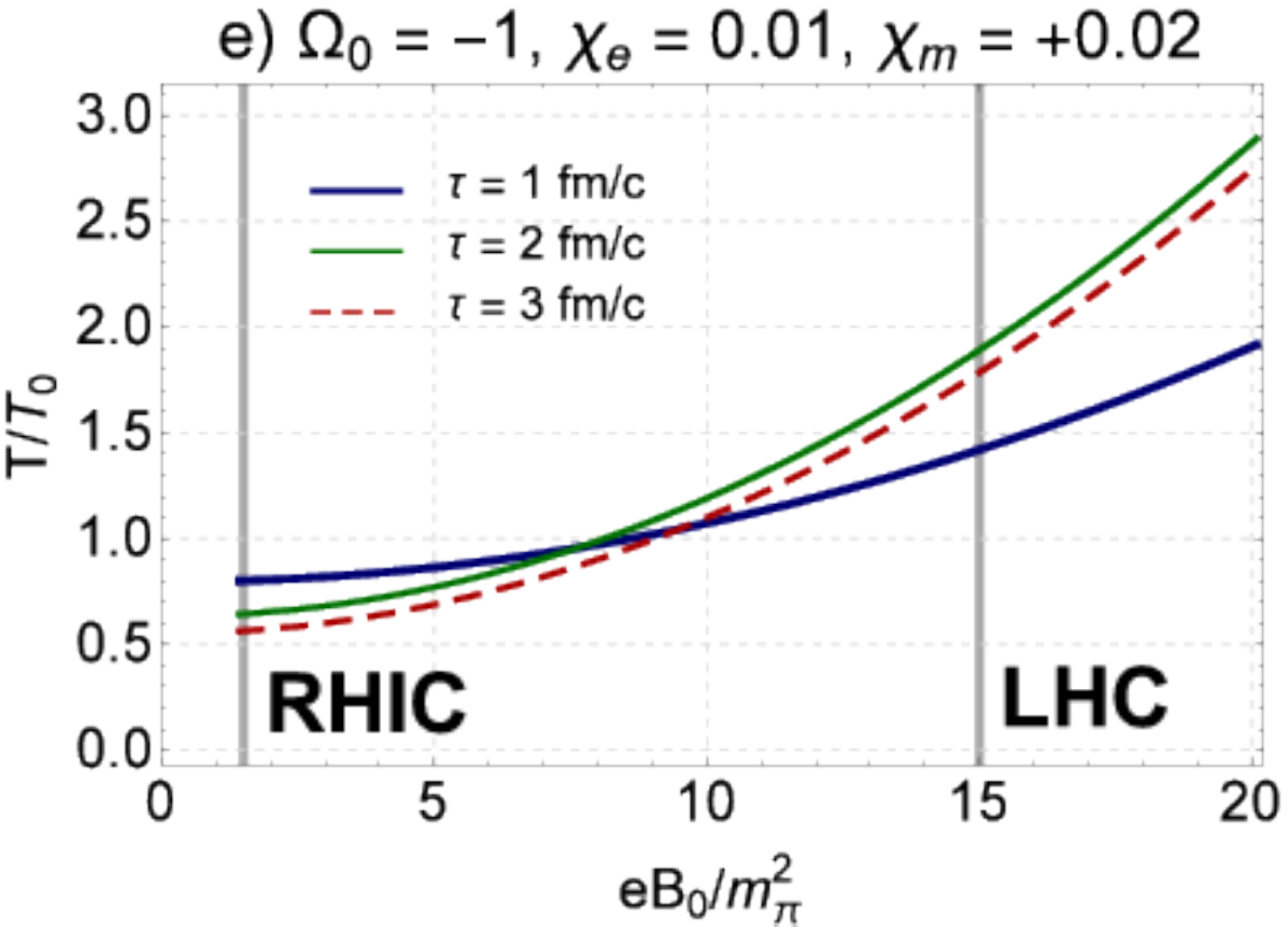}
\includegraphics[width=5.6cm,height=4.4cm]{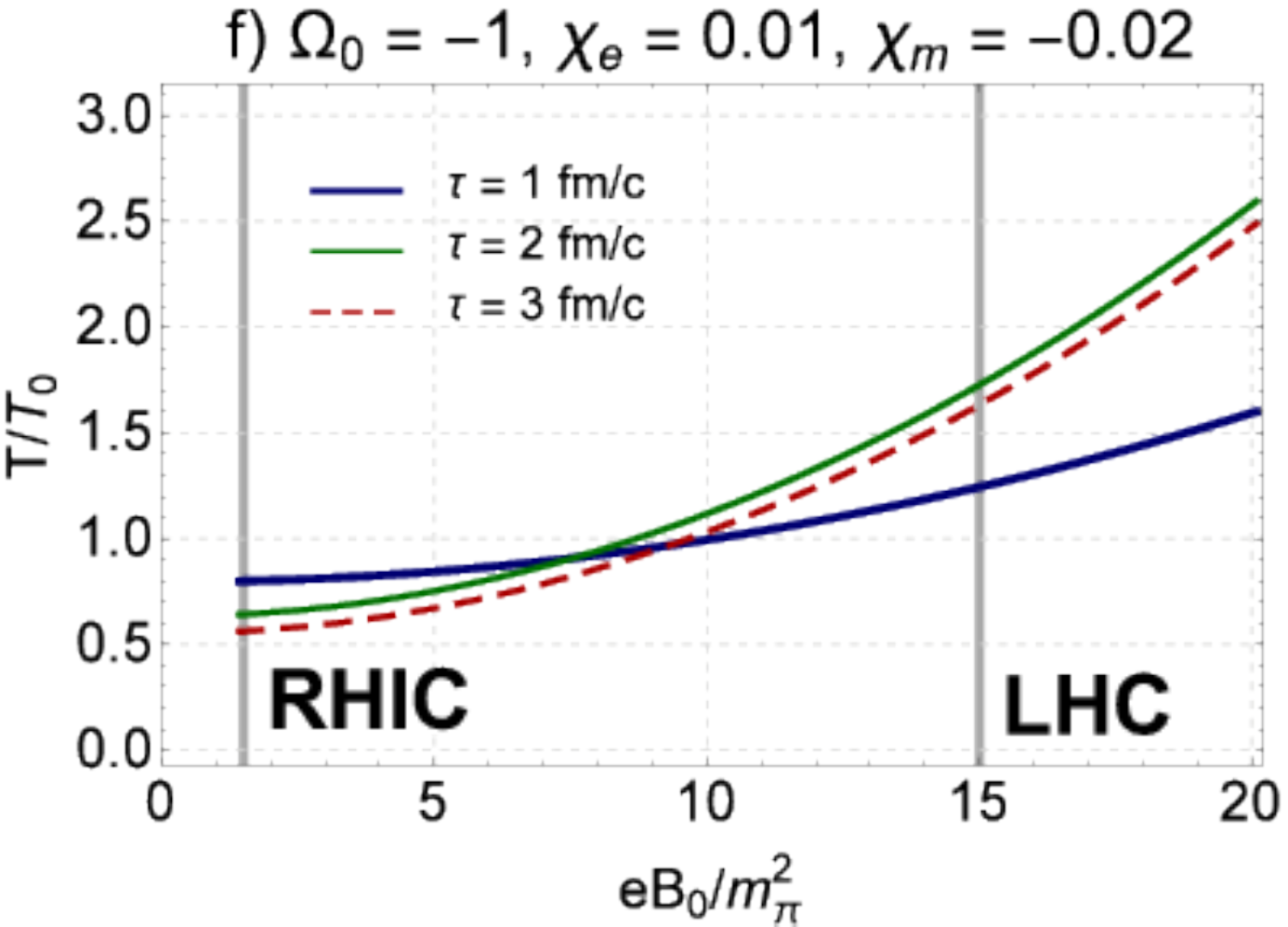}
\caption{(color online). The $eB_{0}/m_{\pi}^{2}$ dependence of $T/T_{0}$ is plotted at $\tau=1,2,3$ fm/c (blue, green solid curves and and red dashed curves). The sets of free parameters  (\ref{OO21}) and (\ref{OO22}) are used to determine $T/T_{0}$ from the numerical solution to (\ref{A37}). The amplitude $T/T_{0}$ is strongly affected by angular velocity $\Omega_0$ as well as electric and magnetic susceptibilities $\chi_{e}$ and $\chi_{m}$.}\label{fig17}
\end{figure*}
%%%%%%%%%%%%%%%% fig 18
\begin{figure*}[hbt]
\includegraphics[width=5.6cm,height=4.4cm]{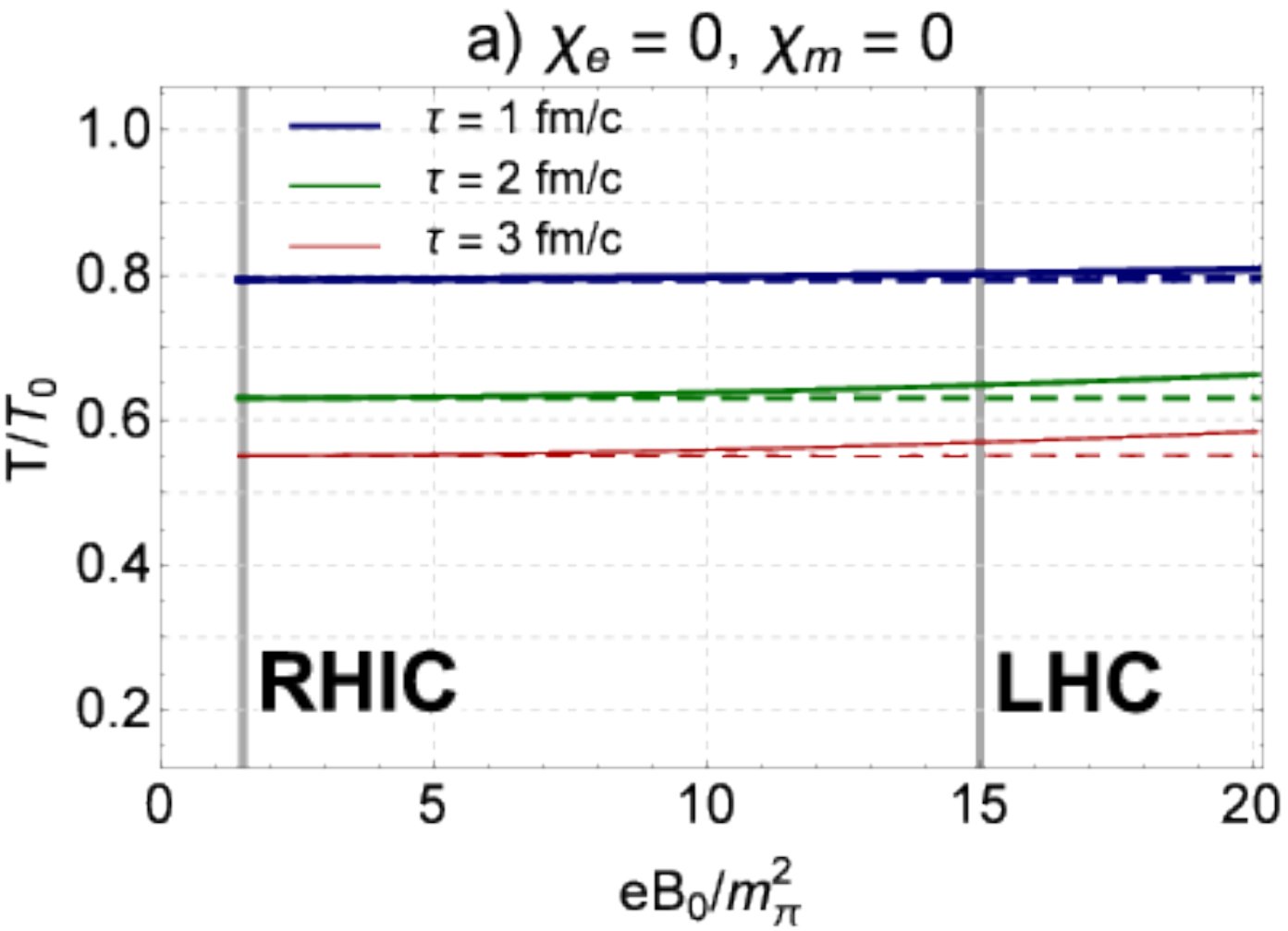}
\includegraphics[width=5.6cm,height=4.4cm]{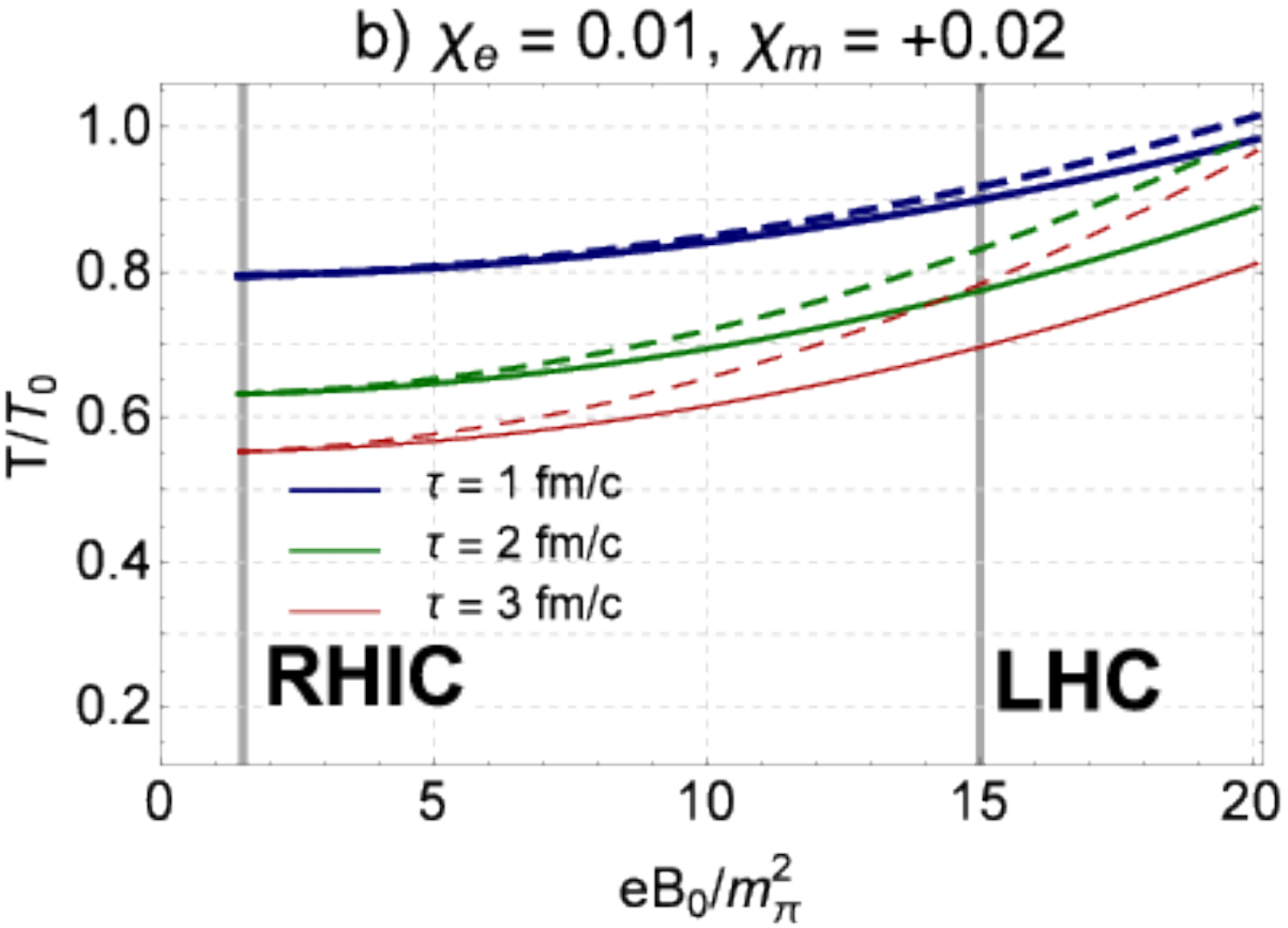}
\includegraphics[width=5.6cm,height=4.4cm]{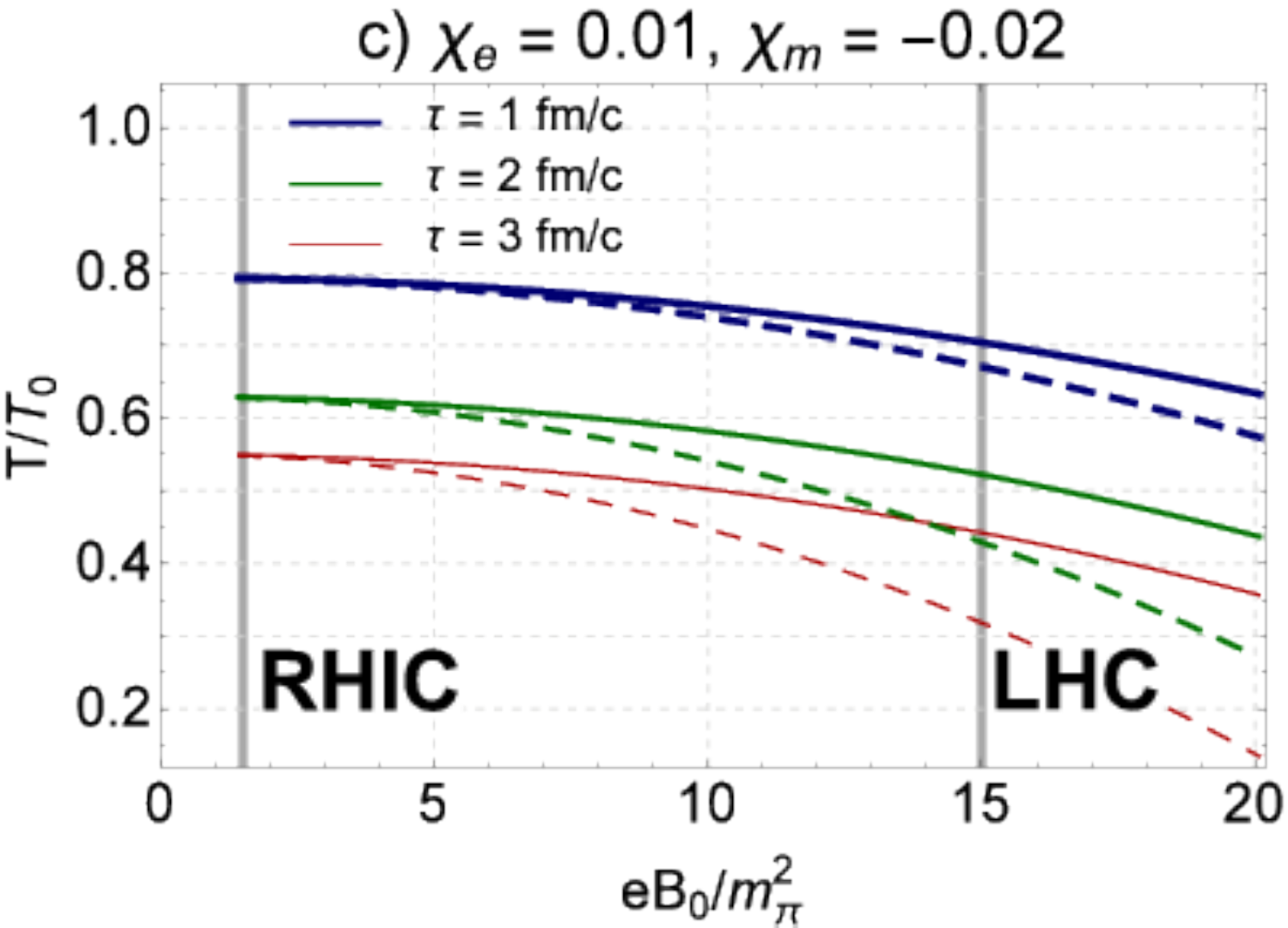}
\caption{(color online). The $eB_{0}/m_{\pi}^{2}$ dependence of $T/T_{0}$ corresponding to rotating (solid curves) and non-rotating (dashed curves) solutions are plotted at $\tau=1,2,3$ fm/c (blue, green and red solid curves). Here, we used the sets of free parameters (\ref{OO21}) and (\ref{OO22}). For the rotating solution, we have particularly used $\Omega_{0}=-0.1$. Apart from different behavior of $T/T_{0}$ for vanishing and non-vanishing susceptibilities $\chi_{e}$ and $\chi_{m}$, the deviation of non-rotating from the rotating solutions strongly depends on whether the fluid is para- or diamagnetic.  } \label{fig18}
\end{figure*}
%%%%%%%%%%%%%%%%%%%%
\subsection{Effects of $\Omega_{0}, \sigma$, $\chi_{m}$ and $\sigma_{0}$ on $B,E$ and $T$ fields}\label{sec5C}
%%%%%%%%%%%%%%%%%%%%
In this section, we will study the dependence of numerical results for $B,E$ and $T$ on the angular velocity $\Omega_{0}=\ell\omega_{0}$, the electric conductivity $\sigma$, the magnetic susceptibility $\chi_{m}$ and $\sigma_{0}=\frac{B_{0}^{2}}{\epsilon_{0}}$. The latter is originally introduced in \cite{rischke2015}, and is a measure for the strength of the magnetic field at $\tau_{0}$. We will focus on the effects of these parameters on the $\tau$ dependence of electromagnetic fields $B,E$ and temperature $T$. We will also study the effect of $\Omega_{0},\sigma,\chi_{m}$ and $\sigma_{0}$ on the behavior of $B/B_{0}, E/E_{0}$ and $T/T_{0}$ for fixed proper time points. As aforementioned, for the choice of free parameters, we have strongly oriented ourselves to sets which may be relevant for QGP.
%%%
\subsubsection{$\Omega_{0}$ dependence of $B,E$ and $T$}\label{sec5C1}
%%%
Let us start by exploring the effect of angular velocity $\Omega_{0}$ on $B,E$ and $T$. In Fig. \ref{fig9}, the $\tau$ dependence of $B/B_{0}$ [Fig. \ref{fig9}(a)], $E/E_{0}$ [Fig. \ref{fig9}(b)], and $T/T_{0}$ [Fig. \ref{fig9}(c)] are plotted for four different sets of free parameters with fixed $\{\tau_{0},\sigma_{0}, \sigma,\beta_{0},\chi_{e},\chi_{m}\}$ and different $\Omega_{0}$s. The $\Omega_{0}$ sets, denoted by O-sets in Fig. \ref{fig9}, are characterized by
\begin{eqnarray}\label{OO9}
\lefteqn{\hspace{-1cm}
\{\tau_{0},\sigma_{0}, \sigma,\Omega_{0},\beta_{0},\chi_{e},\chi_{m}\}
}\nonumber\\
&&\hspace{-1cm}=\{0.5~\mbox{fm/c}, 10, 400~\mbox{MeVc},\Omega_{0},0.01,0,0\},
\end{eqnarray}
with
\begin{eqnarray}\label{OO10}
\left\{
\begin{array}{rl}
\mbox{O-Set 1}:&\Omega_{0}=- 0.1,\\
\mbox{O-Set 2}:&\Omega_{0}=- 0.5,\\
\mbox{O-Set 3}:&\Omega_{0}=- 1.0,\\
\mbox{O-Set 4}:&\Omega_{0}=- 1.5.
\end{array}
\right.
\end{eqnarray}
We exclusively work with negative $\Omega_{0}=\ell\omega_{0}$, because, as it turns out, positive $\Omega_{0}$ leads to unphysical negative amplitudes for $E/E_{0}$ [see, in particular, Fig. \ref{fig11}(b) and our explanations in Appendix]. Assuming $\omega_{0}$ to be positive,  $\Omega_{0}<0$ corresponds to anti-parallel $\mathbf{B}$ and $\mathbf{E}$ fields.
As it is shown in Fig. \ref{fig9}(a), for our specific choice of free parameters, $B/B_{0}$ monotonically decreases with increasing $\tau$, while $E/E_{0}$ exhibits a certain peak in the proper time interval $0.5<\tau<2$ fm/c, and rapidly decreases for $\tau\geq 2$ fm/c [see  Fig. \ref{fig9}(b)]. As concerns the effect of $\Omega_{0}$ on the lifetime of $B/B_{0}$ and $E/E_{0}$, it turns out that the lifetime of $B/B_{0}$ increases with increasing $\Omega_{0}$, or equivalently decreasing $\omega_{0}$, while faster rotating electric fields, with larger $\omega_{0}$, have larger lifetimes. The position and amplitude of $E$-peaks depend also on $\omega_{0}$; as it is shown in Fig. \ref{fig9}(b), for larger $\omega_{0}$, the $E$-peaks arise with larger amplitudes at later proper times.
\par
As concerns the $\tau$ dependence of $T/T_{0}$, demonstrated in Fig. \ref{fig9}(c), $T$-peaks arise only for $\omega_{0}\geq 0.5$. In contrast to $E$-peaks, $T$-peaks occur at $\tau\geq 2$ fm/c. Similar to $E$-peaks, the positions and amplitudes of $T$-peaks are affected by $\omega_{0}$; for larger $\omega_{0}$, the $T$-peaks arise with larger amplitude at later proper times. After these peaks, $T$ decreases slowly at $\tau\geq 3$ fm/c to $T\approx T_{0}$. The slope of this temperature decrease is slightly affected by $\omega_{0}$. However, since $T$-peaks are higher for larger $\omega_{0}$, the system remains longer hot for faster rotating $B$ and $E$ fields, e.g., for $\omega_{0}=1.5$, $T\sim 1.5 T_{0}$ at $\tau\sim 10$ fm/c, while for $\omega_{0}=0.5$, $T\sim 0.5 T_{0}$ at the same $\tau\sim 10$ fm/c. The fact that in the realistic QGP the temperature decreases to values $T<T_{0}$ within $\tau\sim 10$ fm/c indicates that $\omega_{0}$ either vanishes or is small ($\omega_{0}\approx 0.1$). Let us notice that this conclusion is only true for transverse MHD within our above mentioned approximations. In Appendix, we will analyze the general behavior of the solutions of (\ref{A37}), and present a detailed discussion on $E$ and $T$ repeaking.
\par
To study the effect of susceptibilities, $\chi_{e}$ and $\chi_{m}$, the $\tau$ dependence of $T/T_{0}$ is plotted in Fig. \ref{fig10} for the set of free parameters
\begin{eqnarray}\label{OO11}
\hspace{-1cm}\{\sigma_{0},\sigma,\beta_{0},\chi_{e}\}=\{10,400~\mbox{MeVc},0.01,0.01\},
\end{eqnarray}
with three different sets of $\chi_{m}$,
\begin{eqnarray}\label{OO12}
\begin{array}{rclcl}
\chi_{m}&=&0,&\quad& \mbox{(thick solid curves)},\\
\chi_{m}&=&+0.01,&\quad& \mbox{(thin solid curves)},\\
\chi_{m}&=&-0.01,&\quad& \mbox{(dashed curves)},\\
\end{array}
\end{eqnarray}
and $\Omega_{0}=-0.2$ [Fig. \ref{fig10}(a)] and $\Omega_{0}=-1$ [Fig. \ref{fig10}(b)].\footnote{We will later show, that the effect of $\chi_{m}$ on the $\tau$ dependence of $B/B_{0}$ and $E/E_{0}$ can be neglected. This is why we have focused, at this stage, only on the effect of susceptibilities on $T/T_{0}$.} As expected from the results of Fig. \ref{fig9}(c), $T$-peaks appear only for large $\omega_{0}=1$ [Fig. \ref{fig10}(b)]. Moreover, it turns out that for a fixed $\tau$, $T/T_{0}$ increases (decreases) for positive (negative) $\chi_{m}$. The shape of $T/T_{0}$ is, however, not affected by non-vanishing susceptibilities. The results presented in Fig. \ref{fig10}, arising within our aforementioned approximations, thus show that whereas $T$ remains longer high in a paramagnetic fluid, with $\chi_{m}>0$, a diamagnetic fluid, with $\chi_{m}<0$, faster cools. This result is independent on the choice of $\Omega_{0}$.
\par
In Figs. \ref{fig11}(a)-(c), the $\Omega_{0}$ dependence of $B/B_{0}, E/E_{0}$ and $T/T_{0}$ are plotted for fixed $\tau=1$ fm/c (blue curves) $\tau=2$ fm/c (green curves). Here, we have used the set of free parameters
\begin{eqnarray}\label{OO13}
\lefteqn{\hspace{-1cm}
\{\tau_{0},\sigma_{0}, \sigma,\beta_{0},\chi_{e}\}
}\nonumber\\
&&=\{0.5~\mbox{fm/c}, 10, 400~\mbox{MeVc},0.01,0.01\},
\end{eqnarray}
with three different sets of $\chi_{m}$,
\begin{eqnarray}\label{OO14}
\begin{array}{rclcl}
\chi_{m}&=&0,&\quad& \mbox{(thick solid curves)},\\
\chi_{m}&=&+0.01,&\quad& \mbox{(thin solid curves)},\\
\chi_{m}&=&-0.01,&\quad& \mbox{(dashed curves)}.\\
\end{array}
\end{eqnarray}
These plots show that $B/B_{0}$ and $T/T_{0}$ are even in $\Omega_{0}$, while $E/E_{0}$ changes its sign by flipping the sign of $\Omega_{0}$ from negative to positive. This behavior originates from the fact that $E$ is essentially determined by $E=\frac{B}{\Omega_{0}}\frac{d{\cal{M}}}{du}$ [see (\ref{A52})]. Bearing in mind that ${\cal{M}}$ arises from the master equation (\ref{A37}), which is even in $\Omega_{0}$,\footnote{In (\ref{A37}),  $\omega_{0}^{2}=\ell^{2}\Omega_{0}^{2}=\Omega_{0}^{2}$.} $E$ turns out to be odd in $\Omega_{0}$, as it is shown in Fig. \ref{fig11}(b).  Let us notice that since $E=|\mathbf{E}|$ is always positive, the regime $\Omega_{0}>0$, where $E/E_{0}$ becomes negative, is to be excluded [see also Appendix for a more detailed analysis of rotating solutions for $B,E$ and $T$]. A comparison between the curves for positive and negative $\Omega_{0}$ shows that
for negative $\Omega_{0}$, whereas $B/B_{0}$ increases with decreasing $\omega_{0}$, $E/E_{0}$ and $T/T_{0}$ increase with decreasing $\omega_{0}$.
The dependence of $B/B_{0}$ and $E/E_{0}$ on $\Omega_{0}<0$, described above, are indeed expected, because the larger $\omega_{0}$, the faster the $\mathbf{B}$ and $\mathbf{E}$ rotate in a medium with temperature $T$. In this case, whereas the magnitude of the magnetic field $B$ decreases, $E$ becomes larger, and the energy is thus pumped into the medium whose temperature increases consequently.
\par
As it is shown in Fig. \ref{fig10}, the effects of $\chi_{m}$ on $B/B_{0}$ and $T/T_{0}$ are similar, and differ from the effect of $\chi_{m}$ on $E/E_{0}$: For $\Omega_{0}<0$, at each fixed proper time, the amplitudes of  $B/B_{0}$ and $T/T_{0}$ become larger for $\chi_{m}>0$ (paramagnetic fluid) and smaller for $\chi_{m}<0$ (diamagnetic fluid). In contrary, the amplitude of $E/E_{0}$ becomes larger for $\chi_{m}>0$ (paramagnetic fluid) and smaller for $\chi_{m}<0$ (diamagnetic fluid).
\par
Let us notice that in some specific regions of $\Omega_{0}$ and for some specific choices of $\chi_{m}$, $T/T_{0}$ becomes unphysically negative [see Fig. \ref{fig11}(c), where $T/T_{0}$ becomes negative for $\chi_{m}=-0.2$ in the regime $-0.5,\Omega_{0}<+0.5$].\footnote{This does not happen for more relevant values of $\chi_{m}\sim -0.01$.} This regime of parameters is to be excluded from the parameter space.
We shall also notice that $\Omega_{0}=0$ is to be excluded from the plots of Fig. \ref{fig11}, because, as it is argued in Sec. \ref{sec4}, this case corresponds to $\frac{d{\cal{M}}}{du}=0$ from (\ref{A36}), and leads to non-rotating parallel or anti-parallel $\mathbf{B}$ and $\mathbf{E}$ fields. The analytical solution of (\ref{A36}), as well as the $\tau$ and $\eta$ evolutions of $\mathbf{B}$ and $\mathbf{E}$ are already presented in that section. In Fig. \ref{fig11}, we have exclusively demonstrated the results arising from numerical solutions to (\ref{A37}), which lead to rotating parallel or anti-parallel $\mathbf{B}$ and $\mathbf{E}$ fields with $\Omega_{0}\neq 0$.
%%%
\subsubsection{$\sigma$ dependence of $B,E$ and $T$}\label{sec5C2}
%%%
In this part, we will focus on the $\sigma$ dependence of $B, E$ and $T$. In Fig. \ref{fig12}, the $\tau$ dependence of $B/B_{0}$ [Fig. \ref{fig12}(a)], $E/E_{0}$ [Fig. \ref{fig12}(b)] and $T/T_{0}$ [Fig. \ref{fig12}(c)] are plotted for four different sets of free parameters with fixed $\{\tau_{0}, \sigma_{0},\Omega_{0},\chi_{e},\chi_{m}\}$ and different $\{\sigma,\beta_{0}\}$. The latter are chosen in a way that  $\sigma\beta_{0}\sim 4$ MeVc. The $\sigma$ sets, denoted by S-sets in Fig. \ref{fig12}, are characterized by
\begin{eqnarray}\label{OO15}
\hspace{-1cm}
\{\tau_{0},\sigma_{0}, \Omega_{0},\chi_{e},\chi_{m}\}
=\{0.5~\mbox{fm/c}, 10,- 0.2,0,0\},
\end{eqnarray}
with
\begin{eqnarray}\label{OO16}
\hspace{-0.5cm}\left\{
\begin{array}{rrclcrcl}
\mbox{S-Set 1}:&\sigma&=& 4~$MeVc$,&&\beta_{0}&=&1,\\
\mbox{S-Set 2}:&\sigma&=& 40~$MeVc$,&&\beta_{0}&=&0.1,\\
\mbox{S-Set 3}:&\sigma&=& 400~$MeVc$,&&\beta_{0}&=&0.01,\\
\mbox{S-Set 4}:&\sigma&=& 4000~$MeVc$,&&\beta_{0}&=&0.001.\\
\end{array}
\right.
\end{eqnarray}
As it is demonstrated in Fig. \ref{fig12}(a), for our specific choice of free parameters, $B/B_{0}$ monotonically decreases to $B\ll B_{0}$. For fixed $\tau$, $B/B_{0}$ essentially increases with increasing $\sigma$. However, no significant difference occurs between the curves corresponding to S-Set 2 (green dashed curve), S-Set 3  (thick blue curve) and S-Set 4 (thin black curve).  They almost coincide. For larger values of $\sigma\geq 40$ MeVc, the lifetime of $B/B_{0}$ becomes larger, as the slopes of the curves corresponding to S-Set $i$, $i=2,3,4$ are smaller than those corresponding to S-Set 1. This is in contrast to the $\tau$ dependence of  $E/E_{0}$, demonstrated in Fig. \ref{fig12}(b).  Whereas certain peaks occurs in $E/E_{0}$ for large $\sigma=400, 4000$ MeVc (thick blue and thin black curves) in the interval $0.5<\tau<2$ fm/c, for small $\sigma=4,40$ MeVc (green and red dashed curves), $E/E_{0}$ monotonically decreases. The positions of the peaks are slightly affected by $\sigma$. However, the peaks become sharper for large $\sigma=4000$ MeVc and $\beta_{0}=0.001$. The lifetime of $E/E_{0}$ are smaller for smaller values of $\sigma$. The $\tau$ dependence of $T/T_{0}$ is demonstrated in Fig. \ref{fig12}(c). As it turns out, $T/T_{0}$ decreases monotonically for all $\sigma$ sets (\ref{OO16}). Moreover, for a fixed $\tau$, $T/T_{0}$ decreases with increasing $\sigma$, and, as it turns out, a fluid with smaller electric conductivity remains longer hot, as the slope of the curves corresponding to small $\sigma$ are significantly smaller than the slopes of curves corresponding to larger $\sigma$.
\par
In Figs. \ref{fig13}(a), \ref{fig13}(b) and \ref{fig14}, the $\sigma$ dependence of $B/B_{0}$, $E/E_{0}$ and $T/T_{0}$ for $\tau=1$ fm/c (blue curves), $\tau=2$ fm/c (green curves) and $\tau=3$ fm/c are plotted. To determine the amplitudes of $B, E$ and $T$ in Figs. \ref{fig13} and \ref{fig14}, we used following sets of free parameters:
\begin{eqnarray}\label{OO17}
\{\tau_{0},\sigma_{0}, \Omega_{0}\}
=\{0.5~\mbox{fm/c}, 10,- 0.2\},
\end{eqnarray}
with $\sigma\in [4,40]$ MeVc, $\beta_{0}=4/\sigma$ and
\begin{eqnarray}\label{OO18}
\{\chi_{e},\chi_{m}\}&=&\{0,0\},\nonumber\\
\{\chi_{e},\chi_{m}\}&=&\{0.01,+0.2\},\nonumber\\
\{\chi_{e},\chi_{m}\}&=&\{0.01,-0.2\}.
\end{eqnarray}
The amplitude of $B/B_{0}$ for each fixed $\tau$ is almost not affected by $\sigma$ [see Fig. \ref{fig13}(a)]. It increases with $\sigma$ in the regime $\sigma\leq 10$ MeVc, and then remains almost constant. Different choices of $\{\chi_{e},\chi_{m}\}$ have also no effects on the $\sigma$ dependence of $B/B_{0}$, as the curves corresponding to the sets (\ref{OO18}) exactly coincide. This is in contrast to the behavior of $E/E_{0}$ for different sets of parameters (\ref{OO17}) and (\ref{OO18}), as it is demonstrated in Fig.  \ref{fig13}(b). Here, thick solid curves correspond to $\{\chi_{e},\chi_{m}\}=\{0,0\}$, thin solid curves to $\{\chi_{e},\chi_{m}\}=\{0.01,+0.2\}$ and dashed curves to $\{\chi_{e},\chi_{m}\}=\{0.01,-0.2\}$. The amplitudes of $E/E_{0}$ essentially increases with increasing $\sigma$. For a fixed $\sigma$, $E/E_{0}$ decreases (increases) in a para- (dia-) magnetic fluid. This effect enhances for larger values of $\sigma$.
\par
The $\sigma$ dependence of $T/T_{0}$ for different sets of parameters (\ref{OO17}) and (\ref{OO18}) and $\tau=1,2,3$ fm/c is plotted in Fig. \ref{fig14}. Apart from the fact that for $\{\chi_{e},\chi_{m}\}=\{0.01,-0.2\}$, $T/T_{0}$ becomes unphysically negative, the amplitudes of $T/T_{0}$ decrease with increasing $\sigma$ for $4\leq \sigma\leq 20$ MeVc. For $\sigma>20$ MeVc, however, $T/T_{0}$ does not change with increasing $\sigma$.
%%%
\subsubsection{$\chi_{m}$ dependence of $B,E$ and $T$}\label{sec5C3}
%%%
To explore the $\chi_{m}$ dependence of $B,E$ and $T$,  we have plotted  in Fig. \ref{fig15}, $B/B_{0}$, $E/E_{0}$ and $T/T_{0}$ for two sets of free parameters
\begin{eqnarray}\label{OO19}
\lefteqn{\hspace{-0.9cm}
\{\tau_{0}, \sigma_{0},\sigma,\Omega_{0}, \beta_{0},\chi_{e}\}
}\nonumber\\
&&\hspace{-1cm}=\{0.5~\mbox{fm/c},10, 400~\mbox{MeVc}, -0.1, 0.01, 0.01\},
\end{eqnarray}
[see Figs. \ref{fig15}(a)-\ref{fig15}(c)] and
\begin{eqnarray}\label{OO20}
\lefteqn{\hspace{-0.9cm}
\{\tau_{0}, \sigma_{0},\sigma,\Omega_{0}, \beta_{0},\chi_{e}\}
}\nonumber\\
&&\hspace{-0.8cm}=\{0.5~\mbox{fm/c}, 10, 400~\mbox{MeVc}, -1, 0.01, 0.01\},
\end{eqnarray}
[see Figs. \ref{fig15} (d)-\ref{fig15}(f)] and for $\chi_{m}\in [-0.1,+0.1]$. This is the interval which may be relevant for QGP. As it is shown in Figs. \ref{fig15}(a) and \ref{fig15}(d),$B/B_{0}$ is almost not affected by $\chi_{m}$. The same is also true for $E/E_{0}$ [see Figs. \ref{fig15}(b) and \ref{fig15}(e)]. For $\tau=1$ fm/c, $E/E_{0}$ decreases with increasing $\chi_{m}$, but remains almost constant for $\tau=2,3$ fm/c. Apart from the fact that at each fixed $\tau$, $E/E_{0}$ becomes larger when $\omega_{0}$ increases from $\omega_{0}=0.1$ to $\omega_{0}=1$, the $\chi_{m}$ dependence of $E/E_{0}$ remains almost unaffected by $\Omega_{0}$. Comparing with $B/B_{0}$ and $E/E_{0}$, $T/T_{0}$ is strongly affected by $\chi_{m}$. As it is shown in Figs. \ref{fig15}(c) and \ref{fig15}(f),
$T/T_{0}$ increases, in general, with increasing $\chi_{m}$.
Moreover, as expected, the temporal sequence of $T/T_{0}$ changes for faster rotating electromagnetic fields [compare this sequence in Figs. \ref{fig15}(c) and \ref{fig15}(f)].  Let us notice that a change in the temporal sequence of $T/T_{0}$ in Fig. \ref{fig15}(c) comparing with Fig. \ref{fig15}(f) is mainly caused by the appearance of a $T$-peak in the regime $1<\tau<2$ fm/c for large $\omega_{0}$. The same effect is occurred by the plots demonstrated in Fig. \ref{fig10}, where, comparing with the case of $\Omega_{0}=-0.1$ in Fig. \ref{fig10}(a), for $\Omega_{0}=-1$ in Fig. \ref{fig10}(b) a $T$-peak arises.
%%%
\subsubsection{$\sigma_{0}$ dependence of $B,E$ and $T$}\label{sec5C4}
%%%
As it turns out, the numerical solutions of $B$ and $E$ are independent of $\sigma_{0}=\frac{B_{0}^{2}}{\epsilon_{0}}$. We thus focus, in this part, on the dependence of $T/T_{0}$ on $\sigma_{0}$.\footnote{See (\ref{A34}) from which $T/T_{0}$ is determined through (\ref{A29}). Here, it is enough to replace $\frac{E_{0}^{2}}{\epsilon_{0}}$ with $\frac{E_{0}^{2}}{\epsilon_{0}}=\sigma_{0}\beta_{0}^{2}$.} This seems to be interesting also with regard to the evolution of the temperature in HIC experiments: As aforementioned, it is believed that in HIC experiments very strong magnetic fields are created at early stages of the collision (small $\tau$). Depending on the impact parameter and collision energy, their strengths at RHIC and LHC are estimated to be $eB_{0}\sim 1.5 m_{\pi}^{2}$ and $eB_{0}\sim 15 m_{\pi}^{2}$, respectively \cite{warringa2007, skokov2009}. In what follows, after presenting the $\sigma_{0}$ dependence of $T$, we will use the dependence of $\sigma_{0}$ on $B_{0}$, and will plot, in particular, $T/T_{0}$ as a function of  $eB_{0}/m_{\pi}^{2}$ for non-vanishing angular velocity and electric as well as magnetic susceptibilities. To emphasize the effect of rotating electromagnetic fields, we will also compare these results with the corresponding results for $T/T_{0}$ from non-rotating solutions, previously presented in Sec. \ref{sec4A}.
\par
In Fig. \ref{fig16}, the $\sigma_{0}$ dependence of $T/T_{0}$ is plotted for $\tau=1,2,3$ fm/c (blue and green solid and red dashed curves). To solve (\ref{A37}), we have used the set of free parameters
\begin{eqnarray}\label{OO21}
\hspace{-0.5cm}\{\tau_{0}, \sigma, \beta_{0}\}
=\{0.5~\mbox{fm/c}, 400~\mbox{MeVc}, 0.01\},
\end{eqnarray}
with
\begin{eqnarray}\label{OO22}
\{\chi_{e},\chi_{m}\}&=&\{0,0\},\nonumber\\
\{\chi_{e},\chi_{m}\}&=&\{0.01,+0.02\},\nonumber\\
\{\chi_{e},\chi_{m}\}&=&\{0.01,-0.02\},
\end{eqnarray}
and
$\Omega_{0}=-0.1$ [Figs. \ref{fig16}(a)-\ref{fig16}(c)] as well as $\Omega_{0}=-1$ [Figs. \ref{fig16}(d)-\ref{fig16}(f)].
\par
The results presented in Fig. \ref{fig16} show that the $\sigma_{0}$ dependence of $T/T_{0}$ is strongly affected by $\Omega_{0}$, $\chi_{e}$ and $\chi_{m}$; Let us first consider the case of small $\omega_{0}=0.1$ in Figs. \ref{fig16}(a)-\ref{fig16}(c). The slope of $T/T_{0}$ is affected by susceptibilities: Whereas $T/T_{0}$ increases with increasing $\sigma_{0}$ for $\{\chi_{e},\chi_{m}\}=\{0,0\}$ and $\{\chi_{e},\chi_{m}\}=\{0.01,+0.02\}$ in a paramagnetic fluid [see Figs. \ref{fig16}(a) and \ref{fig16}(b)], it decreases with increasing  $\sigma_{0}$ in a diamagnetic fluid with
$\{\chi_{e},\chi_{m}\}=\{0.01,-0.02\}$ [see Fig. \ref{fig16}(c)].
For large $\omega_{0}=1$, a completely different picture arises. Here, the effect of large angular velocity dominates the above mentioned effect of non-vanishing susceptibilities. As it is shown in Figs. \ref{fig16}(d)-\ref{fig16}(f), $T/T_{0}$ increases with increasing $\sigma_{0}$ for all sets of $\chi_{e}$ and $\chi_{m}$ from (\ref{OO22}). Moreover, as expected, a change in the temporal sequence of $T/T_{0}$ amplitudes occurs for large $\omega_{0}=1$ in Figs. \ref{fig16}(d)-\ref{fig16}(f) comparing with small $\omega_{0}=0.1$ from Figs. \ref{fig16}(a)-\ref{fig16}(c). The same behavior was previously observed in Fig. \ref{fig10} and Fig. \ref{fig15}(c) in comparison with Fig. \ref{fig15}(f).
\par
In Fig. \ref{fig17}, we have plotted the same numerical results from Fig. \ref{fig16} in terms of $eB_{0}/m_{\pi}^{2}$ instead of $\sigma_{0}$. The former seems to be a more appropriate measure in relation to HIC experiments. The aim is to look for a possibility to emphasize the effects of $\Omega_{0}$, $\chi_{e}$ and $\chi_{m}$ in a more phenomenological language.
\par
To evaluate $T/T_{0}$ in terms of $eB_{0}/m_{\pi}^{2}$, it is necessary to express $eB_{0}$ in terms of $\sigma_{0}$. This is given by
\begin{eqnarray}\label{OO23}
\frac{eB_{0}}{m_{\pi}^{2}}\sim 1.34\left(\epsilon_{0}\sigma_{0}\right)^{1/2}.
\end{eqnarray}
To arrive at (\ref{OO23}), let us remind that $\sigma_{0}=\frac{(eB_{0})^{2}}{e^{2}\epsilon_{0}}$ is dimensionless, provided $eB_{0}$ is in GeV$^{2}$ and $\epsilon_{0}$ in GeV fm$^{-3}\sim 8\times 10^{-3}$ GeV$^{4}$. Using $m_{\pi}=0.14$ GeV, we have $1$ GeV$^{2}\sim 50 m_{\pi}^{2}$. Replacing $e^{2}=4\pi\alpha_{e}\sim 0.09$ with the fine structure constant $\alpha_{e}=1/137$, we arrive at (\ref{OO23}). Here, $\epsilon_{0}$ is in GeVfm$^{-3}$. For the energy density $\epsilon_{0}\sim 10$ GeVfm$^{-3}$ arising in a typical Au-Au collision with impact parameter $\sim 10$ fm and $\sqrt{s_{NN}}\sim 200$ GeV \cite{rischke2016}, we have
\begin{eqnarray}\label{O10}
\frac{eB_{0}}{m_{\pi}^{2}}\sim 4.5 \sigma_{0}^{1/2}.
\end{eqnarray}
The results presented in Fig. \ref{fig17} have essentially the same feature as the results presented in Fig. \ref{fig16}. Vertical thick solid lines in the plots of Fig. \ref{fig17} denote the values of $eB_{0}$ at RHIC and LHC in $m_{\pi}^{2}$ units (see above).
Remarkable is the difference between the dependence of $T/T_{0}$ on $eB_{0}/m_{\pi}^{2}$ for vanishing and non-vanishing susceptibilities for small angular velocity $\omega_{0}=0.1$. In this case, whereas for a fluid with $\chi_{m}=0$, $T$ remains almost constant with increasing  $eB_{0}/m_{\pi}^{2}$, for a para- (dia-) magnetic fluid with positive (negative) $\chi_{m}$, $T/T_{0}$ increases (decreases) with increasing $eB_{0}/m_{\pi}^{2}$. In contrast for large angular velocity $\omega_{0}=1$, $T/T_{0}$ increases for all values of $\chi_{m}$.
\par
To emphasize the effect of the rotation of electromagnetic fields on the dependence of $T/T_{0}$ on $eB_{0}/m_{\pi}^{2}$, we have plotted in Fig. \ref{fig18}, the $eB_{0}/m_{\pi}^{2}$ dependence of $T/T_{0}$ for vanishing and non-vanishing $\Omega_{0}$. For non-rotating solutions of $T/T_{0}$, we have used the analytical result (\ref{A29}) with $\mathbb{V}=1$ and $\exp\left(\frac{\cal{L}}{\kappa}\right)$  given in (\ref{D10}). For rotating solutions, the same numerical results for $T/T_{0}$ previously demonstrated in Figs. \ref{fig16} and \ref{fig17} are used. In both cases the set of parameters (\ref{OO21}) and (\ref{OO22}) are applied. For the rotating solution, we set, in particular, $\Omega_{0}=-0.1$. The blue, green and red solid (dashed) curves correspond to rotating (non-rotating) solutions. For $\{\chi_{e},\chi_{m}\}=\{0,0\}$ in Fig. \ref{fig18}(a), the non-rotating solution is slightly deviated from the numerical solution with small $\omega_{0}=0.1$. In both cases amplitude $T/T_{0}$ remains almost constant once  $eB_{0}/m_{\pi}^{2}$ is increased. In a paramagnetic fluid with $\{\chi_{e},\chi_{m}\}=\{0.01,+0.02\}$, however, in both rotating and non-rotating cases, $T/T_{0}$ increases with increasing $eB_{0}/m_{\pi}^{2}$, while the deviation of non-rotating solutions [dashed curves in Fig. \ref{fig18}(b)] from rotating solutions [solid curves in Fig. \ref{fig18}(b)]  is positive. The opposite is true for a diamagnetic fluid with $\{\chi_{e},\chi_{m}\}=\{0.01,-0.02\}$. As it is demonstrated in Fig. \ref{fig18}(c), the fluid becomes cooler once $eB_{0}/m_{\pi}^{2}$ increases. However, for non-vanishing $\omega_{0}$, the slope of $T/T_{0}$ as a function of $eB_{0}/m_{\pi}^{2}$ is smaller than that for vanishing $\omega_{0}$ [compare the slope of solid and dashed curves in Fig. \ref{fig18}(c)]. These results, together with the data of $T/T_{0}$ from RHIC and LHC may provide an experimental tool to check whether in HIC experiments, like those at RHIC and LHC, the angular velocity of $\omega_{0}$ vanishes or not.
%%%%%%%%%%%%%%%%%%%%%%%%%%%%%%%%%%%%%%
\section{Summary and conclusions}\label{sec6}
\setcounter{equation}{0}
%%%%%%%%%%%%%%%%%%%%%%%%%%%%%%%%%%%%%%
The search for self-similar analytical solutions of RIHD exhibiting various geometrical symmetry properties has attracted much attention in recent years \cite{kodama2015}. Being, in particular, non-boost-invariant, they represent extensions to the well-known one-dimensional, longitudinally boost-invariant Bjorken flow of RIHD \cite{hwa1974, bjorken1982}. The goal is, among others, to develop new analytical solutions, which overcome the shortcomings of Bjorken flow in reproducing experimental data of RHIC and LHC. Here, very large magnetic fields are shown to be created during early stages of HICs. Numerous attempts have already been undertaken to study the impact of these magnetic fields on electromagnetic and thermal properties of QGP created in HICs. Bearing in mind that the electromagnetic properties of QGP, such as its electric conductivity or its response to external electromagnetic fields, may
elongate the lifetime of the magnetic fields produced in these collisions, they may affect the evolution of thermodynamic quantities, such as the temperature, pressure and energy density of QGP.
\par
Motivated by the above facts, the boost-invariant motion of an ideal magnetized fluid is recently studied in the framework of ideal transverse MHD \cite{rischke2015,rischke2016}. It is shown that the (proper) time evolution of the energy density of the fluid depends on whether the magnetic field decays as  $B(\tau)\sim \tau^{-a}$ with the free parameter $a$ being either $a=1$ or $a\neq 1$.
In the present work, we have extended the studies performed in \cite{rischke2015,rischke2016} to the case of non-ideal transverse MHD, where electric conductivity as well as electric and magnetic susceptibilities of the fluid are assumed to be finite. The aim was to study the evolution of electromagnetic fields satisfying Maxwell and MHD equations. Assuming the electric and magnetic fields to be transverse to the fluid velocity, and parameterizing the corresponding partial differential equations in terms of $B=|\mathbf{B}|$
and $E=|\mathbf{E}|$ as well as $\phi$ and $\zeta$, their angles to a certain fixed axis in the LRF of the fluid, we arrived at two distinct differential equations for a certain function ${\cal{M}}$, which appears in the ``inhomogeneous'' continuity equation $\partial^{\mu}(Bu_{\mu})=B\tder{\cal{M}}$. For any ${\cal{M}}\neq 0$, the latter represents a deviation from the frozen flux theorem.
Whereas boost-invariant solutions for the evolution of $B, E$ and $T$ arise from $\frac{d{\cal{M}}}{du}=0$, another, second-order quadratic differential equation for ${\cal{M}}$ eventually leads to non-boost-invariant solutions to $B,E$ and $T$. The latters are essentially characterized by rotating, parallel or anti-parallel electric and magnetic fields. The rotation occurs with increasing rapidity $\eta$ and with a constant angular velocity $\omega_{0}=\frac{d\phi}{d\eta}=\frac{d\zeta}{d\eta}$. Exact analytical solution for non-rotating electromagnetic fields arises once ${\cal{M}}$ satisfies $\frac{d{\cal{M}}}{du}=0$. Other approximate analytical solutions arise for rotating electric and magnetic fields with $E/B$ being constant (power-law solution) or ${\cal{M}}$ being linear in $\omega_{0} $ (slowly rotating solution). We have, in particular, shown that the power-law decay $B(\tau)\sim \tau^{-a}$ with $a\neq 1$, previously introduced in \cite{rischke2015,rischke2016} as an example for the violation of frozen flux theorem in ideal transverse MHD, naturally arises as one of the approximate solutions of rotating electromagnetic fields in the framework of non-ideal transverse MHD with constant $E/B$. We have also shown that ${\cal{M}}$ plays a major role in determining the thermodynamic fields $T,p$ and $\epsilon$, which exhibit self-similar solutions arising from our method of self-similar solutions of non-conserved charges.
\par
Choosing appropriate sets of free parameters for electric conductivity $\sigma$ and susceptibilities $\chi_{e},\chi_{m}$ of the electromagnetized fluid, we numerically studied the solutions to the second-order, non-linear differential equations for ${\cal{M}}$ with a given $\Omega_{0}=\ell \omega_{0}$. Here, $\ell=\pm 1$ indicates parallel ($\ell=+1$) or anti-parallel ($\ell=-1$) electric and magnetic vectors.
We compared these numerical solutions with the approximate power-law and slowly rotating solutions for $B,E$ and $T$, arising from non-trivial solutions of ${\cal{M}}$, and checked, in this way, the reliability of these approximations.
We further focused on the interplay between the angular velocity $\omega_{0}$ and $\sigma,\chi_{m}$ in connection with their potential effects on the lifetime of $B$ and $E$ fields as well as the evolution of the temperature $T$. We have shown that  for large enough $\omega_{0}$, $E$ and $T$ exhibits certain peaks at early times after the collision, whereas $B$ monotonically decays. In a general analysis of the solutions to the master equation for ${\cal{M}}$, we also discussed the conditions under which these kinds of peaks occur (see Appendix). The effects of $\Omega_{0},\sigma$ and $\chi_{m}$ on the amplitudes $B/B_{0}, E/E_{0}$ and $T/T_{0}$ for fixed proper times $\tau$ have also been separately studied.
We have, in particular, shown that for free parameters chosen according to their relevance for QGP produced in HICs, $\Omega_{0}>0$ leads to unphysical negative values for $E/E_{0}$. This indicates that within our transverse MHD approximations $\mathbf{B}$ and $\mathbf{E}$ in these kinds of experiments have to be anti-parallel to each other.
\par
We have further considered the dependence of $B,E$ and $T$ on the phenomenologically relevant parameter $\sigma_{0}=\frac{B_{0}^{2}}{\epsilon_{0}}$, which appears also in \cite{rischke2015,rischke2016}. Through its dependence on the magnetic field $B_{0}$ and energy density $\epsilon_{0}$ at the initial (proper) time $\tau_{0}$, we  plotted the temperature gradient of the electromagnetized fluid as a function of $eB_{0}/m_{\pi}^{2}$. We have shown that for small values of $\omega_{0}$, the $eB_{0}/m_{\pi}^{2}$ dependence of $T/T_{0}$ at fixed $\tau\gtrsim\tau_{0}$ is significantly affected by $\chi_{m}$, whereas for large values of $\omega_{0}$, $T/T_{0}$ increases with increasing $eB_{0}/m_{\pi}^{2}$ for all values of $\chi_{m}=0, \chi_{m}>0$ and $\chi_{m}<0$. Bearing in mind that the magnetic fields created in HICs are estimated to be  $B\sim 1.5 m_{\pi}^{2}$ at RHIC and $15 m_{\pi}^{2}$ at LHC, a possible difference between the temperature of QGP at RHIC and LHC may provide information about the onset of rotation of electromagnetic fields in these kinds of experiments.
\par
Let us finally note that the method of self-similar solutions for non-conserved charges, developed and used in the present work, is derived under the assumption of a simple EoS, $\epsilon=\kappa p$ with constant $\kappa=c_{s}^{-2}=3$, which is only valid in the ultrarelativistic limit \cite{rischke2015,rischke2016}. The above results can thus be improved by choosing more realistic EoS, arising, e.g. from lattice QCD, where $c_{s}$ turns out to be $T$-dependent.
Apart from the sound velocity $c_{s}$, the electric conductivity and magnetic susceptibility, $\sigma$ and $\chi_m$, can also be chosen to be $T$-dependent.
In a magnetized medium, a dependence of $c_{s}, \sigma$ and $\chi_{m}$ on $B$ and $E$ are also possible. All these may lead to more complicated differential equations for ${\cal{M}}$, which eventually results in more realistic results for the $\tau$ dependence of $B,E$ and $T$. We will postpone these studies to our future works.
%%%%%%%%%%%%%%%%%%%%%%%%%%
\section{Acknowledgments}\label{sec7}
%%%%%%%%%%%%%%%%%%%%%%%%%%
The authors thank F. Taghinavaz and S.M.A. Tabatabaee for useful discussions.
%%%%%%%%%%%%%%%%%%%%%%%%%%%%%%%%%%%%%%
\begin{appendix}
\section*{Appendix: General analysis of the solutions to the master equation for ${\cal{M}}$}\label{appA}
\setcounter{section}{1}
\setcounter{equation}{0}
%%%%%%%%%%%%%%%%%%%%%%%%%%%%%%%%%%%%%%
The analysis of the master equation without actually solving it gives us important insights about the qualitative behavior of $B$, $E$ and $T$ fields. One of the accessible results is the prediction of repeaking in $B,E$ and $T$, i.e. the appearance of maxima after the initial time. These kinds of maxima do not occur in ideal MHD. Interestingly, for $\chi_m>1$, there is also possible for the above fields to have a minimum before rising to a peak. However, as far as the HIC physics is concerned, $\chi_m>1$ is not physically relevant. In what follows, we find the necessary conditions for a repeaking of $E$ and $T$ fields, and prove that in the physically relevant case of $\chi_m<1$ and $\chi_e>-1$,  we must have $\Omega_0\equiv\ell\omega_0<0$. This guarantees $B,E$ and $T$ to be positive. We will, in particular, show that for $\chi_{m}<1$, $B$ is monotonically decreasing, and find certain conditions for which $E$ and $T$ have only one single maximum shortly after the initial time. To this purpose, we will first prove a number of Lemmas.
%\newtheorem{lemma}{Lemma}
%%%%%%%%%%%%%%%%%%%%%%%
\vspace{0.3cm}\\
\textbf{Lemma 1:}\label{lema_1}
	\textit{For $\tilde{u}$ being the extrema of
	\be\label{appA1}
	f(\tau)=f_0\left(\frac{\tau_0}{\tau}\right)e^{\lambda(\tau)},
	\ee
	we have
	\be\label{appA2}
	\deriv{\lambda}{u}\at[\Big]{\tilde{u}} = 1.
	\ee
	In addition, $\tilde{u}$ is a maximum (minimum) if $\derivn[2]{\lambda}{u}\at[\big]{\tilde{u}}$ is negative (positive). Here, $u=\ln\left(\frac{\tau}{\tau_{0}}\right)$. For $f=\{B,E,T\}$, we have $\lambda=\{{\cal{M}},{\cal{N}},{\cal{L}}\}$, respectively. }
\vspace{0.3cm}\\
\textit{Proof:} The proof is straightforward. One first finds the derivative with respect to $\tau$ in terms of derivative with respect to $u$. By setting the first derivative equal to zero, (\ref{appA2}) is derived. The second derivative test then translates into the second claim.
%%%%%%%%%%%%%%%%%%%%%%%
\vspace{0.3cm}\\
\textbf{Lemma 2:}\label{lema_2}
	\textit{A differentiable function $f(u)$ either does not have two subsequent extrema of the same kind (maximum or minimum) or is constant in-between.}
\vspace{0.3cm}\\
\textit{Proof:}
	Consider two extrema of the same kind at points $u_1$ and $u_2$. Then, for some $\epsilon$ we have $f'(u_1+\epsilon)f'(u_2-\epsilon)<0$, and thus another extremum exists in the interval between $u_1$ and $u_2$. This is either of the same or opposite kind. If it is of the same kind, this procedure can be repeated until an extremum of opposite kind is found or $f'(u)=0$ for all points $u\in [u_1,u_2]$.
%%%%%%%%%%%%%%%%%%%%%%%
\vspace{0.3cm}\\
\textbf{Lemma 3:}\label{lema_3}
	\textit{If the sign of second derivative of a differentiable function $f(u)$ in its possible extremum is forced to be negative and non-zero, then
	\begin{enumerate}
		\item If $\left(\deriv{f}{u}\at[\Big]{0}\right)\left(\deriv{f}{u}\at[\Big]{u\gg 1}\right)<0$, $f$ has exactly one maximum somewhere in $[0,\infty)$.
		\item If $\left(\deriv{f}{u}\at[\Big]{0}\right)\left(\deriv{f}{u}\at[\Big]{u\gg 1}\right)>0$, $f$ is monotonically decreasing or increasing.
	\end{enumerate}
}
\vspace{0.3cm}\par\noindent
\textit{Proof:} If the second derivative is negative at any possible extremum, then the function is neither constant nor have a minimum by lemma 2. By Bolzano's theorem, the function has a maximum in $[0,\infty)$ if the first derivative has opposite signs in $0$ and $u\gg 1$. Now consider the case that the derivative is negative both initially ($u=0$) and asymptotically ($u\to \infty$), and assume that $f'(u)=0$ at some point $u^\star$. Then, $u^\star$ needs to be a maximum of $f'(u)$. If it is not, then there exists a point such that $f'(u)>0$, and therefore $f'(u)$ vanishes in another point other than $u^\star$. This is not possible by lemma 2. Being a maximum of $f'(u)$, we have $f''(u)=0$ at $u^\star$. This is again not possible, and therefore $f(u)$ is monotonically decreasing. A similar argument shows that $f(u)$ is monotonically increasing if the first derivative is positive both initially and asymptotically.
%%%%%%%%%%%%%%%%%%%%%%%
\vspace{0.3cm}\\
\textbf{Lemma 4:}\label{lema_4}
	\textit{At the initial time, i.e. $u=0$, the derivatives of functions of interest are given by
\begin{eqnarray}
\lefteqn{\hspace{-0.4cm}
\deriv{\mathcal{M}}{u}\at[\Big]{u=0}=\beta_0\Omega_0,} \label{appA3} \\
\lefteqn{\hspace{-0.4cm}
\derivn[2]{\mathcal{M}}{u}\at[\Big]{u=0}
}\nonumber\\
\hspace{-0.8cm}&&=-\bigg\{\frac{\Omega_0\beta_0\sigma\tau_0+\Omega_0^2[1-\chi_m+
\beta_0^2(1+\chi_e)]}{1+\chi_e}\bigg\},\nonumber\\
\label{appA4}\\
\lefteqn{\hspace{-0.4cm}
\deriv{\mathcal{N}}{u}\at[\Big]{u=0}
=-\bigg[\frac{\Omega_0(1-\chi_m)+\beta_0\sigma\tau_0}{\beta_0(1+\chi_e)}\bigg],}\label{appA5}\\
\lefteqn{\hspace{-0.4cm}
\deriv{\mathcal{L}}{u}\at[\Big]{u=0}
= \frac{\sigma_0}{c_s^2}\bigg\{\bigg[\sigma\tau_0
}
\nonumber\\
&& +
\chi_e\left(\frac{\Omega_0(1-\chi_m)+\beta_0(1+\sigma\tau_0+\chi_e)}{\beta_0(1+\chi_e)}\right)\bigg]\beta_0^2\nonumber\\
&&+\chi_m\left(1-\beta_0\Omega_0\right)\bigg\}
\label{appA6}.
\end{eqnarray}
}
\vspace{0.3cm}\\
\textit{Proof:}
	The first relation \eqref{appA3} was already used  in Sec. \ref{sec3} [see (\ref{A52}) and evaluate it at $u=0$]. Plugging $\deriv{\cal{M}}{u}$ from (\ref{appA3}) into (\ref{A37}), (\ref{appA4}) is found. As concerns (\ref{appA5}, one uses
	\begin{eqnarray}\label{appA7}
	\deriv{\mathcal{M}}{u}\deriv{\mathcal{N}}{u} = \derivn[2]{\mathcal{M}}{u}+\left(\deriv{\mathcal{M}}{u}\right)^2,
		\end{eqnarray}
	along with earlier results, and arrive at (\ref{appA5}).
	Finally, from (\ref{A27}),	one finds
	\begin{eqnarray}\label{appA8}
	\deriv{\mathcal{L}}{u} &=& \frac{\kappa B^2}{\epsilon}\bigg\{\bigg[\sigma\tau-\chi_e\left(\deriv{\mathcal{N}}{u}-1\right)\bigg]\frac{E^2}{B^2}\nonumber\\
&&+\chi_m\left(1-\frac{E}{B}\Omega_0\right)\bigg\},
	\end{eqnarray}
	which at $u=0$ yields the desired relation (\ref{appA6}).
%%%%%%%%%%%%%%%%%%%%%%%
\vspace{0.3cm}\\
\textbf{Lemma 5:}\label{lema_5}
	\textit{The asymptotic behavior of quantities of interest at $u\gg 1$ are given by
	\begin{eqnarray}	
	\hspace{-0.8cm}\deriv{\mathcal{M}}{u}\bigg|_{u\gg 1}&\sim& -\frac{\Omega_0^2(1-\chi_m)}{\sigma\tau_0}e^{-u},\label{appA9}\\
		\hspace{-0.8cm}\derivn[2]{\mathcal{M}}{u}\bigg|_{u\gg 1}&\sim&-\deriv{\mathcal{M}}{u},\label{appA10}\\
		\hspace{-0.8cm}\mathcal{M}(u)\bigg|_{u\gg 1}&\sim&-\frac{\Omega_0^2(1-\chi_m)}{\sigma\tau_0}(1-e^{-u}),\label{appA11}\\
		\hspace{-0.8cm}\deriv{\mathcal{N}}{u}\bigg|_{u\gg 1}&\sim&-1-\frac{\Omega_0^2(1-\chi_m)}{\sigma\tau_0}e^{-u},\label{appA12}\\
		\hspace{-0.8cm}\mathcal{N}(u)\bigg|_{u\gg 1}&\sim&-u-\frac{\Omega_0^2(1-\chi_m)}{\sigma\tau_0}\left(1-e^{-u}\right),\label{appA13}
\end{eqnarray}
and
\begin{eqnarray}\label{appA14}
\lefteqn{e^{\mathcal{L}/\kappa}\bigg|_{u\gg 1}\sim 1+\sigma_0\exp\left(-\frac{2(1-\chi_m)\Omega_0^2}{\sigma\tau_0}\right)
}\nonumber\\
&&\times\bigg\{\frac{\chi_m}{1-c_s^2}\left(1-e^{-(1-c_s^2)u}\right)\nonumber\\
&&+\inv{2-c_s^2}\left(\beta_0^2\sigma\tau_0+\frac{\chi_m\Omega_0^2(1-\chi_m)}{\sigma\tau_0}\right)\nonumber\\
%&&\times\nonumber\\
&&\times \left(1-e^{-(2-c_s^2)u}\right)+\frac{2\beta_0^2\chi_e}{3-c_s^2}\left(1-e^{-(3-c_s^2)u}\right)\nonumber\\
&&+\frac{\beta_0^2\chi_e\Omega_0^2(1-\chi_m)}{\sigma\tau_0(4-c_s^2)}\left(1-e^{-(4-c_s^2)u}\right)\bigg\},
\end{eqnarray}
as well as
\begin{eqnarray}\label{appA15}
	\derivv{u}\left(e^{\mathcal{L}/\kappa}\right)\bigg|_{u\gg 1}&\sim& \chi_m\sigma_0\exp\left(-\frac{2(1-\chi_m)\Omega_0^2}{\sigma\tau_0}\right)\nonumber\\
&&\times e^{-(1-c_s^2)u}.
	\end{eqnarray}
}
\vspace{0.3cm}\\
\textit{Proof:}
	We start by inspecting the master equation (\ref{A37}) at $u\gg 1$. To do so, we re-evaluate it in terms of $w\equiv \inv{u}$. Using
	\be\label{appA16}
	\deriv{\mathcal{M}}{u}=-w^2\deriv{\mathcal{M}}{w},\quad\frac{\rd{}^2 \mathcal{M}}{\rd{u}^2}=w^3\left(2\deriv{\mathcal{M}}{w}+w\frac{\rd{}^2 \mathcal{M}}{\rd{w}^2}\right),
	\ee
     keeping non-vanishing terms in $w\to 0$, and rewriting the result in terms of $u$, we find the asymptotic equation as
	\be\label{appA17}
	\deriv{\mathcal{M}}{u}e^{u}\sigma\tau_0+\omega_0^2(1-\chi_m)=0.
	\ee
This immediately leads to  (\ref{appA9}). Equations (\ref{appA10}) and (\ref{appA11}) are found by differentiating and integrating (\ref{appA9}) with respect to $u$. Here, ${\cal{M}}(0)=0$ is assumed. Using (\ref{appA7}), and earlier results, we arrive at  (\ref{appA12}). The result is then integrated to give (\ref{appA13}). We use ${\cal{N}}(0)=0$.
	In order to find (\ref{appA14}), we first rewrite \eqref{A34} as
	\begin{eqnarray}\label{appA18}
e^{\mathcal{L}/\kappa}&=&1+\sigma_0\bigg[\sigma\tau_0\beta_0^2
\int_{0}^{u}\rd{u'}e^{2\mathcal{N}(u')+c_s^2 u'}\nonumber\\
&&+\chi_e\beta_0^2\int_{0}^{u}\rd{u'}\left(1-\deriv{\mathcal{N}}{u'}\right)e^{2\mathcal{N}(u')-(1-c_s^2) u'}
	\nonumber\\
	&&+\chi_m\int_{0}^{u}\rd{u'}\left(1-\deriv{\mathcal{M}}{u'}\right)e^{2\mathcal{M}(u')-(1-c_s^2) u'}\bigg].\nonumber\\
	\end{eqnarray}
Then, plugging earlier results into (\ref{appA18}), and performing the corresponding integrals, we arrive at (\ref{appA14}). Here,
\begin{eqnarray*}
e^{2{\cal{N}}(u')}\sim e^{-2u'-\frac{2\Omega_{0}^{2}(1-\chi_{m})}{\sigma\tau_{0}}},
\end{eqnarray*}
is used. Taking finally the derivative of (\ref{appA14}), all terms except the first one are suppressed at $u\gg 1$. We thus arrive at (\ref{appA15}).
\par
We are now in the position to analyze the rotating solutions for $B,E$ and $T$ in following theorems:
%%%%%%%%%%%%%%%%%%%%%%%
\vspace{0.3cm}\\
\textbf{Theorem 1:}
	\textit{In order for the system to be physical (i.e. $E>0$)
	\be\label{appA19}
	\Omega_0(1-\chi_m) < 0.
	\ee	
	In other words, the sign of $\deriv{\mathcal{M}}{u}$ does not change during the time evolution.
}
\vspace{0.3cm}\\
\textit{Proof:}
From (\ref{A52}), it is evident that, in order for $E$ to be non-negative, we must have
  \be\label{appA20}
 \frac{1}{\Omega_0} \deriv{\cal{M}}{u} > 0,
  \ee
provided $B>0$.
In other words, $\deriv{\cal{M}}{u}$ does not change sign in the whole interval of $u$. Moreover, according to (\ref{appA9}), for $\chi_m < 1$ ($\chi_m > 1$), $\deriv{\cal{M}}{u}$ becomes asymptotically negative (positive). Multiplying (\ref{appA9}) with $1/\Omega_{0}$, we obtain
$$
\frac{1}{\Omega_0} \frac{d{\cal{M}}}{du}\sim -\Omega_{0}(1-\chi_{m})\xi,\quad \mbox{with }\quad \xi\equiv\frac{e^{-u}}{\sigma\tau_{0}}>0,
$$
which leads to (\ref{appA19}), upon using (\ref{appA20}).
%%%%%%%%%%%%%%%%%%%%%%%
\vspace{0.3cm}\\
\textbf{Theorem 2:}
	\textit{For $\chi_m < 1$, the magnetic field monotonically decreases.}
\vspace{0.3cm}\\
\textit{Proof:}
According to theorem 1, $\chi_{m}<1$ leads to $\Omega_0 < 0$. Negative $\Omega_{0}$ thus leads to $\deriv{\cal{M}}{u}<0<1, \forall u\in [0,\infty)$ [see (\ref{appA20})]. This shows that $B$ has always a negative derivative, and is thus monotonically decreasing.
%%%%%%%%%%%%%%%%%%%%%%%
\vspace{0.3cm}\\
\textbf{Theorem 3:}
	\textit{For $\chi_m<1$ and $\chi_{e}>-1$, the electric field repeaks exactly once if
	\be\label{appA21}
	\Omega_0<-\frac{\beta_0(1+\chi_e+\sigma\tau_0)}{1-\chi_m}.
	\ee
	Otherwise, it is monotonically decreasing.}
\vspace{0.3cm}\\
\textit{Proof:}
Let us determine the sign of
$$\frac{dE}{du}=E\left(\frac{d{\cal{N}}}{du}-1\right),$$
at $u=0$ and $u\gg 1$ by separately inspecting the sign of $\frac{d{\cal{N}}}{du}-1$ at $u=0$ and $u\gg 1$. Using (\ref{appA5}) and (\ref{appA12}), we have
\begin{eqnarray}\label{appA22}
\frac{d{\cal{N}}}{du}\bigg|_{u=0}-1&\sim&-\frac{[\Omega_{0}(1-\chi_{m})+\beta_{0}\sigma\tau_{0}+\beta_{0}(1+\chi_{e})]}{\beta_{0}(1+\chi_{e})},\nonumber\\
\frac{d{\cal{N}}}{du}\bigg|_{u\gg 1}-1&\sim&-2-\frac{\Omega_{0}^{2}(1-\chi_{m})}{\sigma\tau_{0}}e^{-u}.
\end{eqnarray}
For $\chi_{m}<1$ and $\chi_{e}>-1$, $\frac{d{\cal{N}}}{du}\big|_{u\gg 1}-1$ and thus $\frac{dE}{du}|_{u\gg 1}$ are always negative. Hence, in order for $E$ to repeak only once, $\frac{d{\cal{N}}}{du}\big|_{u=0}-1$ and thus $\frac{dE}{du}|_{u=0}$ are to be positive (see lemma 3). This fixes $\Omega_{0}$ to be
\begin{eqnarray*}
	\Omega_0<-\frac{\beta_0(1+\chi_e+\sigma\tau_0)}{1-\chi_m},
\end{eqnarray*}
as claimed in (\ref{appA21}). Moreover, according to lemma 3, $E$ is monotonically decreasing for
\begin{eqnarray*}
	\Omega_0\geq -\frac{\beta_0(1+\chi_e+\sigma\tau_0)}{1-\chi_m}.
\end{eqnarray*}
This completes the proof.
%%%%%%%%%%%%%%%%%%%%%%%
\vspace{0.3cm}\\
\textbf{Theorem 4:}
\textit{Causality ensures $T\to 0$ when $\tau\gg \tau_{0}$. However, the parameters have to be constrained in order for $T$ to be positive. The constraint is roughly given by $\chi_m \gtrsim -{\sigma_0^{-1}}$.}
\vspace{0.3cm}\\
\textit{Proof:}
According to (\ref{A29}), for $\mathbb{V}=1$, the self-similar solution of $T/T_{0}$ is given by
\begin{eqnarray}\label{appA23}
\frac{T}{T_{0}}=\left(\frac{\tau_{0}}{\tau}\right)^{\frac{1}{\kappa}}e^{\frac{\cal{L}}{\kappa}}.
\end{eqnarray}
As we have seen in (\ref{appA14}), the asymptotic form of $T/T_0$ contains terms of $1-e^{-(k-c_s^2)u}$, with $k=1,\cdots,4$. Since $c_s\leq 1$ by causality, for all $k=1,\cdots, 4$ we have $1-e^{-(k-c_s^2)u}\to 1$ as $u\to\infty$. The factor  $e^{\mathcal{L}/\kappa}$ in (\ref{appA23}) becomes therefore constant at $u\gg 1$. Hence, the behavior of $T/T_0$ is exclusively dictated by the Bjorken factor $\tau^{-c_s^2}$, that vanishes at $\tau\gg \tau_{0}$. We therefore have $T\to 0$ at large $\tau\gg \tau_{0}$, as claimed.
\par
Inspecting now the limit of $e^{\mathcal{L}/\kappa}$ at $u\gg 1$, it may becomes negative, especially for $\chi_m,\chi_{e}< 0$. An exact constraint, which ensures $T$ to be positive reads
\begin{eqnarray}\label{appA24}
\sigma_0\left(\frac{\chi_m}{1-c_s^2}+{\cal{R}}\right)\geq -1,
\end{eqnarray}
with ${\cal{R}}$ defined by
\begin{eqnarray}\label{appA25}
{\cal{R}}&\equiv&\exp\left(-\frac{2(1-\chi_m)\Omega_0^2}{\sigma\tau_0}\right)\nonumber\\
&&\times\bigg[\inv{2-c_s^2}\left(\beta_0^2\sigma\tau_0+\frac{\chi_m\Omega_0^2(1-\chi_m)}{\sigma\tau_0}\right)\nonumber\\
&&+\frac{\beta_0^2\chi_e}{3-c_s^2}+\frac{\beta_0^2\chi_e\Omega_0^2(1-\chi_m)}{\sigma\tau_0(4-c_s^2)}\bigg].
\end{eqnarray}
Assuming
\be\label{appA26}
\frac{\Omega_0^2}{\sigma\tau_0}\ll 1,
\ee
the second and the last term in the bracket appearing in ${\cal{R}}$ are suppressed, and it becomes always positive for  $\chi_{e},\sigma>0$. In this case, neglecting ${\cal{R}}$, the positivity condition (\ref{appA24}) is roughly given by $\chi_m \gtrsim -{\sigma_0^{-1}}$, as claimed.
%%%%%%%%%%%%%%%%%%%%%%%
\vspace{0.3cm}\\
\textbf{Theorem 5:}
\textit{For $\chi_m<1$, the temperature repeaks exactly once if
\begin{eqnarray}\label{appA27}
\lefteqn{
\frac{d{\cal{L}}}{du}\bigg|_{u=0}=\frac{\sigma_0}{c_s^2}\bigg\{\bigg[\sigma\tau_0
}\nonumber\\
&&+\chi_e\left(\frac{\Omega_0(1-\chi_m)+
\beta_0(1+\sigma\tau_0+\chi_e)}{\beta_0(1+\chi_e)}\right)\bigg]\beta_0^2
\nonumber\\
&&+\chi_m\left(1-\beta_0\Omega_0\right)\bigg\}>1.
\end{eqnarray}
}
\vspace{0.3cm}\\
\textit{Proof:}
To determine the sign of
$$\frac{dT}{du}=T\left(\frac{d{\cal{L}}}{du}-1\right),$$
at $u=0$ and $u\gg 1$, we separately inspect the sign
 of $\frac{d{\cal{L}}}{du}-1$
at $u=0$ and $u\gg 1$. According to (\ref{appA15}), $\frac{d}{du}\left(e^{\cal{L}}\right)\to 0$ for $u\gg 1$. This translates into $\frac{d{\cal{L}}}{du}=0$. Hence, in the limit of large $u\gg 1$,  $\frac{dT}{du}|_{u\gg 1}$ is always negative. According to lemma 3, $T$ peaks only once if $\frac{dT}{du}|_{u=0}>0$. Using (\ref{appA6}), we thus arrive at (\ref{appA27}), as claimed.
\end{appendix}

\end{document}